\documentstyle[12pt,titlepage,graphicx,twoside,axodraw]{book}
\begin{document}

\renewcommand{\thepage}{}

\newcommand{\lsim}{\raisebox{-0.13cm}{~\shortstack{$<$ \\[-0.07cm] $\sim$}}~}
\newcommand{\gsim}{\raisebox{-0.13cm}{~\shortstack{$>$ \\[-0.07cm] $\sim$}}~}
\newcommand{\ra}{\rightarrow}
\newcommand{\ee}{e^+e^-}
\newcommand{\s}{\\ \vspace*{-3mm} }
\newcommand{\nn}{\noindent}
\newcommand{\non}{\nonumber}
\newcommand{\beq}{\begin{eqnarray}}
\newcommand{\eeq}{\end{eqnarray}}
\newcommand{\tb}{\tan\beta}
\newcommand{\mx}{m_{\tilde{\chi}_1^0}}
\newcommand{\msqa}{m_{\tilde{f}_1}}
\newcommand{\msqb}{m_{\tilde{f}_2}}
\newcommand{\sqa}{\tilde{f}_1}
\newcommand{\sqb}{\tilde{f}_2}
\newcommand{\sq}{s_{2\theta_{\tilde{f}}}}
\newcommand{\cq}{c_{2\theta_{\tilde{f}}}}
\newcommand{\lsp}{\mbox{$\tilde\chi_1^0$}}


\pagestyle{empty}

\begin{center}
\includegraphics{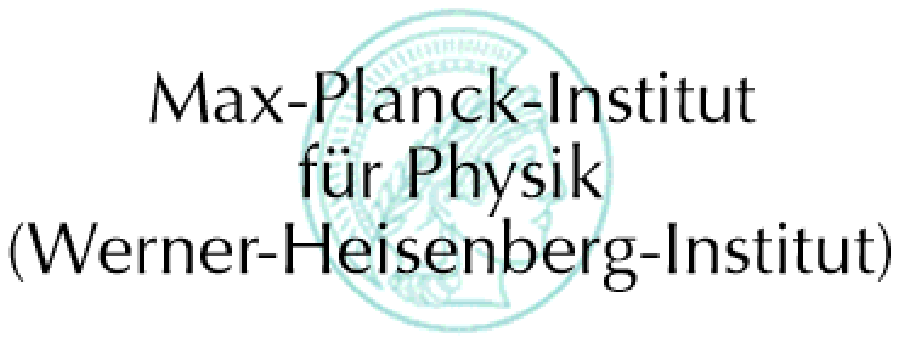}
\end{center}

\vspace{1.0cm}

\begin{center}
{\huge\bf Phenomenological Aspects of }
\\
{\huge \bf Supersymmetric Gauge Theories}
\end{center}

\vspace*{0.5cm}

\begin{center}
{\Large  Pavel Fileviez P\'erez}
\end{center}

\vspace{1.0cm}

\begin{center}
Advisors:
\vspace{0.2cm}\\ 
Prof. Dr. Manuel Drees (TUM, Munich)\\ 
Prof. Dr. Goran Senjanovi\'c (ICTP, Trieste)
\end{center}

\vfill

\begin{center}
\includegraphics{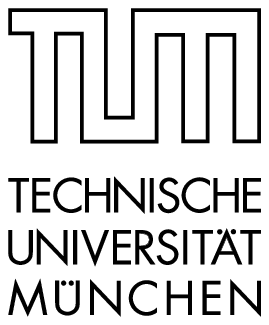}
\hfill 
\includegraphics{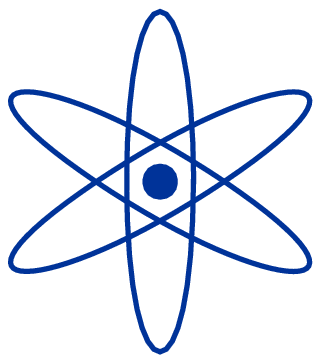}
\end{center}

\newpage

\begin{center}
\includegraphics{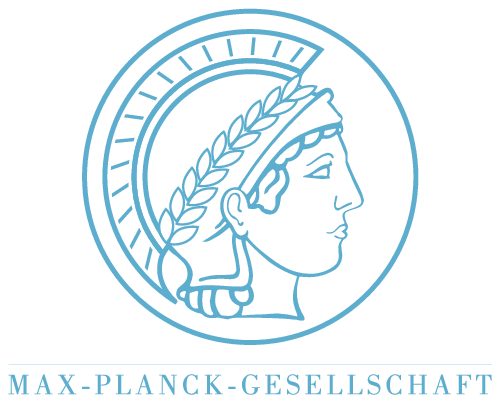}
\end{center}

\newpage

\cleardoublepage

\markboth{}{}

\thispagestyle{empty}

\begin{center}
{\bf Technische Universit\"at M\"unchen}
\\
{\bf Physik Department}
\\
Institut f\"ur Theoretische Physik T30e, Prof. Dr. Manuel Drees.

\vfill

{\huge\bf Phenomenological Aspects of }
\\
{\huge \bf Supersymmetric Gauge Theories}

\vfill

{\Large\sc Pavel Fileviez P\'erez}

\vfill

\end{center}
Vollst\"andiger Abdruck der von der Fakult\"at f\"ur Physik der Technischen Universit\"at M\"unchen zur Erlangung des akademischen Grades eines

\begin{center}
\bf Doktors der Naturwissenschaften (Dr. rer. nat.)
\end{center}
genehmigten Dissertation.

\begin{center}
\begin{tabular}{lll}
Vorsitzender:              &    & Univ.-Prof. Dr. O. Zimmer\\[0.3cm]
Pr\"ufer der Dissertation: & 1. & Univ.-Prof. Dr. M. Drees\\[0.1cm]
		  	   & 2. & Univ.-Prof. Dr. A. J. Buras
\end{tabular}
\end{center}

\noindent 
Die Dissertation wurde am 20. 06. 2003 bei der Technischen
Universit\"at M\"unchen eingereicht und durch die Fakult\"at f\"ur
Physik am  31. 07. 2003 angenommen.
\newpage

\vspace*{\fill}

\newpage

\vspace*{\fill}
\begin{flushright}
{\bf \textit{A mi Madre}}
\end{flushright}

\vspace*{\fill}
\newpage

\vspace*{\fill}

\newpage

\pagestyle{empty}

\begin{center}
\bf Abstract
\end{center}
In this thesis we study two important phenomenological
issues in the context of supersymmetric gauge theories. In the
Minimal Supersymmetric Standard Model (MSSM) we analyze properties  
of the neutral Higgs boson decays into two neutralinos, taking into
account the effect of new quantum corrections. In the second part
of our work we study in detail the Proton decay in the 
Minimal Supersymmetric $SU(5)$ Theory. 
\\
\\
The main new results obtained in our studies are:

\begin{itemize}

\item We compute one-loop corrections to the neutralino
couplings to the Higgs bosons, taking into account contributions 
with fermions and sfermions inside the loop. Our analytical results 
are valid for arbitrary momenta and general sfermion mixings. 
For the neutralino couplings to Higgs bosons, we find in all 
cases corrections of up to a factor of two for reasonable 
values of the input parameters. 

\item The contribution of Bino--like lightest supersymmetric particles 
to the invisible decay width of the lightest MSSM Higgs boson might 
be measurable at future high--energy and high--luminosity $e^+e^-$
colliders, when the new quantum corrections are present. 

\item The masses ($M_T$) of the heavy triplets $T$ and
$\bar T$ responsible for $d=5$ proton decay are computed, when we allow 
for arbitrary trilinear coupling of the heavy fields in $\Sigma$ and 
use higher dimensional terms as a possible source of their masses. 
In this case $M_T$ may go up naturally by a factor of
thirty, which would increase the proton lifetime by a factor of $10^3$.

\item The relation between fermion and/or sfermion masses, and proton
decay is studied in detail. We find the conditions needed to suppress the
$d=5$ contributions to the decay of the proton, in
the context of the Minimal Supersymmetric $SU(5)$ model.

\item We point out that the Minimal Supersymmetric Grand Unified
Theory $SU(5)$ is not ruled out as claimed before. 

\end{itemize}

\tableofcontents

\pagestyle{headings}

\chapter{Introduction}
\pagenumbering{arabic}
In Nature the fundamental interactions are described by gauge
theories. The Standard Model based on the gauge group 
$SU(3)_C \times SU(2)_L \times U(1)_Y$ explains all the
properties of the electroweak and strong interactions, while 
Einstein's Theory of General Relativity describes the
gravitational interaction. 
\\
\\
Grand Unified Theories are the main theories beyond the Standard
Model. They explain the quantization of the electric charge, 
predict the  weak mixing angle, the decay of the Proton, 
the bottom-tau Yukawa coupling unification and the existence of 
magnetic monopoles. At the same time, they provide a natural 
framework for understanding Baryogenesis and/or Leptogenesis, 
and for the implementation of the see-saw mechanism of neutrino masses.
\\
\\
It has been shown that \textit{Supersymmetry} (SUSY) \cite{SUSY1}, a symmetry
between fermions and bosons, plays an important role in the
development of Unified Theories. There are many motivations 
for considering SUSY, the most important one is the possibility to
cancel quadratic divergencies in the self energy of the Standard Model Higgs
boson. If we consider radiative corrections to the Higgs
mass, we see that these are proportional to the fundamental scale
 $( M_{Planck} \sim 10^{18}$ GeV$)$ square,
therefore these corrections can change its value by many orders of magnitude. 
This is the so-called \textit{Hierarchy Problem}. Also 
it is possible to unify at the high scale $M_{GUT} \sim 10^{16}$ GeV 
all the gauge coupling constants of the Minimal Supersymmetric 
Standard Model \cite{Dimop1,Ibanez1,Einhorn1,Marciano}. The possibility to 
break the electroweak $SU(2)_L \times U(1)_Y$  symmetry
of the Standard Model radiatively \cite{Alvarez} is widely regarded as one of the main arguments
in favor of SUSY, since it offers a dynamical explanation for the
mysterious negative mass square of the Higgs boson. SUSY in the context of
the Minimal Supersymmetric Standard Model provides us a candidate to
describe the Non-Baryonic Dark Matter present in our
Universe \cite{Goldberg,Ellis1,SDM}. Another important motivation
is the possibility to cancel the Tachyonic states in String Theory \cite{Polchinski}, the
most popular scenario where all the fundamental interactions are unified. 
\\
\\
Another popular scenario for the solution of the
\textit{Hierarchy Problem} is large extra dimensions, which
for two such new ones may be as large as a fraction of a mm
\cite{Arkani-Hamed:1998rs,Antoniadis:1998ig,Arkani-Hamed:1998nn}.
In this case the field-theory cutoff ($\Lambda_F$) must be
low and experiments demand: $\Lambda_F>(10-100)$ TeV. Clearly,
one then must fine-tune (somewhat) the Higgs mass, since

\begin{equation}
\label{higgs}
m_h^2\approx m_0^2+{3 y_t^2\over 16\pi^2}\Lambda_F^2
\approx({\rm few}\; 100\; {\rm GeV})^2
\end{equation}
where $m_0$ and $y_t$ are the tree level Higgs mass and 
the top Yukawa coupling respectively.
\\
We believe this is acceptable; compared to the fine-tuning problem
when $\Lambda_F$ is pushed to $M_{Planck}$ (or $M_{GUT}$), this is
negligible. What is missing in this program is some serious physical
reason to have $\Lambda_F$ so low. In low-energy 
supersymmetry, where $\Lambda_F$ gets
traded for $\Lambda_{SUSY}$ (here defined as the mass difference
between particles and superparticles of the MSSM). This can be 
as low as a few hundred GeV, therefore no fine-tuning whatsoever is needed. 
\\
\\
In our work we study two important phenomenological issues in the context
of supersymmetric gauge theories. In the Minimal Supersymmetric 
Standard Model, we study the invisible decays of neutral Higgs 
bosons into two neutralinos at one-loop level. Proton decay
in the context of Minimal Supersymmetric $SU(5)$ is our second major objective. 
\\
\\
In the first part of the thesis we investigate the invisible Higgs decays 
into two neutralinos in the Minimal Supersymmetric Standard Model, 
taking into account new one-loop corrections to the neutral Higgs boson
couplings with neutralinos. Since the CP-odd Higgs boson does not 
couple to identical sfermions we expect that these corrections  
will be suppressed, while for the CP-even states there is the
possibility to get large corrections. The possible 
impact of the corrections on the invisible width of the 
lightest CP-even Higgs boson is very important, because these decays
could be enhanced to a level that should be easily measurable 
at future high-energy $e^+ e^-$ colliders.
\\
\\
In the second part of the thesis we focus on the Proton decay in the 
Minimal Supersymmetric Grand Unified Theory $SU(5)$. We will study in 
detail the $d=5$ operators contributing to the decay of
the proton, writing down the possible contributions for each
decay channel\footnote{Here $d$ refers to the mass dimension of the
operator, not to the dimension of spacetime.}. We point out 
the major sources of uncertainties in estimating the proton decay 
lifetime. We compute the masses of the color octet and 
weak triplet supermultiplets in the adjoint Higgs, in a general
model where non-renormalizable operators are present in order
to correct the relation between fermion masses. 
We study the effect of the mixings between fermion and
sfermions in proton decay. Finally we will see if it is 
possible to satisfy the experimental bounds on proton decay.
\\
\\
The present thesis contains seven chapters. In the second chapter we
review all the basics for Supersymmetry, we define the SUSY algebra
and introduce all the needed tools to write down the supersymmetric
version of gauge field theories. In chapter 3, the minimal supersymmetric
extension of the Standard Model is introduced, all the interactions
and relevant mass matrices for our analysis are studied. In the fourth
chapter we start with the study of our first objective, the invisible
Higgs decays into two neutralinos. We show how to compute the one-loop
corrections, and give several numerical examples to show the
effect of our quantum corrections. In chapter 5 we outline all the
important aspects of the Minimal Supersymmetric Grand Unified Theory 
$SU(5)$. In Chapter 6, we study our second important phenomenological
issue, Proton decay. We discuss all the relevant operators
contributing to the decay of the proton in supersymmetric theories,
and we focus our analysis in the Minimal Supersymmetric
$SU(5)$. Finally in Chapter 7 we conclude, pointing out possible future directions.

\chapter{Basics of Supersymmetry}

\section{SUSY Algebra}

Supersymmetry is a symmetry between fermions and
bosons, which is generated by a fermionic generator $Q$.

\begin{displaymath}
\ \ Q
\end{displaymath}
\vspace{-1.0cm}
\begin{displaymath}
\Psi_{fermionic} \ \Longleftrightarrow \ \Psi_{bosonic} 
\end{displaymath}
In general we could define a Supersymmetric Field Theory, as a theory 
which is invariant under SUSY transformation:

\begin{displaymath}
\delta_{\small SUSY}({\small Q}) {\cal S} = \delta_{\small
SUSY}({\small Q}) \int d^{D} x
{\cal L}(\Psi) \equiv 0
\end{displaymath}
With the usual Poincar\'e and internal symmetry algebra, it is
possible to define the Super-Poincar\'e Lie algebra, which contains the 
additional SUSY generators $Q_{\alpha}^i$ and $\bar Q_{\dot
\alpha}^i$, where $ \bar Q_{\dot \alpha}^i = ( Q_{\alpha}^i )^{\dagger}$\cite{WessBagger}\cite{Sohnius}:

\begin{equation} 
{[} P_{\mu},P_{\nu}{]}  =   0
\end{equation}

\begin{equation}
{[} P_{\mu},M_{\rho\sigma}{]} =
i(g_{\mu \rho}P_{\sigma} -
 g_{\mu \sigma}P_{\rho}) 
\end{equation}

\begin{equation}
{[} M_{\mu \nu} , M_{\rho \sigma} {]}   =    i(g_{\nu \rho}M_{\mu
\sigma} - g_{\nu \sigma}M_{\mu \rho} - g_{\mu \rho}M_{\nu \sigma}
+ g_{ \mu \sigma}M_{\nu \rho})
\end{equation}

\begin{equation}
 {[} B_r , B_s {]}    =    i c_{rs}^t B_{t}
\end{equation}

\begin{equation}
 {[} B_r , P_{\mu} {]}    =   0
\end{equation}

\begin{equation}
 {[} B_r ,M_{\mu \sigma} {]} = 0
\end{equation}

\begin{equation}
 {[} Q_{\alpha}^i , P_{\mu} {]} =  {[} \bar Q_{\dot \alpha}^i , P_{\mu} {]} = 0
\end{equation}

\begin{equation}
 {[} Q_{\alpha}^i , M_{\mu \nu} {]}  =   \frac{1}{2} (\sigma_{\mu \nu})_{\alpha}
^{\beta}Q_{\beta}^i 
\end{equation}

\begin{equation}
{[}\bar Q_{\dot \alpha}^i ,M_{\mu
\nu}{]} = - \frac{1}{2} \bar Q_{\dot \beta}^i (\bar \sigma_{\mu
\nu})_{\dot \alpha} ^{\dot \beta} 
\end{equation}
 
\begin{equation}
 {[} Q_{\alpha}^i , B_r {]}  =  (b_r)_{j}^i Q_{\alpha}^j 
\end{equation}

\begin{equation}
{[}\bar Q_{\dot \alpha}^i , B_r {]} = - \bar Q_{\dot \alpha}^j
 (b_r)_j^i 
\end{equation}

\begin{equation}
\{ Q_{\alpha}^i , \bar Q_{\dot \beta}^j \}
  =   2 \delta^{ij} (\sigma ^{\mu})_{\alpha \dot \beta }P_{\mu} 
\end{equation}

\begin{equation}
 \{ Q_{\alpha}^i , Q_{\beta}^j \}  =   2 \epsilon_{\alpha \beta}Z^{ij}
\end{equation}

\begin{equation}
Z_{ij} = a_{ij}^r B_r , \ \ \ \ Z^{ij} = Z_{ij}^{\dagger} 
\end{equation}

\begin{equation}
 \{ \bar Q_{\dot \alpha}^i , \bar Q_{\dot \beta}^j \}  =  - 2 \epsilon
_{\dot \alpha \dot \beta}Z^{ij} 
\end{equation}

\begin{equation}
{[}Z_{ij} , anything {]} = 0
\end{equation} 
where $\alpha$ , $\dot \alpha  =  1, 2; \ \ i,j = 1, 2, \ldots , N$,
with $N$ as the number of supersymmetries.
\\
Here $P_{\mu}$ is the four-momentum operator, $M_{ij}$ and $M_{0i}$
are the angular momentum and boost operators respectively, $B_r$ the internal symmetry
generators, $g_{\alpha \beta}$ is the metric, $c^{t}_{rs}$ and
$a^{r}_{ij}$ are structure constants and $Z_{ij}$ are the so-called central charges; $\alpha , \dot
\alpha, \beta , \dot \beta $ are spinorial indices. In the
simplest case one has one spinor generator $Q_\alpha$ (and the
conjugated one $\bar Q_{\dot{\alpha}}$) that corresponds to an
ordinary or N=1 supersymmetry. It is has been proved that the 
Super-Poincar\'e Lie algebra contains all possible symmetry generators for
symmetries of the S-matrix. It is the so-called the Coleman-Mandula 
Theorem\cite{Coleman}. 
\\
\\
There are many important conclusions coming from the SUSY algebra.
We see from equations 2.8 and 2.9, that the SUSY generators change the
spin by a half-odd amount and change the statistics. While from 
equation 2.7 we can conclude that a fermionic (or bosonic) field
and its superpartner in a theory with exact
supersymmetry must have the same mass. It is the reason
why SUSY must be broken in order to get a realistic spectrum in
particle physics.

\section{Superspace and Superfields}

An elegant formulation of supersymmetric transformations and
invariants can be achieved in the framework of superspace \cite{StradSalam}.
Superspace differs from the ordinary Euclidean (Minkowski)
space by adding two new coordinates, $\theta_{\alpha}$ and
$\bar \theta_{\dot \alpha}$, which are Grassmannian, 
i.e. anti\-com\-muting, variables
$$\{ \theta_{\alpha}, \theta_{\beta} \} = 0 , \ \ \{\bar
\theta_{\dot \alpha}, \bar \theta_{\dot \beta} \} = 0, \ \
\theta_{\alpha}^2 = 0,\ \ \bar \theta_{\dot \alpha}^2=0, \ \
\alpha,\beta, \dot\alpha, \dot\beta =1,2.$$
Thus, we go from space to superspace
$$\begin{array}{cc} Space & \ \Rightarrow \ \ Superspace \\
x_{\mu} & \ \ \ \ \ \ \ x_{\mu}, \theta_{\alpha} , \bar \theta_
{\dot \alpha} \end{array}$$
\\ 
A SUSY group element can be constructed in
superspace in the same way as an ordinary translation in the usual
space
\begin{equation}
 G(x,\theta ,\bar \theta ) = e^{\displaystyle
i(-x^{\mu}P_{\mu} + \theta Q + \bar \theta \bar Q)}\label{st}
\end{equation}
It leads to a supertranslation in superspace
\begin{equation}
 \begin{array}{ccl}
x_{\mu} & \rightarrow & x_{\mu} + i\theta \sigma_{\mu} \bar
\varepsilon
 - i\varepsilon \sigma_{\mu} \bar \theta \\
\theta & \rightarrow & \theta + \varepsilon \label{sutr} \\ \bar
\theta & \rightarrow & \bar \theta + \bar \varepsilon 
\end{array}
 \end{equation}
where $\varepsilon $ and $\bar \varepsilon $ are Grassmannian
transformation parameters. From eq.(\ref{sutr}) one can easily
obtain the representation for the supercharges acting on the superspace
\begin{equation}
Q_\alpha =\frac{\partial }{\partial \theta_\alpha
}-i\sigma^\mu_{\alpha \dot \alpha}\bar{\theta}^{\dot \alpha
}\partial_\mu  \ \ \ \bar{Q}_{\dot \alpha} =-\frac{\partial
}{\partial \bar{\theta}_{\dot \alpha}
}+i\theta_\alpha\sigma^\mu_{\alpha \dot \alpha}\partial_\mu 
\label{q}
\end{equation}
We are now ready to introduce the superfields. The superfields can be 
defined as functions in Superspace, $F = F (x, \theta, \bar{\theta})$. 
However, these superfields are in general reducible representations of the SUSY
algebra. To get an irreducible one, we
define a chiral superfield $\Phi$ which obeys the equation:
\begin{equation}
\bar D \Phi = 0  \ \ \ \ \ \ \mbox{where} \ \ \bar D = -\frac{\partial}{
\partial \overline{ \theta}} - i \theta \sigma^{\mu}
\partial_{\mu} \label{k}
\end{equation}
is a superspace covariant derivative.
For the chiral superfield, the Grassmannian Taylor expansion looks like
($y =x + i\theta \sigma \bar \theta $)
\begin{eqnarray}
\Phi (y, \theta ) & = & A(y) + \sqrt{2} \theta \psi (y) + \theta
 \theta F(y) \nonumber \\
  & = & A(x) + i\theta \sigma^{\mu} \bar \theta \partial_{\mu}A(x)
 + \frac{1}{4} \theta \theta \bar \theta \bar \theta \Box A(x) \nonumber\\
  & + & \sqrt{2} \theta \psi (x) - \frac{i}{\sqrt{2}} \theta
\theta \partial_{\mu} \psi (x) \sigma^{\mu} \bar \theta + \theta
 \theta F(x) \label{field}
\end{eqnarray}
The coefficients are ordinary functions of $x$, being the usual
fields. There are two physical fields, a bosonic one $A$ and
a fermionic $\psi$, while $F(x)$ is  an auxiliary field 
without physical meaning, needed to close the SUSY algebra. 
\\
\\
Under SUSY transformation the fields convert into one another
\begin{eqnarray}
\delta_\varepsilon A &=& \sqrt 2 \varepsilon \psi \nonumber \\
\delta_\varepsilon \psi &=& i \sqrt 2 \sigma^\mu \bar \varepsilon
\partial_\mu A + \sqrt 2 \varepsilon F \\
\delta_\varepsilon F &=&i \sqrt 2  \bar \varepsilon\sigma^\mu\partial_\mu\psi
\nonumber
\end{eqnarray}
Note that the variation of the $F$-component is a total derivative, i.e.
, with appropriate boundary conditions it vanishes when integrated over the space-time.
\\
One can also construct an antichiral superfield $\Phi^{\dagger}$ obeying
the equation:
$$D\Phi^{\dagger} = 0, \ \ \ \ \mbox{with}  \ \ \
 D = \frac{\partial}{\partial \theta} + i \sigma^{\mu}\bar \theta
\partial_{\mu} $$
\\
The product of chiral (antichiral)  superfields $\Phi^2 , \Phi^3$, 
etc is also a chiral (antichiral) superfield, while the product
of chiral and antichiral ones $\Phi^{\dagger} \Phi$ is a general
superfield, it is not a chiral superfield, its $\theta \theta
\bar{\theta} \bar{\theta}$ component transforms under SUSY as a 
total divergence.
\\
To construct the gauge invariant interactions, one needs a real
vector superfield, which is defined as $V = V^{\dagger}$. The
explicit form of $V$ is: 

\begin{eqnarray}
V(x, \theta, \bar \theta) & = & C(x) + i\theta \chi (x) -i\bar
\theta \bar \chi (x)  \nonumber \\
 & + & \frac{i}{2} \theta \theta [M(x) + iN(x)] - \frac{i}{2} \bar
 \theta \bar \theta [M(x) - iN(x)]  \nonumber \\
 & - & \theta \sigma^{\mu} \bar \theta v_{\mu}(x) + i \theta \theta
\bar \theta [\lambda (x) + \frac{i}{2}\bar \sigma^{\mu} \partial
_{\mu} \chi (x)]  \nonumber \\
 & - & i\bar \theta \bar \theta \theta [\lambda + \frac{i}{2}
\sigma^{\mu} \partial_{\mu} \bar \chi (x)] + \frac{1}{2} \theta
\theta \bar \theta \bar \theta [D(x) + \frac{1}{2}\Box C(x)]
\nonumber\\
& &
\end{eqnarray}
The physical degrees of freedom corresponding to a real vector
superfield $V$ are the vector gauge field $v_{\mu}$ and the
Majorana spinor field $\lambda$. All other components are
unphysical and can be eliminated. 
\\
\\
Under the Abelian (super)gauge transformation the superfield $V$ is 
transformed as $V\ \ \ \rightarrow \ \ \ V + \Phi + \Phi^{\dagger}$,
 where $\Phi$ and $\Phi^{\dagger}$ are some chiral superfields. In 
components it looks like \cite{WessBagger}
\begin{eqnarray}
C \ \ \ & \rightarrow & \ \ \ C + A + A^* \nonumber \\ \chi \ \ \
& \rightarrow & \ \ \ \chi -i\sqrt{2} \psi, \nonumber \\ M + iN \
\ \ & \rightarrow & \ \ \ M + iN - 2iF \nonumber \\ v_{\mu} \ \ \
& \rightarrow & \ \ \ v_{\mu} -i\partial_{\mu} (A - A^*)
\label{n}  \\ \lambda \ \ \ & \rightarrow & \lambda  \nonumber \\
D \ \ \ & \rightarrow & D  \nonumber
\end{eqnarray}
where $A^*$ is the complex conjugate of $A$. According to 
eq.(\ref{n}), one can choose a gauge (the Wess-Zumino gauge) 
where $C = \chi = M = N =0 $, leaving one with only physical 
degrees of freedom except for the auxiliary field $D$. In this gauge

\begin{eqnarray}
V & = & - \theta \sigma^{\mu} \bar \theta v_{\mu}(x) + i \theta
\theta \bar \theta \bar \lambda (x) -i\bar \theta \bar \theta
\theta \lambda (x) + \frac{1}{2} \theta \theta \bar \theta \bar
\theta D(x)  \nonumber\\ V^2 & = & - \frac{1}{2} \theta \theta
\bar \theta \bar \theta v_{\mu}(x)v^{\mu}(x)  \nonumber\\ V^3 & =
& 0 \ \ \ etc.
\end{eqnarray}

\section{Supersymmetric Lagrangians}

Using the rules of Grassmannian integration:
$$\int \ d\theta_\alpha =
0  \ \ \ \ \int \theta _\alpha\ d\theta _\beta=
\delta_{\alpha\beta} $$
we can define the general form of a SUSY and gauge invariant
lagrangian as \cite{WessBagger}:
\begin{eqnarray}
{\cal L}_{SUSY}^{YM} & = & \frac{1}{4}\int d^2 \theta
~Tr(W^{\alpha}W_{\alpha})   + \frac{1}{4}\int d^2 \bar{\theta}
~Tr(\bar{W}^{\alpha}\bar{W}_{\alpha})  \label{nonab}\\ &+& \int
d^2 \theta d^2 \bar \theta  \ \Phi^{\dagger}_{ia} \ (e^{gV})_b^a \ \Phi_i^b
+\int d^2 \theta ~{\cal W}(\Phi_i)   +\int d^2 \bar{\theta}
~\bar{{\cal W}}(\bar{\Phi}_i)   \nonumber
 \end{eqnarray}
$\Phi_i$ are chiral superfields which transform as:
\begin{displaymath}
\Phi_i \to e^{-i g \Lambda} \Phi_i 
\end{displaymath}
and
\begin{displaymath}
e^{g V } \to  e^{i g \Lambda^{\dagger}} e^{g V} e^{-i g \Lambda}
\end{displaymath}
where, both $\Lambda$ and $V$ are matrices:

\begin{displaymath}
\Lambda_{ij}= \tau^a_{ij} \Lambda_a \ \ \ \ \ V_{ij}= \tau^a_{ij} V_a
\end{displaymath}
with $\tau^a$ the gauge generators. The supersymmetric field strength $W_{\alpha}$ is equal to

\begin{displaymath}
W_{\alpha}= - \frac{1}{4} \bar{D} \bar{D} e^{-V} D_{\alpha} e^{V}
\end{displaymath}
and transforms as: $W \to e^{-i \Lambda} W e^{i \Lambda}$
\\
\\
${\cal W}$ is the superpotential, which should be invariant
under the group of symmetries of a particular model.
\\
In terms of component fields the above Lagrangian takes the form \cite{SUSYexplicit}

\begin{eqnarray}
{\cal L}_{SUSY}^{YM} & = & -\frac{1}{4}F^a_{\mu \nu }F^{a\mu \nu
}-i\lambda^a\sigma^\mu D_\mu \bar{\lambda}^a+\frac{1}{2}D^aD^a
\nonumber\\ &+&(\partial_\mu A_i -igv^a_\mu \tau{^a} A_i)^\dagger
(\partial_\mu A_i -igv^{a\mu} \tau^a A_i)
-i\bar{\psi}_i\bar{\sigma}^\mu (\partial_\mu \psi_i
-igv^{a\mu} \tau^a \psi_i) \nonumber \\ &-& D^a A^\dagger_i
\tau^a A_i-i\sqrt{2}A^\dagger_i \tau^a \lambda^a\psi_i +
i\sqrt{2}\bar{\psi}_i \tau^a A_i\bar{\lambda}^a + F^\dagger_iF_i
\nonumber \\ &+& \frac{\partial {\cal W}}{\partial A_i} F_i+
\frac{\partial \bar{{\cal W}}}{\partial A_i^\dagger}F^\dagger_i
-\frac{1}{2}\frac{\partial^2 {\cal W}}{\partial A_i \partial
A_j}\psi_i\psi_j -\frac{1}{2}\frac{\partial^2 \bar{{\cal
W}}}{\partial A_i^\dagger \partial
A_j^\dagger}\bar{\psi}_i\bar{\psi}_j  
 \end{eqnarray}
Integrating out the auxiliary fields $D^a$ and $F_i$, one
reproduces the usual Lagrangian.
\\
\\
Contrary to the SM, where the scalar Higgs potential is arbitrary and is
defined only by the requirement of the gauge invariance, in
supersymmetric theories it is completely  defined by the
superpotential. It consists of the contributions from the
$D$-terms and $F$-terms. The kinetic energy of the gauge fields
yields the $\frac{1}{2} D^aD^a$ term, and the
matter-gauge interaction yields the
$gD^a \tau^a_{ij}A^*_iA_j$ one. Together they give
\begin{equation}
{\cal L}_D=\frac{1}{2}D^aD^a + gD^a \tau^a_{ij}A^*_iA_j \label{d}
\end{equation}
The equation of motion reads
\begin{equation}
D^a=-g \tau^a_{ij}A^*_iA_j \label{sol}
\end{equation}
Substituting it back into eq.(\ref{d}) yields the $D$-term part of
the potential
\begin{equation} {\cal L}_D=-\frac{1}{2}D^aD^a \ \ \ \
\Longrightarrow V_D=\frac{1}{2}D^aD^a
\end{equation}
where $D$ is given by eq.(\ref{sol}).
\\
\\
The $F$-term contribution can be derived from the matter field
self-in\-ter\-action. For a general type
superpotential ${\cal W}$ one has
\begin{equation}
{\cal L}_F=F^*_iF_i+(\frac{\partial {\cal W}}{\partial A_i}F_i + h.c.)
 \end{equation}
Using the equations of motion for the auxiliary field $F_i$
 \begin{equation}
F^*_i=-\frac{\partial {\cal W}}{\partial A_i} \label{solf}
 \end{equation}
yields
 \begin{equation} {\cal L}_F=-F^*_iF_i \ \ \ \
\Longrightarrow V_F= F^*_iF_i 
\end{equation}
where $F$ is given by eq.(\ref{solf}). The full potential is the
sum of the two contributions
\begin{equation}
V=V_D+V_F
\end{equation}
Thus, the form of the Lagrangian is constrained by symmetry
requirements. The only freedom is the field content, the value of
the gauge coupling $g$, Yukawa couplings $y_{ijk}$ and the masses.
Because of the renormalizability constraint $V \leq A^4 $ the
superpotential should be limited by ${\cal W} \leq \Phi^3 $. 
All members of a supermultiplet have the same
masses, i.e. bosons and fermions are degenerate in masses. This
property of SUSY theories contradicts the phenomenology and
requires supersymmetry breaking.

\section{SUSY Breaking}

Since the supersymmetric algebra leads to mass degeneracy in a
supermultiplet, it should be broken to explain the absence of
superpartners at accessible energies. There are several ways of supersymmetry breaking. 
It can be broken either explicitly or spontaneously. 
Performing SUSY breaking one has to be careful not
to spoil the cancellation of quadratic divergencies which allows
one to solve the \textit{Hierarchy problem}. This is achieved by
spontaneous breaking of SUSY.
\\
It is possible show that in SUSY models the energy is always
nonnegative definite. According to quantum mechanics the energy is equal 
to:

\beq
 E = \langle 0 \vert \ \widehat{H} \  \vert 0 \rangle 
\eeq
where $\widehat{H}$ is the Hamiltonian and due to the SUSY algebra:

\beq
\{Q_{\alpha}, \bar Q_{\dot \beta} \} =
2(\sigma^{\mu}) _{\alpha \dot \beta}P_{\mu} 
\eeq
taking into account that $Tr(\sigma^{\mu}P_{\mu}) = 2P_0 $
 one gets

\beq
E = \frac{1}{4} \sum_{\alpha = 1,2}<0| \{Q_{\alpha}, \bar
Q_{\alpha} \} |0> = \frac{1}{4} \sum_{\alpha} \|Q_{\alpha}
  |0> \|^2 \geq 0 
\eeq
Hence $$ E = <0| \ \widehat{H} \ |0> \neq 0 \ \ \ \ if \ and \ only \ if \ \
\ Q_{\alpha}|0> \neq 0 $$
Therefore, supersymmetry is spontaneously broken, i.e. the vacuum is
not invariant under $Q$ $(Q_{\alpha} |0 \rangle \neq 0 )$, {\em if and only if} the
minimum of the potential is positive $(i.e.\ E \geq 0)$ .
\\
\\
Spontaneous breaking of supersymmetry is achieved in the same way
as electroweak symmetry  breaking. One introduces a field
whose vacuum expectation value is nonzero and breaks the symmetry.
However, due to the special character of SUSY, this should be a
superfield whose auxiliary $F$ or $D$ component acquires nonzero
v.e.v.'s. Thus,  among possible spontaneous SUSY breaking
mechanisms one distinguishes the $F$ and $D$ ones.
\\
\\
i) Fayet-Iliopoulos ($D$-term) mechanism \cite{SUSYexplicit}. \\
 In this case the, the linear $D$-term is added to the Lagrangian
\begin{equation} 
\Delta {\cal L} = \xi V\vert_{ \theta \theta \bar \theta \bar \theta} = \xi\int
d^2\theta \ d^2 \bar{\theta} \ V
\end{equation}
It is $U(1)$ gauge and SUSY invariant by itself; however, it may
lead to spontaneous breaking of both of them depending on the value of
$\xi$. The drawback of this mechanism is the necessity of $U(1)$ gauge
invariance. It can be used in SUSY generalizations of the SM but
not in GUTs.
The mass spectrum also causes some troubles since the
following sum rule is always valid

\begin{equation}
STr {\cal{M}}^2= \sum_{J}(-1)^{2J}(2J+1)m^2_J=0 \label{sumrule}
\end{equation}
which is bad for phenomenology.
\\
\\
ii) O'Raifeartaigh ($F$-term) mechanism \cite{SUSYexplicit}. \\
\\
In this case, several chiral fields are needed and the superpotential should be
chosen in such way that trivial zero v.e.v.s for the auxiliary
$F$-fields are forbidden. For instance, choosing the superpotential
to be:

\beq
{\cal W}(\Phi)=
\lambda \Phi_3 +m\Phi_1\Phi_2 +g \Phi_3\Phi_1^2
\eeq
one gets the equations for the auxiliary fields

\begin{eqnarray}
F^*_1&=&mA_2+2gA_1A_3 \\
F^*_2 &=& mA_1 \\
F^*_3 &=& \lambda +gA^2_1
\end{eqnarray}
which have no solutions with $\langle F_i \rangle =0$ and SUSY is spontaneously broken.
\\
\\
The drawback of this mechanism is, that there is a lot of arbitrariness in the
choice of potential. The sum rule (\ref{sumrule}) is also valid
here.
\\
\\
Unfortunately, none of these mechanisms explicitly works in SUSY
generalizations of the SM. None of the fields of the SM can
develop nonzero v.e.v.s for their $F$ or $D$ components without
breaking  $SU(3)$ or $U(1)$ gauge invariance since they are not
singlets with respect to these groups. This requires the presence
of extra sources for spontaneous SUSY breaking \cite{SUSYbreak, 
Barbieri:1982eh,Dine:1981gu,Giudice:1998bp, Dvali:1996rj, Dvali:1996bg}.

\chapter{The MSSM}

\begin{figure}[ht]
 \begin{center}
 \leavevmode
  \epsfxsize=10cm
 \epsffile{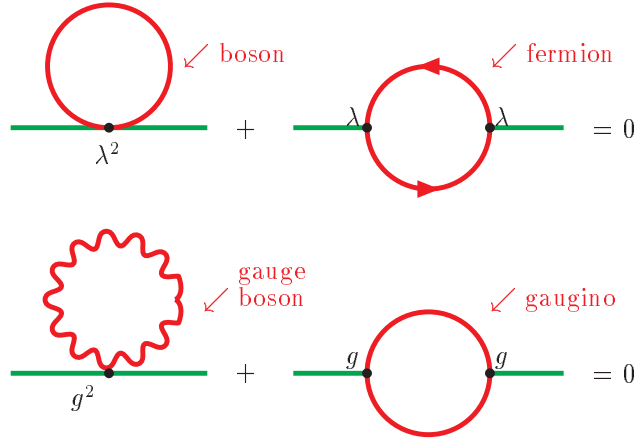}
 \end{center}\vspace{-1cm}
 \caption{Cancellation of quadratic divergences (from reference \cite{Kazakov})}\label{fig:cancel}
 \end{figure}

The Standard Model (SM) describes with a very good precision all electroweak and strong
processes. It is based on gauge invariance under the symmetry group:

\begin{equation}
G_{SM}=SU(3)_{C} \times  SU(2)_{L} \times U(1)_Y
\end{equation}
and its partial spontaneous symmetry breaking. In Table 1 we show
all its constituents, the elementary fermions (quarks and leptons), the
scalar Higgs boson, the gauge bosons, and their transformation properties
under  $G_{SM}$. We use the relation $Q=T_3+Y/2$, where $Q$, $T_3$ and
$Y$ are the electric charge, isospin and hypercharge respectively.
\\
\\
The lagrangian of the Standard Model has the following form:
\begin{equation}
{\cal L}_{SM}= {\cal L}_{fermions} + {\cal L}_{gauge \ bosons} + {\cal
L}_{scalars}+ {\cal L}_{Yukawa}
\end{equation}
The explicit form of the SM lagrangian is well known, for our objectives
we will write explicitly only the expression of the Yukawa interactions:

\begin{equation}
{\cal L}_{Yukawa} = (\bar{u}^i \ \bar{d}^i )_L \ h_{ij}^d \ \Phi \ d^j_R + 
(\bar{u}^i \ \bar{d}^i)_L \ h_{ij}^u i \ \sigma_2 \Phi^{*} \ u^j_R 
+ (\bar{ \nu}^i \ \bar{e}^i)_L \ h_{ij}^e \ \Phi \ e^j_R
+ h.c
\end{equation}
$u^i$, $d^i$, $h^u$ and $h^d$ are the quarks with isospin $1/2$ and
$-1/2$, and their Yukawa matrices respectively, $e^i$ and $h^e$ stand for the charged 
leptons and their Yukawa matrices, while $\nu^i$ are the neutrinos for
each family. The subscripts $L$ and $R$ refer to right and left
chirality respectively, while i and j are the generation indices.  
\\
\\
\begin{array}[b]{cccc}
\underline{Table \ 1. \
Standard \ Model \ Particles} & & & \\ \\
 & & & SU(3)_C,\ SU(2)_L, U(1)_Y \\
           \underline{ \mbox{Quarks: }} \\ \\
\pmatrix{u^{\alpha} \cr d^{\alpha}}_L & \pmatrix{c^{\alpha} \cr
s^{\alpha}}_L 
& \pmatrix{t^{\alpha} \cr b^{\alpha}}_L
& (3_C,2_L,1/3) \\ \\
 u^{\alpha}_R  & c^{\alpha}_R & t^{\alpha}_R & (3_C,1_L,4/3) \\ \\
d^{\alpha}_R & s^{\alpha}_R & b^{\alpha}_R & (3_C,1_L,-2/3) \\ \\
where \ \alpha=1, 2, 3 \ (colors) & & & \\ \\
            \underline{\mbox{Leptons:}} \\ \\
\pmatrix{\nu_e \cr e}_L & \pmatrix{\nu_\mu \cr \mu}_L & \pmatrix{\nu_\tau
\cr \tau}_L & (1_C,2_L,-1) \\ \\
            e_R & \mu_R & \tau_R & (1_C,1_L,-2) \\ \\

            \underline{\mbox{Scalars:}} \\ \\
           \Phi=\pmatrix{\phi^+ \cr \phi^0} & & & (1_C,2_L,1) \\ \\
            \underline{\mbox{Gauge bosons:}} \\ \\
          G^a_\mu \ with \ a=1,2..8 &  &  &  (8_C,1_L,0) \\ \\
            W^b_\mu \ with \ b=1, 2, 3 &  &  &  (1_C,3_L,0) \\ \\
            B_\mu &  &  &  (1_C,1_L,0) \\ \\
\end{array}
\\
Note that we can write these terms
using only one scalar field $\Phi$, which after electroweak
symmetry breaking can generate mass for all the quarks and leptons
in the Standard Model.	
\\
An important free parameter in the Standard Model is the Weinberg angle
$\theta_W$, which is defined as: 

\beq
\sin \theta_W = \frac{g_{U(1)}}{\sqrt{g^2_{SU(2)} + g^2_{U(1)}}}
\eeq
where $g_{U(1)}$ and $g_{SU(2)}$ are the gauge couplings for the
$U(1)$ and $SU(2)$ gauge groups. 
\\
\\
As was mentioned above, the standard model has an extremely economical
Higgs sector, which accounts for all the particle masses. Baryon (B) and
Lepton (L) numbers are automatically conserved and it is an
anomaly free Quantum Field Theory. However not all is perfect, at
present there is no evidence for Higgs.
\\
\\
A dramatic problem in the Standard Model is present in the Higgs
sector. If we consider radiative corrections to the Higgs
mass, we see that it has quadratic divergencies, which
can change its value by many orders of magnitude
 \cite{DreesReview}. In Fig.\ref{fig:cancel} we show two of the problematic contributions,
these are the contributions with fermions and gauge bosons inside the loops. 
The only known way to cancel these divergencies is supersymmetry. 
SUSY automatically cancels quadratic corrections 
to all orders of perturbation theory. This is due to
the contributions of superpartners of  ordinary particles. The
contribution from boson loops cancel those from the fermion ones
because of an additional factor (-1) coming from Fermi statistics.
\\
\\
The first line of Figure \ref{fig:cancel} shows the contribution 
of an SM fermion and its superpartner. The strength of interaction 
is given by the Yukawa coupling $\lambda$. The second line 
represents the gauge interaction proportional to the gauge 
coupling constant $g$ with the contribution from the
gauge boson and its superpartner.
\\
\\
In both cases, cancellation of quadratic terms takes place.
This cancellation occurs up to the SUSY breaking scale, 
$\Lambda_{SUSY}$, since

\begin{equation}
\sum_{bosons} m^2 - \sum_{fermions} m^2= \Lambda_{SUSY}^2
\end{equation}
which should be around $\sim$ 1 TeV to make the
fine-tuning natural. Indeed, let us take the Higgs boson mass.
Requiring for consistency of perturbation theory that the
radiative corrections to the Higgs  boson mass do not  exceed the
mass itself gives

\begin{equation}
\delta M_h^2 \sim \frac{3 y_t^2}{16 \pi^2} \Lambda^2_{SUSY} \sim M_h^2 \label{del}
\end{equation}
So, if $M_h \sim 10^2$ GeV one needs $\Lambda_{SUSY} \sim 10^3$ GeV in order that the relation
(\ref{del}) is valid. Thus, we again get more or less the same rough estimate
of $\Lambda_{SUSY} \sim $ 1 TeV as from the gauge coupling unification. 
Two requirements match together. However as we mentioned before,
$\Lambda_{SUSY}$ could be in the range $1-10$ TeV if we accept some fine-tuning.

\section{Particles and their Superpartners}

In the MSSM, we add a superpartner for each SM particle in the same
representation of the gauge group. Usually we use the SUSY operators 
in the left-chiral representation,
therefore it is convenient to rewrite the SM particles (Table 1.) 
in the left-chiral representation and define the superpartners accordingly.
 This leads to the superfield formalism, which makes it easier to
construct SUSY invariant lagrangians. In this case we have to introduce for each
SM particle one superfield, which contains the SM particle, its
superpartner and an auxiliary unphysical field.
In Table 3 we show the third generation of the SM particles, their
superpartners and the superfields needed to write our lagrangian. Note
that we have an extended Higgs sector and the color index is omitted. 
\\
In Table 2 we show the names of the superpartners.
\\
\begin{array}[b]{cccc}
\\ \\ \underline{ Table \ 2.\  Names \ of \ superpartners.} & & & \\ \\ \\
\underline{\rm{Matter \ Fermions}} & \iff  & \underline{\rm{Sfermions}} & \\ \\
(quarks, leptons) & &  (squarks, sleptons) & \\ \\
\ s=1/2 & &  \ s=0 \\ \\
\underline{\rm{Gauge \ Bosons}} & \iff & \underline{\rm{Gauginos}} & \\ \\
(W^\pm, Z, \gamma, gluons) & &  (Wino, Zino, photino, gluinos) & \\ \\
\ s=1 & &  \ s=1/2 & \\ \\
\underline{\rm{Higgs}} & \iff &  \underline{\rm{Higgsinos}} & \\ \\
\ s=0 & &  \ s=1/2 & \\ \\
\end{array}
\\
For example the superpartner of the top quark is called 
stop, of the photon the superpartner is the photino, and similarly for 
the other particles. However, in the Higgs sector we must add another 
new Higgs boson and its superpartner to write the Yukawa interactions 
needed to generate masses
for all quarks and to obtain an anomaly free model.
\\
\section{MSSM Lagrangian}

We can divide the lagrangian of the minimal
extension of the Standard Model into two
fundamental parts, the SUSY invariant and
the Soft breaking term \cite{Haber}:

\begin{equation}
{\cal L}_{MSSM}= {\cal L}_{SUSY} + {\cal L}_{Soft}
\end{equation}
In general we can write the SUSY invariant
term as:
\begin{equation}
{\cal L}_{SUSY}= {\cal L}_{gauge}+
{\cal L}_{leptons}+ {\cal L}_{quarks}+
{\cal L}_{Higgs}+ \int{d^2 \theta} \ {\cal W} + h.c
\end{equation}
Defining the content of the MSSM in Table 3, we can 
write the different terms of the lagrangian. The term 
${\cal L}_{gauge}$ has the following form:
\begin{equation}
{\cal L}_{gauge}=\frac{1}{4} \int{d^2
\theta}[2 \ Tr(W^3W^3) + 2 \ Tr(W^2W^2)+W^1W^1]
\end{equation}
with:
\begin{equation}
W^3_\alpha=-\frac{1}{4} \overline{D}
\  \overline{D} \exp (-G_3) \ D_\alpha
 \exp(G_3)
\end{equation}
\begin{equation}
G_3 = \sum_{a=1}^8 \frac{ \lambda ^a}{2}
G^a_3 \\
\end{equation}
\begin{equation}
W^2_\alpha =-\frac{1}{4} \overline{D}
\ \overline{D} \exp(-G_2) \ D_\alpha
 \exp(G_2)
\end{equation}
\begin{equation}
G_2=\sum_{b=1}^3 \frac{\sigma^a}{2} G^b_2
\end{equation}
\\
\\
\begin{array}[t]{cccc}
       \underline{Table \ 3.\ Content \ of \ the \ MSSM.}  & & &    \\ \\

\underline{\rm{Superfields}} & & & \\ \\
            \underline{\rm{Vector \ Superfields}} & & & \\ \\
& \underline{\rm{Bosonic \ Fields}} & \underline{\rm{Fermionic \ Fields}} &
G_{SM} \\ \\ \\

          G^a_3 & G^a_\mu \ with \ a=1,2..8 & \widetilde {G}^a &
(8_C,1_L,0)
\\ \\

             G^b_2 & W^b_\mu \ with \ b=1,2,3 &  \widetilde{W}^b  &
(1_C,3_L,0) \\ \\

            G_1 & B_\mu &  \widetilde{B} &  (1_C,1_L,0) \\ \\

            \underline{\rm{Chiral \ Superfields}} & & & \\ \\

\underline{\rm{Leptons}} & & & \\ \\

L=\pmatrix{ N \cr E} & \widetilde{L}= \pmatrix{ \widetilde{\nu} \cr \widetilde{e}}
& \pmatrix{ \nu \cr e} & (1_C,2_L,-1) \\ \\

            E^C & \widetilde{e}^C & e^C & (1_C,1_L,2) \\ \\

\underline{\rm{Quarks}} & & & \\ \\
  Q=\pmatrix{ U \cr D} &  \widetilde{Q}= \pmatrix{ \widetilde{u} \cr \widetilde{d}} &
\pmatrix{u \cr d} & (3_C,2_L,1/3) \\ \\
            U^C & \widetilde{u}^C & u^C & (\bar{3}_C,1_L,-4/3) \\ \\
            D^C & \widetilde{d}^C & d^C & (\bar{3}_C,1_L,2/3) \\ \\

\underline{\rm{Higgs}} \\ \\

  \bar{H} &  H_1 = \pmatrix{H^0_1 \cr H^-_1} & \pmatrix{ \widetilde{H}^0_1 \cr
\widetilde{H}^-_1} & (1_C,2_L,-1) \\ \\
H &  H_2 = \pmatrix{H^+_2 \cr H^0_2} & \pmatrix{ \widetilde{H}^+_2 \cr
\widetilde{H}^0_2} & (1_C,2_L,1) \\ \\
\end{array}
\\ \\ 
and
\begin{equation}
W^1_\alpha=-\frac{1}{4} \overline{D}
\ \overline{D} D_\alpha G_1
\end{equation} 
where  $\lambda^a$ \ and \  $\sigma^a$
\ are the Gell-Mann and Pauli matrices respectively, and $G$'s are
defined in Table 3. 
\\
\\
The SUSY covariant derivatives D and $\overline{D}$ are used in the
left chiral representation:
\begin{equation}
D_L = \frac{\partial}{\partial \theta}+
2i\sigma^\mu \overline{\theta}\partial_\mu
\end{equation}
and
\begin{equation}
\overline{D}_L = - \frac{ \partial}
{ \partial  \overline{\theta}}
\end{equation}
The lagrangians for gauge interactions of leptons, quarks and
the Higgs bosons are:
\begin{displaymath}
{\cal L}_{leptons}= \int{d^2 \theta
\ d^2 \overline{\theta}} \ L^{\dagger} \ \exp (2g_2 G_2 + g_1
\frac{Y_L}{2} G_1)  \ L \ +
\end{displaymath}
\begin{equation}
+ \int{d^2 \theta \ d^2 \overline{\theta}}
\ {E^C}^{\dagger} \ \exp(2 g_2 G_2 + g_1
\frac{Y_{E^C}}{2} G_1)  \ E^C
\end{equation}
\begin{displaymath}
{\cal L}_{quarks}= \int{d^2 \theta d^2
\ \overline{\theta}} \ Q^{\dagger}
 \ \exp(2g_3G_3+ 2 g_2 G_2 + g_1 \frac{Y_Q}{2} G_1) \ Q \ +
\end{displaymath}  
\begin{displaymath}
+ \int{d^2 \theta \ d^2 \overline{\theta}}
\ {U^C}^{\dagger} \ \exp(g_1\frac{Y_{U^C}}{2}
G_1 - g_3 (\lambda^a)^* G^a_3)  \ U^C \ +
\end{displaymath}
\begin{equation}
+\int{d^2 \theta \ d^2 \overline{\theta}}
\ {D^C}^{\dagger} \ \exp(g_1\frac{Y_{D^C}}{2}
G_1- g_3 (\lambda^a)^* G^a_3) \ D^C
\end{equation}
\begin{displaymath}
{\cal L}_{Higgs}= \int{d^2 \theta \ d^2
\overline{\theta}} \ H^{\dagger} \ \exp(2g_2G_2+g_1
\frac{Y_{H}}{2} G_1) \ H \ +
\end{displaymath}  
\begin{equation}
+ \int{d^2 \theta \ d^2 \overline{\theta}} \ \bar{H}^{\dagger}
\ \exp(2g_2G_2+g_1 \frac{Y_{\bar{H}}}{2} G_1) \ \bar{H}
\end{equation}
where $g_3, g_2, g_1$ are the SU(3), SU(2) and U(1) coupling
constants, respectively. The $Y_i$ represent the hypercharges of the
different superfields.
\\
\\
We can write the superpotential as the sum of two
terms, $\mathcal{W}$=${\mathcal W}_R$+${\mathcal W}_{NR}$. The first conserves 
lepton (L) and baryon (B) numbers:
\begin{equation}
{\cal W}_R= \epsilon_{ij}
[- \mu  \bar{H}^i  H^j \ + \ \bar{H}^i  L^j  Y_E  E^C \ 
+ \ \bar{H}^i  Q^j  Y_D  D^C \ + \ H^j  Q^i  Y_U  U^C ] 
\end{equation}
where $\epsilon_{ij}$ is the antisymmetric tensor,
$\mu$ the Higgs mass parameter and
$Y_U$, $Y_D$ and $Y_E$ are the different
Yukawa matrices. The term ${\mathcal{W}}_{NR}$, which
explicitly breaks  L and B numbers, is:
\begin{equation}
\label{WNR}
{\cal W}_{NR}= 
\epsilon_{ij} [- \mu^{'}  H^i L^j \ + \ 
\lambda L^i L^j E^C \ + \ \lambda^{'} L^i Q^j  D^C ]\ + \ \lambda^{''}
D^C D^C U^C 
\end{equation}
In ${\cal W}_{NR}$ the first three terms break lepton number, while the last
term breaks baryon number. From these terms we find $d=4$ operators
contributing to the decay of the proton, which will be analyzed in the
next chapters. In the last section of this chapter we will analyze the R-symmetry
related with the L and B number conservation and
its implications.
\\
\\
SUSY is broken explicitly if we introduce the following terms:
\begin{displaymath}
- {\cal L}_{soft} = m^2_1 \vert H_1
\vert^2 \ + \ m^2_2 \vert H_2 \vert^2 \ + \ m^2_{12}(H_1 H_2+H^*_1
 H^*_2) 
\end{displaymath}
\begin{displaymath}
\ + \ \widetilde{Q}^{\dagger} M^2_{\widetilde{Q}}
\widetilde{Q} \ + \ {\widetilde{u^C}}^{\dagger}
m^2_{\widetilde{u}^C} \widetilde{u}^C \ + \
{\widetilde{d^C}}^{\dagger} m^2_{\widetilde{d}^C}
{\widetilde{d}^C} \ + \ {\widetilde{L}}
^\dagger M^2_{\widetilde{L}} \widetilde{L} 
\ + \ {\widetilde{E^C}}^{\dagger} m^2_{\widetilde{E}^C}
\widetilde{E}^C 
\end{displaymath}
\begin{displaymath}
\ + \ [ H_2 \widetilde{Q}(Y_U A_U) \widetilde{u}^C \ + \
 H_1 \widetilde{Q}(Y_D A_D) \widetilde{d}^C \ + \
 H_1 \widetilde{L}(Y_E A_E) \widetilde{E}^C + h.c ] 
\end{displaymath}
\begin{equation}
\ + \ \frac{1}{2} [M_1 \overline{\widetilde{B}}   
\widetilde{B} \ + \ M_2 \overline{\widetilde{W^b}}
\widetilde{W^b} \ + \ M_3 \overline{\widetilde{g^a}}
\widetilde{g^a}]
\label{l1}
\end{equation}
\\
Note that in order to describe SUSY breaking we
introduce many free parameters, and several 
terms have mass dimension less than
4 (super-renormalizable, but not SUSY invariant). 
The different mass terms remove the degeneracy between
particles and their superpartners.

\section{Neutralinos and Charginos}

Once $SU(2)_{L} \times U(1)_Y$ is broken in the MSSM, fields with different
$SU(2)_{L} \times U(1)_Y$ quantum numbers can mix, if they have the same
$SU(3)_{C} \times U(1)_{em}$ quantum numbers, and the same spin. 
\\
\\
The neutralinos are mixtures of the
$ \widetilde{B} $, the neutral $ \widetilde{W}_3 $ and the
two neutral Higgsinos. 
\\
\\
The mass term for the neutralinos is equal to:

\begin{equation}
- {\cal L}^{mass}_{\psi^0} = \frac{1}{2}(\psi^0)^T \ {\cal M}_N \ \psi^0 + h.c
\end{equation}
If we define the physical states as $\widetilde{\chi^0_i}$ = $N_{ij} \psi^0_j$, 
the diagonal mass matrix is ${\cal M}_D$ = $N^* {\cal M}_N
N^{\dagger}$. 
\\
\\
In general these states form
four distinct Majorana fermions, which are eigenstates
of the symmetric mass matrix [in the basis ($ \widetilde{B} $, $
\widetilde{W_3} $, $ \widetilde{H^0_1} $, $
\widetilde{H^0_2} $)] \cite{Haber}:
\\
\\
\begin{eqnarray} \label{matrix}
{\cal M}_N = \left[ \begin{array}{cccc}
M_1 & 0 & -m_Z s_W c_\beta & m_Z  s_W s_\beta \\
0   & M_2 & m_Z c_W c_\beta & -m_Z  c_W s_\beta \\
-m_Z s_W c_\beta & m_Z  c_W c_\beta & 0 & -\mu \\
m_Z s_W s_\beta & -m_Z  c_W s_\beta & -\mu & 0
\end{array} \right] 
\end{eqnarray}
where $M_1$ and $M_2$ are the SUSY breaking masses for the U(1)$_Y$
and SU(2)$_L$ gauginos, $\mu$ is the higgsino mass parameter, and
$s_\beta \equiv \sin \beta$, $s_W \equiv \sin \theta_W$, etc.
\\
\\
We will assume that all soft breaking parameters as well as $\mu$ are real,
i.e. conserve CP. We can then work with a
real, orthogonal neutralino mixing matrix $N$ if we allow the
eigenvalues $m_{\tilde \chi_i^0}$ to be negative.
\\
\\
This matrix can be diagonalized analytically, but the 
expressions of the neutralino masses and the $N_{ij}$ matrix elements
are rather involved. However, if the entries in the off--diagonal 
$2 \times 2$ submatrices are small compared to the diagonal entries, one can expand the 
eigenvalues in powers of $m_Z$ \cite{loop}:
\begin{eqnarray} \label{massapp}
m_{\tilde{\chi}_{1}^0} &\simeq&
M_1 - \frac{m_Z^2}{\mu^2-M_1^2} \left( M_1 +\mu s_{2\beta} \right) s_W^2
\\
m_{\tilde{\chi}_{2}^0} &\simeq&
M_2 - \frac{m_Z^2}{\mu^2-M_2^2} \left( M_2 +\mu s_{2\beta} \right) c_W^2
\\
m_{\tilde{\chi}_{3}^0} &\simeq&
-\mu - \frac{m_Z^2(1- s_{2\beta})} {2} \left( \frac{s_W^2} {\mu+M_1}
+ \frac{c_W^2} {\mu+M_2} \right)
\\
m_{\tilde{\chi}_{4}^0} &\simeq&
\mu + \frac{m_Z^2(1+ s_{2\beta})} {2} \left( \frac{s_W^2} {\mu-M_1}
+ \frac{c_W^2} {\mu-M_2} \right)
\end{eqnarray}
In our analysis, we are interested in the situation $|\mu| > M_1, \ M_2$ and
$\mu^2 \gg m_Z^2$. In this case the lighter of the two neutralinos will be
gaugino--like. If $|M_1| < |M_2|$, the lightest state will be 
bino--like, and the next--to--lightest state will be wino--like. 
The two heaviest states will be dominated by their higgsino
components. The components of the mass eigenvectors can also be 
expanded in powers of $m_Z$. We find for the bino--like state:
\begin{eqnarray} \label{binostate}
N_{11} &=& \left[ 1+ \left( N_{12}/N_{11} \right)^2+
\left( N_{13}/N_{11} \right)^2+
\left( N_{14}/N_{11} \right)^2 \right]^{-1/2} 
\\
\frac{N_{12}}{N_{11}} &=& \frac{m_Z^2 s_W c_W} {\mu^2 - M_1^2}
\frac { (s_{2\beta} \mu + M_1)} {(M_1-M_2)} +{\cal O}(m_Z^3) 
\\
\frac{N_{13}}{N_{11}} &=&  m_Z s_W
\frac {(s_\beta \mu + c_\beta M_1)}{\mu^2 - M_1^2} + 
{\cal O}(m_Z^2) \\
\frac{N_{14}}{N_{11}} &=&  - m_Z s_W  \frac{ ( c_\beta \mu + s_\beta M_1)}
{\mu^2 - M_1^2} 
+ {\cal O}(m_Z^2)
\end{eqnarray}
The corresponding expressions for the wino--like state read:
\begin{eqnarray} \label{winostate}
N_{22} &=& \left[ 1+ \left( N_{21}/N_{22} \right)^2+
\left( N_{23}/N_{22} \right)^2+
\left( N_{24}/N_{22} \right)^2 \right]^{-1/2} 
\\
\frac{N_{21}}{N_{22}} &=& \frac{m_Z^2 s_W c_W} {\mu^2 - M_2^2}
\frac {(s_{2\beta} \mu + M_2)} {(M_2-M_1)} + {\cal O}(m_Z^3) 
\\
\frac{N_{23}}{N_{22}} &=&  -m_Z c_W  
\frac {(s_\beta \mu + c_\beta M_2)}{\mu^2 - M_2^2} +
{\cal O}(m_Z^2) \\
\frac{N_{24}}{N_{22}} &=&  m_Z c_W  \frac{(c_\beta \mu + s_\beta M_2)}
{\mu^2 - M_2^2}
+ {\cal O}(m_Z^2)
\end{eqnarray}
Note that the higgsino components of the gaugino--like states start at
${\cal O}(m_Z)$, whereas the masses of these states deviate from their 
$|\mu| \rightarrow \infty$ limit ($M_1$ and $M_2$) only at ${\cal
O}(m_Z^2)$. 
\\
\\
The charginos are mixtures of the $\widetilde{W}^{\pm}$ and
$\widetilde{H}^{\pm}$. The chargino mass matrix [in
the basis ($\widetilde{W^{\pm}}, \widetilde{H^{\pm}}$) ] \cite{Haber}is:
\begin{eqnarray} \label{charmat}
{\cal M}_C = \left[ \begin{array}{cc} M_2 & \sqrt{2}m_W s_\beta
\\ \sqrt{2}m_W c_\beta & \mu \end{array} \right]
\end{eqnarray}
if we expand in powers of $m_W$, the two chargino masses are:
\begin{equation}
m_{\tilde{\chi}_{1}^\pm}  \simeq   M_2 - \frac{m_W^2}{\mu^2-M_2^2}
\left( M_2 +\mu s_{2\beta} \right) \simeq m_{\tilde \chi_2^0}\ \  \ \ 
\end{equation}

\begin{equation}
m_{\tilde{\chi}_{2}^\pm}  \simeq  \mu + 
\frac{m_W^2}{\mu^2-M_2^2} \left( M_2 s_{2 \beta} +\mu \right) 
\end{equation}
so that for $|\mu| \rightarrow \infty$, the lightest chargino
corresponds to a pure wino state while the heavier chargino
corresponds to a pure higgsino state. 
\\
\\
Usually the neutralino is considered the lightest 
supersymmetric particle (LSP) in models where R-parity is 
conserved. It has been realized many years ago that they are good
candidates to describe the Non-Baryonic Dark Matter present
in the Universe \cite{Goldberg,Ellis1,SDM}. There is no direct
experimental limit for the neutralino mass, however
from LEP experiments we know that the chargino mass
$m_{\tilde{\chi_1^{\pm}}}$ must be bigger than $103.5$ GeV\cite{LEP2}.

\section{Squarks and Sleptons}

After the electroweak symmetry breaking, several terms in the MSSM
lagrangian contribute to the sfermion mass matrices. Ignoring flavor mixing
between sfermions, the mass matrix for charged matter sfermion is[in the
basis $(\tilde{f}, \tilde{f}^C)$] \cite{Haber}:
\\
\beq \label{sqmass_matrix}
{\cal M}^2_{\tilde{f}} =
\left(
  \begin{array}{cc} m_f^2 + m_{LL}^2 & m_f \, \tilde{A}_f  \\
                    m_f\, \tilde{A}_f    & m_f^2 + m_{RR}^2
\end{array} \right)
\eeq
\\ \\
with
\beq
\begin{array}{l}
\ m_{LL}^2 =m_{\tilde{f}}^2 + (I_3^f - Q_f \sin^2 \theta_W)\, m_Z^2\,
\cos 2\beta \\ \\
m_{RR}^2 = m_{\tilde{f}^C}^2 + Q_f \sin^2 \theta_W \, m_Z^2\, \cos
2\beta \\ \\
\ \tilde{A}_f  = A_f - \mu (\tb)^{-2 I_3^f}
\end{array} 
\eeq
where $f$ represents the different charged fermions $u^i$, $d^i$ and
$e^i$.
\\ 
\\
The charged sfermions mass matrices are diagonalized 
by $ 2 \times 2$ rotation matrices described by 
the angles $\theta_{\tilde{f}}$, which turn the current eigenstates,
$\tilde{f}$ and $\tilde{f}^C$, into the mass eigenstates
$\tilde{f}_1$ and $\tilde{f}_2$; the mixing angle and sfermion masses
are then given by
\beq \label{sqstate}
\sin 2 \theta_{\tilde{f}} = \frac{2 m_f \tilde{A}_f} { m_{\tilde{f}_1}^2
-m_{\tilde{f}_2}^2 } \ \ , \ \
\cos 2 \theta_{\tilde{f}} = \frac{m_{LL}^2 -m_{RR}^2}
{m_{\tilde{f}_1}^2 -m_{\tilde{f}_2}^2 }, \hspace*{0.8cm}  \\
m_{\tilde{f}_{1,2}}^2 = m_f^2 +\frac{1}{2} \left[
m_{LL}^2 +m_{RR}^2 \mp \sqrt{ (m_{LL}^2
-m_{RR}^2 )^2 +4 m_f^2 \tilde{A}_f^2 } \ \right] 
\eeq
The physical states are defined as:

\begin{equation}
\widetilde{f}_1 = \widetilde{f} \cos \theta_{ \widetilde{f}}
+ \widetilde{f}^C \sin \theta_{ \widetilde{f}}
\end{equation}
and
\begin{equation}
\widetilde{f}_2 = - \widetilde{f} \sin \theta_{ \widetilde{f}}
+ \widetilde{f}^C \cos \theta_{ \widetilde{f}}
\end{equation}
In the case of sneutrinos we have:

\begin{equation}
{\cal M}^2_{\tilde{\nu}}= M^2_{\tilde{L}}+ \frac{1}{2} m^2_Z \cos 2 \beta
\end{equation}
Note the contributions of the different soft breaking parameters in the
mass matrices.

\section{Higgs Bosons}
As we have mentioned before, the existence of a scalar Higgs boson is the main
motivation to introduce SUSY. In the MSSM the
tree-level Higgs potential is given by \cite{Haber}\cite{HHG}:

\begin{equation}
V_{Higgs} = m^2_{H_1} \vert H_1 \vert^2 + m^2_{H_2} \vert H_2 \vert^2 +
m^2_{12} (H_1 H_2 +h.c)+
\end{equation}

\begin{equation}
+ \frac{g^2_1 + g^2_2}{8}(\vert H_1 \vert  ^2 - \vert
H_2 \vert ^2) ^2 + \frac{g^2_2}{2} \vert H^{\dagger}_1  H_2 \vert^2
\end{equation}
where $m^2_{H_i} = m^2_i + \vert \mu \vert ^2$ with  $i=1, 2$
\\
\\
Note that in this equation the strength of the quartic interactions is
determined by the gauge couplings.
\\
\\
After electroweak symmetry breaking, three of the eight degrees of
freedom contained in the two Higgs boson doublets get eaten by the
$W^{ \pm}$ and $Z$ gauge bosons. The five physical degrees of freedom that
remain form a neutral pseudoscalar boson $A^0$, two neutral scalar Higgs
bosons $h^0$ and $H^0$, and two charged Higgs bosons $H^+$ and $H^-$.
\\
\\
The physical pseudoscalar Higgs boson $A^0$ is a linear combination of
the imaginary parts of $H^0_1$ and $H^0_2$, which have the mass matrix
[ in the basis  $( \frac{Im \ H^0_1}{\sqrt{2}}, \frac{Im \ H^0_2}{\sqrt{2}}) $]:

\begin{equation}
\small{M^2_I = \left( \begin{array}{cc}
-m^2_{12} \ \tan \beta & -m^2_{12} \\
-m^2_{12}  & -m^2_{12} \ \cot \beta
\end{array} \right)}
\end{equation}

\begin{equation}
m^2_{A^0}= tr M^2_I = -2 m^2_{12} \ / \sin(2 \beta)
\end{equation}
The other neutral Higgs bosons are mixtures of the real parts of
$H^0_1$ and $H^0_2$, with tree-level mass matrix [$ \frac{Re \ H^0_1}{\sqrt{2}},
\frac{Re \ H^0_2}{ \sqrt{2}}$]:

\begin{equation}
\small{M^2_R = \left( \begin{array}{cc}
-m^2_{12} \tan \beta + m^2_Z \cos^2 \beta & m^2_{12} - \frac{1}{2} m^2_Z \sin 2 \beta
\\
m^2_{12} - \frac{1}{2} m^2_Z \sin 2 \beta & -m^2_{12} \cot \beta + m^2_Z \sin^2 \beta
\end{array} \right)}
\end{equation}
In this case the eigenvalues are:
\begin{equation}
m^2_{H^0,  h^0} = \frac{1}{2}[ m^2_{A^0} + m^2_Z \pm \sqrt{(m^2_{A^0}
+m^2_Z)^2 - 4m^2_{A^0}m^2_Z \cos^2(2 \beta)}]
\end{equation}
Explicitly the mass eigenstates are:

$$\begin{array}{l}
\left\{
\begin{array}{lllr}
G^0 \ &=& \frac{1}{\sqrt 2}[-\cos\beta \ Im \ H^0_1 + \sin \beta \ Im
\ H^0_2 ] , & \ \ \
Goldstone \ boson \to Z^0, \\
A^0 \ &=& \frac{1}{\sqrt 2}[ \sin\beta \ Im \ H^0_1 + \cos \beta \ Im
\ H^0_2 ] , & \
\ \ \ \ \ \ \ Pseudoscalar \ Higgs, \end{array}\right.\\  \\
\left\{
\begin{array}{lllr}
G^+ &=& \frac{1}{\sqrt 2}[ -\cos\beta \ (H^-_1)^* + \sin \beta \ H^+_2
] , & \ \ \ \ \ \ \
Goldstone \ boson \to W^+, \\
H^+ &=& \frac{1}{\sqrt 2} [ \sin\beta \ (H^-_1)^* + \cos \beta
\ H^+_2 ] , &\ \ \ Charged \ Higgs, \end{array}\right.\\ \\
\left\{
\begin{array}{lllr}
h^0 \ &=& \frac{1}{\sqrt 2}[ -\sin\alpha \ Re \ H^0_1 + \cos\alpha \
Re \ H^0_2 ] , & \ \ \ \ \ \
\ light \ Higgs, \\
H^0 \ &=& \frac{1}{\sqrt 2} [ \cos\alpha \ Re \ H^0_1 + \sin\alpha \
Re \ H^0_2 ] , &
\ \ \ \ \ \ \ heavy \ Higgs, \end{array}\right.
\end{array}$$
\\
\\
where the mixing angle $\alpha $ is given by:
\beq
\tan 2\alpha = -\tan 2\beta
\left(\frac{m^2_{A^0} + M^2_Z}{m^2_{A^0} - M^2_Z}\right)
\eeq
From these equations we can see that at tree level,
the MSSM predict that $m_{h^0} \leq m_Z$, 
however when we consider one-loop corrections, 
the mass of the
light Higgs boson is modified significantly.
For example assuming that the stop masses
do not exceed 1 TeV, $m_{h^0} \lsim 130 \ \rm{GeV}$ \cite{CarenaHaber}.

\section{The R-symmetry and its Implications}

In the Standard Model  the conservation of Baryon (B) and Lepton (L)
number is automatic, this is an accidental consequence of the gauge
group and matter content. In the MSSM, as we showed in the second section of
this chapter, we can separate the most general gauge invariant superpotential into two fundamental
parts, where the first term conserves B and L, while the second breaks
these symmetries.
\\
\\
In the MSSM, B and L conservation can be related to a new discrete symmetry, which can
be used to classify the two kinds of contributions to the superpotential.
This symmetry is the matter-parity, defined as:
\begin{equation}
M = (-1)^{3(B-L)}
\label{l2}
\end{equation}
Quark and lepton supermultiplets have $M=-1$, while the Higgs
 and gauge supermultiplets have $M= +1$. The symmetry principle in this case
will be that a term in the lagrangian is allowed only if the product of
the M parities is equal to 1.
\\
\\
The conservation of matter-parity as defined 
in equation (\ref{l2}), together with spin conservation, also implies
the conservation of another discrete symmetry called R-parity,
defined such that it will be +1 for the SM particles
and -1 for all the sparticles:

\begin{equation}
R =(-1)^{2S}M
\end{equation}
These two symmetries are equivalent, since in the superpotential only
the scalar fields get v.e.v. However only M commutes with the SUSY 
generators.
\\
\\
Now if we impose the conservation of $R$, we will have some important
phenomenological consequences in SUSY models:

\begin{itemize}

\item  The lightest particle with $R=-1$, called the lightest
supersymmetric particle (LSP), must be stable.

\item Each sparticle other than the LSP must decay into a state with an
odd number of LSPs.

\item Sparticles can only be produced in even numbers from SM
particles.

\item There are not $d=4$ operators contributing to the decay of the proton.

\end{itemize}
It is important to mention that the conservation of $R$ parity is
predicted in a large class of Supersymmetric Grand Unified Theories 
as Minimal SUSY $SO(10)$\cite{SO101,SO102}.

\chapter{SUSY decays of neutral Higgs bosons}

In this chapter we will analyze different aspects related to 
supersymmetric decays of the Higgs bosons in the Minimal
Supersymmetric Standard Model (MSSM), studying the Higgs decays into two
neutralinos. In particular we will compute and show
the effect of new loop corrections to the Higgs-neutralino-neutralino
couplings and to the invisible branching ratios.

\section{Higgs decays in the MSSM}

In order to provide a complete analysis of the most important aspects
of Higgs decays in the Minimal Supersymmetric Standard Model (MSSM),
we will start with the properties of the SM Higgs boson. The Higgs mass
$m^2_{h_{SM}} = \frac{1}{2} \lambda  v^2$ is a free parameter, since
$\lambda$ is unknown at present. However the theory predicts the Higgs
couplings to fermions and gauge bosons as:

\begin{equation}
g^{SM}_{h f \bar{f}} = \frac{m_f}{v}  \qquad \qquad g^{SM}_{h V V} = \frac{2 m^2_V}{v}
\end{equation}
where $f$ is used for any fermion, and $V$ for $W^{\pm}$ and $Z^0$.
In Fig 4.1 the different SM Higgs branching ratios versus $m_{h_{SM}}$ is plotted.
\begin{figure}
\begin{center}
\includegraphics[width=.7\textwidth]{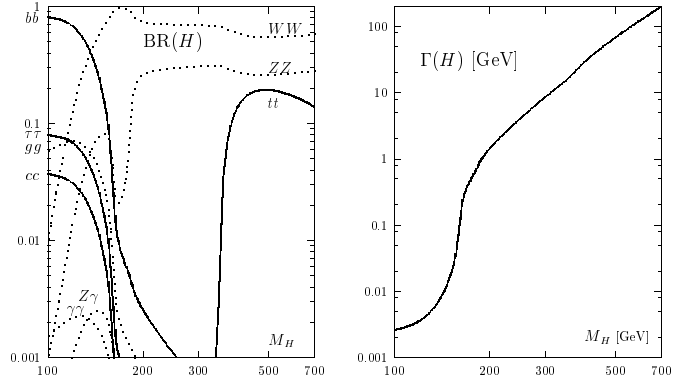}
\caption[]{\small The decay branching ratios (left) and the total decay width (right)
of the SM Higgs boson as a function of its mass.[from reference \cite{Djouadi}]}
\end{center}
\end{figure}
From this figure we can appreciate that there are two important
intervals, for $m_{h_{SM}} \lsim 135 \ GeV$ the most important channel
is $h_{SM} \to b \ \bar{b}$  with branching ratio close to $\sim 90\%$,
while for  $m_{h_{SM}} \gsim 135 \ GeV$ the dominant decay mode is
$h_{SM} \to W \ W^{*}$ (where one of the gauge bosons may be
virtual). This behaviour is easy to understand if we take a look at
the couplings listed above. There are other important channels such as
$h_{SM} \to \tau^{+} \ \tau^{-}$, $h_{SM} \to c \ \bar{c}$, and
at one-loop we have the decays $h_{SM} \to g \ g$ and
$h_{SM} \to \gamma \ \gamma$, which are important for Higgs searches \cite{CarenaHaber}.
In the low mass range, the Higgs width is very narrow, with $\Gamma <
10 \ MeV$, but increasing $m_{h_{SM}}$ we reach $1 \ GeV$ at the $Z^0
\ Z^0$ threshold (see Fig 4.1).
\\
\\
The same analysis in the context of the MSSM is more difficult, due to
the presence of three neutral Higgs bosons and one charged pair. There
are new decay modes which modify the branching ratios of all the
channels, in particular the decay into supersymmetric particles could
play an important role \cite{Gunion1,Gunion2,Zerwas1,Zerwas2}.
\\
\\
In order to understand how the branching ratio are modified, we list
the couplings of the neutral Higgs bosons to $f \ \bar{f}$ relative to
the Standard Model values:
\\
\\
\\
for the light CP-even Higgs $h^0$:
\begin{equation}
h^0 \ d_i \ \bar{d_i} ( \ or \ e_i \ \bar{e_i} \ ): - \frac{\sin \alpha }{\cos
\beta} \qquad \qquad
h^0 \ u_i \ \bar{u_i} : \frac{\cos \alpha }{\sin \beta}
\end{equation}
for the heavy CP-even $H^0$
\begin{equation}
H^0 \ d_i \ \bar{d_i} ( \ or \ e_i \ \bar{e_i} \ ): \frac{\cos \alpha }{\cos
\beta} \qquad \qquad
H^0 \ u_i \ \bar{u_i} : \frac{\sin \alpha }{\sin \beta}
\end{equation}
and for the CP-odd $A^0$ we have:
\begin{equation}
A^0 \ d_i \ \bar{d_i} ( \ or \ e_i \ \bar{e_i} \ ): \gamma_5 \tan \beta
\qquad \qquad
A^0 \ u_i \ \bar{u_i} : \gamma_5 \cot \beta
\end{equation}
\begin{figure}
\begin{center}
\includegraphics[width=.9\textwidth]{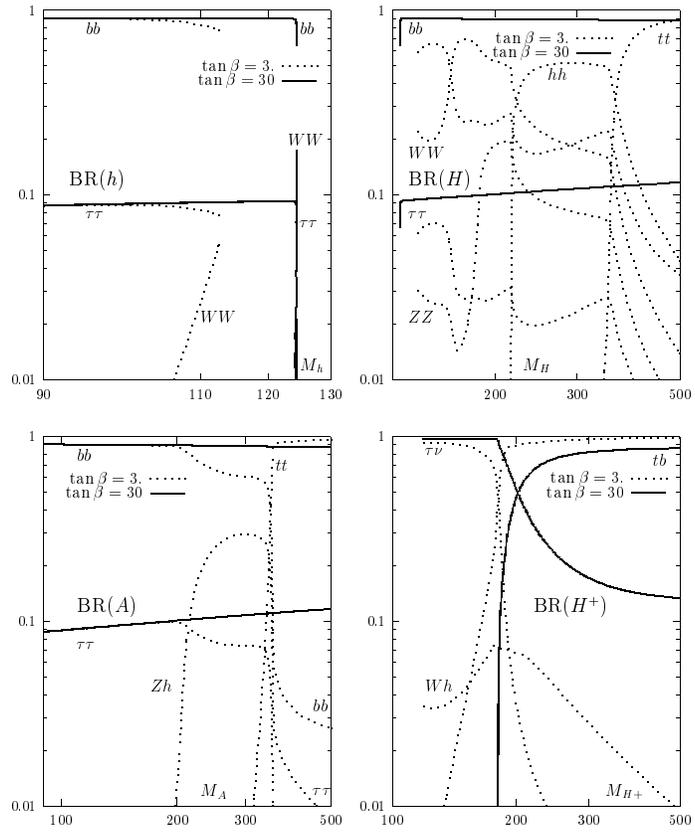}
\caption{\small Dominant MSSM Higgs bosons decay branching ratios as functions of
the Higgs boson masses for $\tb=3$ and 30.[from reference \cite{Djouadi}]}
\end{center}
\end{figure}
\\
while the couplings of the two CP-even Higgs bosons to $W^{\pm}$
and $Z^0$ pairs are given by:
\begin{equation}
g_{h^0 V   V} = g^{SM}_{h^0  V  V} \sin(\beta - \alpha)
\qquad \qquad
g_{H^0  V  V} = g^{SM}_{H^0  V  V} \cos(\beta - \alpha)
\end{equation}
\\
From the couplings listed above, we note that there are two new
parameters which will play an important role in the prediction of the
branching ratios of neutral Higgs bosons. For example the decay mode
$h^0 \to b \ \bar{b}$ could be significantly modified at large values
of $\tan \beta$ and/or small values of $\sin \alpha$. The prediction
of the branching ratios depends on the set of MSSM parameters, in
particular the spectrum of SUSY particles change appreciably the SUSY
Higgs decays.
\\ \\
In Fig 4.2 we show the different decay modes of the neutral Higgs
bosons for two different values of $\tan \beta $ as functions of the Higgs masses. The branching ratio of the
charged Higgs boson is also shown in order to complete our analysis.
As we know there is a limit for the light Higgs mass in the MSSM,
$m_{h^0} \lsim 130 \ GeV$\cite{CarenaHaber}, therefore $h^0$ will decay mainly into
fermion pairs, in particular the most important channel is $h^0 \to b
\ \bar{b}$. This is in general also the dominant decay mode of
the $H^0$ and $A^0$ bosons, since for $\tan \beta \gg 1$ the decay
rate into $b \bar{b}$ and $\tau^- \tau^+$ pairs are of the order
of $90 \%$ and $10 \%$, respectively. For large masses the
top decay channels $H^0, A^0 \to t \ \bar{t}$ are suppressed for
large $\tan \beta$.
\\
\\
In order to complete our analysis, the SUSY decays of the neutral Higgs
bosons must be considered, which could be dominant in different
regions of the parameter space. In general any Higgs boson could decay
into sfermions (squarks and sleptons), charginos or neutralinos.
However taking into account the lastest results of the SUSY searches
experiments \cite{LEP2}, we know that in the case of the light Higgs boson, the
only allowed SUSY channels are two neutralinos or two 
sneutrinos. For the heavy CP-even and CP-odd Higgs bosons the decays into
squarks or charginos are also allowed, excluding the decay of $A^0$
into two sneutrinos.
\\\\
These SUSY decays will be dominant, of course, when the
channels present in the Standard Model are suppressed. As we already
noted the most important decay models are the decays into  $b$ or
$\tau$ pairs, these channels are suppressed in the case of low or
moderate $\tan \beta$, or when the mixing angle $\alpha$ of the
Higgs sector is quite small. Combining these two scenarios, we will
able to get significant branching ratios for these channels.
\\
\\
The various decay widths and branching ratios of the SM and MSSM can be
calculated in a very precise way with the Fortran code HDECAY
\cite{HDECAY}, where all the relevant experimental constraints are
taken into account. The subroutines of HDECAY dealing with the decays
of neutral Higgs bosons into neutralinos use the results of reference \cite{loop}.

\section{Higgs boson decays into two Neutralinos}

In the Standard Model there are no invisible decays of the Higgs
boson, since no $\nu_R$ is present in the model. In the MSSM
the situation is quite different, there are new couplings which allow
new decays of the Higgs bosons. The MSSM neutral Higgs bosons $h^0, H^0$ and
$A^0$ could decay into invisible neutralino $\tilde{\chi}^0_1$ or
sneutrino $\tilde{\nu}$ pairs. These decay modes are
invisible in the case that the neutralinos or sneutrinos are 
the lightest SUSY particles (LSP) and $R$-parity is
conserved. Note that in the case of the CP-odd Higgs field, there is
only one possibility, the decays into two neutralinos, since the
coupling to two sneutrinos does not exist.  In this section we will
describe in detail neutral Higgs decays into two neutralinos
in the gaugino limit, considering quantum corrections at one-loop level.
\\
\\
Before computing and discussing the partial widths for the decays of the
neutral Higgs bosons into pairs of identical neutralinos, let us
discuss the properties of the couplings $g^0_{\Phi \ \tilde{\chi}_1^0
\ \tilde{ \chi}_1^0}$,
where $\Phi = h^0, H^0$ and $A^0$ and
$\tilde{\chi}_1^0$ is the lightest neutralino, in our case the
lightest supersymmetric particle (LSP).
\\
\\
At tree level, the couplings of the neutralinos $\tilde{\chi}_i^0$ to
the neutral CP-even Higgs bosons $\phi = h^0, H^0$  and to the CP-odd
boson $A^0$ are given by:

\begin{equation} \label{hcoup0}
  g^0_{\phi \tilde \chi_i^0 \tilde \chi_j^0}=  \frac{g}{2}
\left[ (N_{i2} - \tan\theta_W N_{i1} )
( d_\phi N_{j3} + e_\phi N_{j4} ) \ + \ i \leftrightarrow j \right]
\end{equation}
\begin{equation}\label{acoup0}
 g^0_{A^0 \tilde \chi_i^0 \tilde \chi_j^0} = \frac{g}{2}
\left[ (N_{i2} - \tan\theta_W N_{i1})
(d_{A^0} N_{j3} + e_{A^0} N_{j4}) \ + \ i \leftrightarrow j \right]
\end{equation}
where the quantities $d_{\Phi}$ and $e_{\Phi}$ are:

\begin{equation} \label{deh}
d_{H^0}= - \cos\alpha \ , \ d_{h^0} = \sin\alpha \ , \ d_{A^0}=
\sin\beta
\end{equation}
\begin{equation}
e_{H^0} = \sin\alpha \ , \  e_{h^0} = \cos\alpha \ , \  e_{A^0}= -\cos \beta .
\end{equation}
and $N_{ij}$ are the components of the matrix which diagonalizes the
four dimensional neutralino mass matrix.
\\
\\
Now using these equations we find the expressions for $g^0_{\Phi \
\tilde{\chi}_1^0 \ \tilde{\chi}_1^0}$ :
\begin{equation} \label{hcoup0}
 g^0_{h^0 \tilde \chi_1^0 \tilde \chi_1^0}=  g
\left[ (N_{12} - \tan\theta_W N_{11} )
( \sin \alpha N_{13} + \cos \alpha N_{14} ) \right]
\end{equation}
\begin{equation} 
 g^0_{H^0 \tilde \chi_1^0 \tilde \chi_1^0}=  g
\left[ (N_{12} - \tan\theta_W N_{11} )
( - \cos \alpha N_{13} + \sin \alpha N_{14} ) \right]
\end{equation}
\begin{equation}\label{acoup0}
 \gamma_5 \ g^0_{A^0 \tilde \chi_1^0 \tilde \chi_1^0} = \gamma_5 \ g
\left[ (N_{12} - \tan\theta_W N_{11})
( \sin \beta N_{13} - \cos \beta N_{14}) \right]
\end{equation}
we see that all these couplings are exactly zero in the \textit{pure}
gaugino $(N_{13}=N_{14}=0)$ or higgsino $(N_{11}=N_{12}=0)$ limit.
\\
\\
Inserting eqs.~(\ref{binostate}) into eqs.~(\ref{hcoup0}) to 
(\ref{acoup0}), we see that the LSP couplings to the Higgs bosons
$\Phi = h^0,H^0, A^0$ already receive contributions at ${\cal O}(m_Z)$:
\begin{equation}
g^{0}_{\Phi\tilde{\chi}_1^0\tilde{\chi}_1^0} \sim
d_\Phi N_{13} + e_\Phi N_{14}
 \sim  s_W  m_Z \left[ \frac{(d_\Phi s_\beta - e_\Phi c_\beta) \mu}
{\mu^2 - M_1^2} + \frac{ ( d_\Phi c_\beta - e_\Phi s_\beta ) M_1 }
{\mu^2 - M_1^2} \right]
\end{equation}
Similar expressions can be given for the couplings of the wino--like
state.
\\
\\
This suppression of the tree--level couplings follows from the
fact that, in the neutralino sector, the Higgs boson couples only to one
higgsino and one gaugino current eigenstate, together with the fact
that mixing between current eigenstates is suppressed if $|\mu| \gg
m_Z$. These couplings thus {\em vanish} as $|\mu| \rightarrow
\infty$.
\\
\\
Knowing all the properties of the couplings, and taking into account
that neutralinos are Majorana particles, we are able to compute
the partial widths for the decays of the neutral Higgs bosons,
$\Phi = h^0,H^0$, and  $A^0$, into pairs of identical neutralinos:
\begin{equation}
\Gamma (\Phi \rightarrow \tilde{\chi}_1^0 \tilde{\chi}_1^0) =
\frac{\beta_\Phi^n M_\Phi}{16\pi}
\left| g^0_{\Phi\tilde{\chi}_1^0\tilde{\chi}_1^0} +
 g^1_{\Phi \tilde{\chi}_1^0 \tilde{\chi}_1^0} \right|^2.
\end{equation}
where $n=3$ for the CP--even fields $h^0$ and $H^0$, while for the
CP--odd $A^0$ Higgs boson  $n=1$. $M_{\Phi}$ is the Higgs mass, and 
${\beta_\Phi}^2 = ( 1  -  \frac{4 \
m^2_{\tilde{\chi}_1^0}}{M^2_{\Phi}})$. We include the possible one-loop
corrections to the couplings $g^1_{\Phi \tilde{\chi}_1^0
\tilde{\chi}_1^0}$, which will be discussed in the next section.

\section{Higgs decays into Neutralinos at one-loop}
At the one-loop level, the couplings of the lightest neutralinos to
the Higgs bosons can be generated, in principle, by diagrams
with the exchange of either sfermions and fermions, or of charginos or
neutralinos together with gauge or Higgs bosons, in the loop. However,
the latter class of diagrams can contribute to the couplings of Higgs bosons
to neutralinos only if one of the particles in the loop is a
higgsino. These loop contributions will thus be suppressed by inverse
powers of $|\mu|$, in addition to the usual loop suppression factor,
since the couplings $g^0(H^\pm \tilde{\chi}_1^0 \tilde{\chi}_{1}^\mp)
\sim {\cal O}(m_W/\mu), \ g^0(W^\pm \tilde{\chi}_1^0
\tilde{\chi}_{1}^\mp) \sim {\cal O}(m^2_W/\mu^2)$ in the gaugino limit. We
therefore do not expect them to be able to compete with the
tree--level couplings that exist for finite $|\mu|$.
\\
\\
We consider diagrams with fermions and sfermions in the
loop, as shown below. For the $\Phi \tilde{\chi}_1^0
\tilde{\chi}_1^0$ couplings, only the third generation (s)particles,
which have large Yukawa couplings, can give significant contributions
to the amplitudes. Note that in the bino limit there is no wave
function renormalization to perform, since the tree--level couplings
are zero. Off--diagonal wave function
renormalization diagrams could convert one of the gaugino--like
neutralinos into a higgsino--like state, but this kind of contribution
is again suppressed by $1/|\mu|$, and can thus not compete with the
tree--level coupling.
\vspace*{-.5cm}
\begin{center}
\begin{picture}(1000,200)(70,0)
\DashArrowLine(80,100)(140,100){4}
\Text(110,110)[]{$\Phi$}
\ArrowLine(180,140)(140,100)
\ArrowLine(140,100)(180,60)
\DashArrowLine(180,60)(180,140){4}{}
\ArrowLine(230,140)(180,140)
\ArrowLine(180,60)(230,60)
\Text(240,130)[]{$\tilde\chi_1^0 (p_1)$}
\Text(240,70)[]{$\tilde\chi_1^0 (p_2)$}
\Text(190,100)[]{$\tilde{f}_i$}
\Text(160,130)[]{$f$}
\Text(160,70)[]{$f$}
\DashArrowLine(250,100)(310,100){4}
\Text(280,110)[]{$\Phi$}
\DashArrowLine(310,100)(350,140){4}{}
\DashArrowLine(350,60)(310,100){4}{}
\ArrowLine(350,140)(350,60)
\ArrowLine(400,140)(350,140)
\ArrowLine(350,60)(400,60)
\Text(410,130)[]{$\tilde\chi_1^0(p_1)$}
\Text(410,70)[]{$\tilde\chi_1^0 (p_2)$}
\Text(110,140)[]{${\bf a)}$}
\Text(280,140)[]{${\bf b)}$}
\Text(360,100)[]{$f$}
\Text(330,130)[]{$\tilde{f}_i$}
\Text(330,70)[]{$\tilde{f}_j$}
\end{picture}
\end{center}
\vspace*{-1.5cm}
{\small{The Feynman diagrams contributing to the one--loop couplings
of the lightest neutralinos to the $\Phi = h^0, H^{0}$ and $A^0$ Higgs
bosons. Diagrams with crossed neutralino lines have to be added.}}
\\
\\
We have calculated the contributions of these diagrams for arbitrary
momentum square of the Higgs, finite masses for the
internal fermions and sfermions as well as for the external LSP
neutralinos, and taking into account the full mixing in the sfermion
sector. The amplitudes are ultra--violet finite as it should be. The
contributions from diagrams a) and b) to the $\Phi \tilde{\chi}_1^0
\tilde{\chi}_1^0$ couplings are separately finite for each fermion
species.\footnote{The contribution of diagram a) is finite only after
summation over both sfermion mass eigenstates.} We have performed
the calculation in the dimensional reduction scheme \cite{Siegel,Jones1};
since the one--loop couplings are finite
and do not require any renormalization, the result should be scheme 
independent. The results are given below for a general gaugino limit
($|\mu| \gg M_1, M_2$, for arbitrary ordering of $M_1$ and $M_2$).
\\
\\
The one--loop Higgs boson couplings to the LSP
neutralinos in the gaugino limit are given by:
\begin{equation} \label{goneh}
g^{1}_{\phi\tilde{\chi}_1^0\tilde{\chi}_1^0} = \frac{g}{4\pi^2} \bigg[
\sum_f N_c \delta_\phi^{(f)} \bigg]
\end{equation}
\begin{equation}
g^{1}_{A^0 \tilde{\chi}_1^0\tilde{\chi}_1^0} =
\frac{g}{4\pi^2} \bigg[ \sum_f N_c \delta_{A^0}^{(f)} \bigg]
\end{equation}
where
\beq \label{hloop}
\delta_\phi^{(f)} &=& \frac{m_f g_{\phi ff}}{2m_W} \left\{
\sq (v_1+v_3)\left[ - \left(\msqa^2 + m_f^2 + \mx^2 \right) C_0(\sqa) +
4 \mx^2 C_1^+(\sqa) \right.\right. \non\\
&& \ \ \ + \left. \left(\msqb^2 + m_f^2 + \mx^2 \right) C_0(\sqb) -
4 \mx^2 C_1^+(\sqb) \right] \non\\
&& + \ 2(v_1+v_2 c^2_{\theta_{\tilde{f}}}) m_f\mx
\left[ C_0(\sqa) - 2C_1^+(\sqa) \right] \non \\
&& + \left. 2(v_1+v_2 s^2_{\theta_{\tilde{f}}}) m_f\mx \left[ C_0(\sqb)
- 2C_1^+(\sqb) \right] \right\} \non \\
&-& C_{\phi\sqa\sqa} \left\{ -\sq (v_1+v_3) m_f C_0(\sqa,\sqa) +
2(v_1+v_2 c^2_{\theta_{\tilde{f}}}) \mx C_1^+(\sqa,\sqa) \right\} \non\\
&-& C_{\phi\sqb\sqb} \left\{ \sq (v_1+v_3) m_f C_0(\sqb,\sqb) +
2(v_1+v_2 s^2_{\theta_{\tilde{f}}}) \mx C_1^+(\sqb,\sqb) \right\} \non\\
&-& C_{\phi\sqa\sqb} \left\{ -2 \cq (v_1+v_3) m_f C_0(\sqa,\sqb)
 -2 \sq v_2\mx C_1^+(\sqa,\sqb)
 \right\}
\eeq

\begin{displaymath}
\delta_{A^0}^{(f)} = \frac{m_f g_{A^0 ff}}{2 m_W} \left\{ (v_1+v_3) \sq
\left[\left( \msqa^2 - m_f^2 -\mx^2 \right) C_0(\sqa) -
\left( \msqb^2 - m_f^2 -\mx^2 \right) C_0(\sqb) \right]\right\} 
\end{displaymath}

\begin{displaymath}
+ \frac{m_f g_{A^0 ff}}{2 m_W} \left\{ 2(v_1+v_2
c^2_{\theta_{\tilde{f}}}) \mx m_f C_0(\sqa) +
2(v_1+v_2 s^2_{\theta_{\tilde{f}}})\mx m_f C_0(\sqb) \right\} 
\end{displaymath}

\beq
&+& C_{A^0 \sqa\sqb} \left\{ -2 m_f (v_1+v_3) C_0(\sqa,\sqb)
+ 2\mx (2v_1 + v_2) \sq C_1^-(\sqa,\sqb) \right\} \non \\
&&
\eeq
The Higgs--fermion--fermion coupling constants are given by
\beq \label{hffcoup}
g_{h^0 uu} &=& \frac{\cos\alpha}{\sin\beta} \qquad , \qquad
g_{h^0 dd} = - \frac{\sin\alpha}{\cos\beta}\ , \non\\
g_{H^0 uu} &=& \frac{\sin\alpha}{\sin\beta} \qquad , \qquad
g_{H^0 dd} = \frac{\cos\alpha}{\cos\beta} \ , \non\\
g_{A^0 uu} &=& \cot\beta \qquad , \qquad g_{A^0 dd} = \tan\beta \ ,
\eeq
and 
\beq \label{gauginocoup}
v_1 & = & \ \frac{1}{2} \left( g Q_f N_{11} \tan\theta_W \right)^2, \non \\
v_2 &=& \frac{I_3^f}{2 Q_f} v_0 \left[ -2 +\frac{I_3^f}{Q_f}
 + 2 \left( 1 - \frac{I_3^f}{Q_f} \right) \frac{N_{12}} {N_{11} \tan \theta_W}
 + \frac{I_3^f}{Q_f} \left( \frac{N_{12}} {N_{11} \tan \theta_W} \right)^2
 \right] , \non \\
v_3 &=&  \frac{I_3^f}{2 Q_f} v_0 \left( \frac {N_{12}} {N_{11}
\tan \theta_W} -1 \right).
\eeq

%
while the Higgs--sfermion--sfermion coupling constants read:
\begin{displaymath} \label{hsfsfcoup}
C_{h^0 \tilde{u}_1\tilde{u}_1} = \frac{m_Z}{c_W}
s_{\beta+\alpha} \left[ I_{3}^u c^2_{\theta_{\tilde{u}}}
-Q_u s^2_W c_{2\theta_{\tilde{u}}}\right]
-\frac{m_u^2 g_{h^0 uu}}{m_W} -
\frac{m_u s_{2\theta_{\tilde{u}}}}{2m_W}
\left[ A_u g_{h^0 uu} + \mu g_{H^0 uu} \right] 
\end{displaymath}

\begin{displaymath}
C_{h^0 \tilde{u}_2\tilde{u}_2} = \frac{m_Z}{c_W}
s_{\beta+\alpha} \left[ I_{3}^u s^2_{\theta_{\tilde{u}}}
+ Q_u s^2_W c_{2\theta_{\tilde{u}}}\right]
-\frac{m_u^2 g_{h^0 uu}}{m_W} +
\frac{m_u s_{2\theta_{\tilde{u}}}}{2m_W} \left[ A_u g_{h^0 uu} +
\mu g_{H^0 uu} \right] 
\end{displaymath}

\begin{displaymath}
C_{h^0 \tilde{u}_1\tilde{u}_2} = \frac{m_Z}{c_W}
s_{\beta+\alpha} \left[ Q_u s^2_W
-I_{3}^u/2 \right] s_{2\theta_{\tilde{u}}}
-\frac{m_u}{2m_W} \left[ A_u g_{h^0 uu} +
\mu g_{H^0 uu} \right] c_{2\theta_{\tilde{u}}} 
\end{displaymath}

%

\begin{displaymath}
C_{h^0 \tilde{d}_1\tilde{d}_1} = \frac{m_Z}{c_W}
s_{\beta+\alpha} \left[ I_{3}^d c^2_{\theta_{\tilde{d}}}
-Q_d s^2_W c_{2\theta_{\tilde{d}}}\right]
-\frac{m_d^2 g_{h^0 dd}}{m_W} -
\frac{m_d s_{2\theta_{\tilde{d}}}}{2m_W}
\left[ A_d g_{h^0 dd} - \mu g_{H^0 dd} \right] 
\end{displaymath}


\begin{displaymath}
C_{h^0 \tilde{d}_2\tilde{d}_2} = \frac{m_Z}{c_W}
s_{\beta+\alpha} \left[ I_{3}^d s^2_{\theta_{\tilde{d}}}
+ Q_d s^2_W c_{2\theta_{\tilde{d}}}\right]
-\frac{m_d^2 g_{h^0 dd}}{m_W} +
\frac{m_d s_{2\theta_{\tilde{d}}}}{2m_W} \left[ A_d g_{h^0 dd} -
\mu g_{H^0 dd} \right] 
\end{displaymath}

\begin{displaymath}
C_{h^0 \tilde{d}_1\tilde{d}_2} = \frac{m_Z}{c_W}
s_{\beta+\alpha} \left[ Q_d s^2_W-I_{3}^d/2 \right] s_{2\theta_{\tilde{d}}}
-\frac{m_d}{2m_W} \left[ A_d g_{h^0 dd} -
\mu g_{H^0 dd} \right] c_{2\theta_{\tilde{d}}}
\end{displaymath}
%

\begin{displaymath}
C_{H^0 \tilde{u}_1\tilde{u}_1} = -\frac{m_Z}{c_W}
c_{\beta+\alpha} \left[ I_{3}^u c^2_{\theta_{\tilde{u}}}
-Q_u s^2_W c_{2\theta_{\tilde{u}}}\right]
-\frac{m_u^2 g_{H^0 uu}}{m_W} -
\frac{m_u s_{2\theta_{\tilde{u}}}}{2m_W} \left[ A_u g_{H^0 uu} -
\mu g_{h^0 uu} \right] 
\end{displaymath}

\begin{displaymath}
C_{H^0 \tilde{u}_2\tilde{u}_2} = -\frac{m_Z}{c_W}
c_{\beta+\alpha} \left[ I_{3}^u s^2_{\theta_{\tilde{u}}}
+Q_u s^2_W c_{2\theta_{\tilde{u}}}\right]
-\frac{m_u^2 g_{H^0 uu}}{m_W} +
\frac{m_u s_{2\theta_{\tilde{u}}}}{2m_W} \left[ A_u g_{H^0 uu} -
\mu g_{h^0 uu} \right]
\end{displaymath}

\begin{displaymath}
C_{H^0 \tilde{u}_1\tilde{u}_2} = -\frac{m_Z}{c_W}
c_{\beta+\alpha} \left[ Q_u s^2_W-I_{3}^u/2 \right] s_{2\theta_{\tilde{u}}}
-\frac{m_u}{2m_W} \left[ A_u g_{H^0 uu} -
\mu g_{h^0 uu} \right] c_{2\theta_{\tilde{u}}} 
\end{displaymath}


\begin{displaymath}
C_{H^0 \tilde{d}_1\tilde{d}_1} = -\frac{m_Z}{c_W}
c_{\beta+\alpha} \left[ I_{3}^d c^2_{\theta_{\tilde{d}}}
-Q_d s^2_W c_{2\theta_{\tilde{d}}}\right]
-\frac{m_d^2 g_{H^0 dd}}{m_W} -
\frac{m_d s_{2\theta_{\tilde{d}}}}{2m_W} \left[ A_d g_{H^0 dd} +
\mu g_{h^0 dd} \right] 
\end{displaymath}

\begin{displaymath}
C_{H^0 \tilde{d}_2\tilde{d}_2} = -\frac{m_Z}{c_W}
c_{\beta+\alpha} \left[ I_{3}^d s^2_{\theta_{\tilde{d}}}
+Q_d s^2_W c_{2\theta_{\tilde{d}}}\right]
-\frac{m_d^2 g_{H^0 dd}}{m_W} +
\frac{m_d s_{2\theta_{\tilde{d}}}}{2m_W} \left[ A_d g_{H^0 dd} +
\mu g_{h^0 dd} \right] 
\end{displaymath}

\begin{displaymath}
C_{H^0 \tilde{d}_1\tilde{d}_2} = -\frac{m_Z}{c_W}
c_{\beta+\alpha} \left[ Q_d s^2_W-I_{3}^d/2 \right] s_{2\theta_{\tilde{d}}}
-\frac{m_d}{2m_W} \left[ A_d g_{H^0 dd} +
\mu g_{h^0 dd} \right] c_{2\theta_{\tilde{d}}} 
\end{displaymath}

%

\beq
C_{A^0 \tilde{u}_1\tilde{u}_2} = \frac{m_u}{2m_W} (A_u \cot\beta
+\mu)
\eeq
\beq
C_{A^0 \tilde{d}_1\tilde{d}_2} = \frac{m_d}{2m_W} (A_d \tan\beta
+\mu) 
\eeq
with $s_{\alpha + \beta} = \sin (\alpha + \beta)$, $c_{\alpha + \beta}
= \cos (\alpha + \beta)$, $s_{\theta_{\tilde {f}}} = \sin
\theta_{\tilde{f}}$, $c_{\theta_{\tilde {f}}} = \cos
\theta_{\tilde{f}}$ and $I^{u(d)}_3 = (-) \frac{1}{2}$.
The Passarino--Veltman three--point functions, are defined as
\beq \label{loopfun}
C_{0,1}^{+,-} (\tilde{f}) &\equiv & C_{0,1}^{+,-}
(q^2, m_{\tilde\chi_1^0}^2, m_f^2, m_f^2, m_{\tilde{f}}^2) ; \non \\
C_{0,1}^{0,+,-} (\tilde{f}_1, \tilde{f}_2) & \equiv & C_{0,1}^{+,-}
(q^2, m_{\tilde\chi_1^0}^2, m_{\tilde{f}_1}^2, m_{\tilde{f}_2}^2, m_f^2),
\eeq
see Appendix A for the explicit form of these functions.
\\
\\
Now knowing all the details about the tree level couplings and the
quantum corrections considered, we are able to show few numerical
examples to see the effect of the quantum corrections to the branching
ratios of the neutral Higgs bosons decays into two neutralinos.
\\
\\
The real part of the coupling of an on--shell lightest $h^0$ boson to a
LSP pair is displayed, at the Born and one--loop level, in the
Figs.~4.3 and 4.4. It is shown as a function of the
sfermion masses, for $\tan \beta = 15$, $\mu=1$ TeV and pseudoscalar
mass input values $M_{A^0}=200$ GeV (Fig.~4.3) and 1 TeV (Fig.~4.4). 
Top quarks couple with ${\cal O}(1)$ Yukawa coupling to the Higgs bosons, and for
the given choice of large $|A_t|$ the (dimensionful) $h^0 \tilde t_1 \tilde t_1$ coupling
significantly exceeds the $\tilde t_1$ mass. Moreover, due to $\tilde
t_L - \tilde t_R$ mixing the lighter $\tilde t$ mass eigenstate is
often not only lighter than the other squarks, but also lighter than
the sleptons. If $|A_t|$ is large, as in the present
example, the loop corrections to the $h^0 \ \tilde{\chi}_1^0 \ \tilde{\chi}_1^0 $
coupling can even exceed the tree-level contribution.
%
%
\begin{figure}
\begin{center}
\includegraphics[width=.5\textwidth]{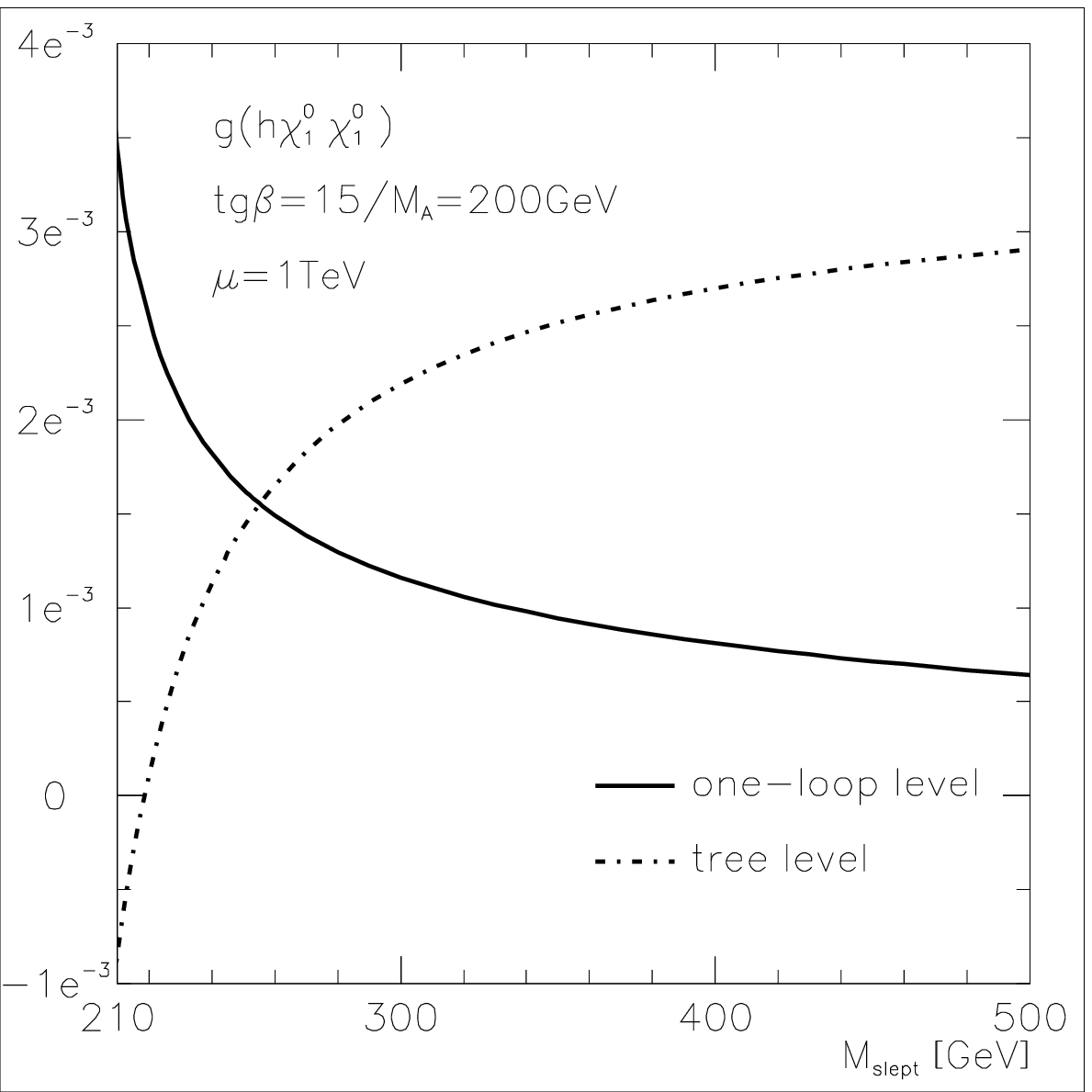}
\includegraphics[width=.5\textwidth]{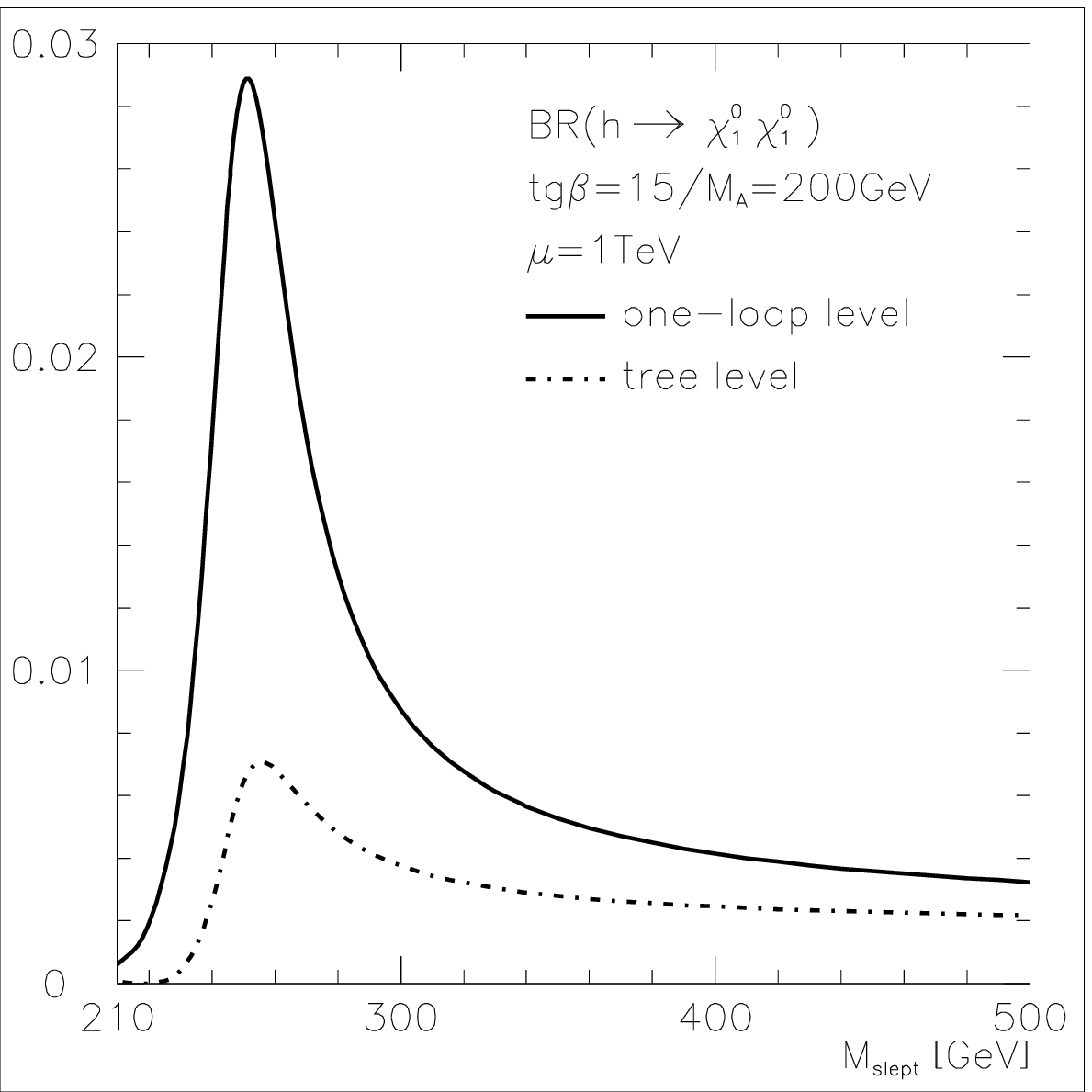}
\caption{\small The lightest $h^0$ boson couplings (top) and branching
ratios (bottom) to pairs of the lightest neutralinos as functions of
the common slepton mass. These results are given for $\tan \beta = 15, \mu
=1$ TeV, $m_{\tilde{q}} = 2 m_{\tilde{l}}, A_t = 2.9 m_{\tilde{q}}$
and gaugino masses $ M_1 = 30 $ GeV, $ M_2 = 120 $ GeV, and we took
$M_{A^0} = 200 $ GeV}
\end{center}
\end{figure}
%
%
\begin{figure}
\begin{center}
\includegraphics[width=.5\textwidth]{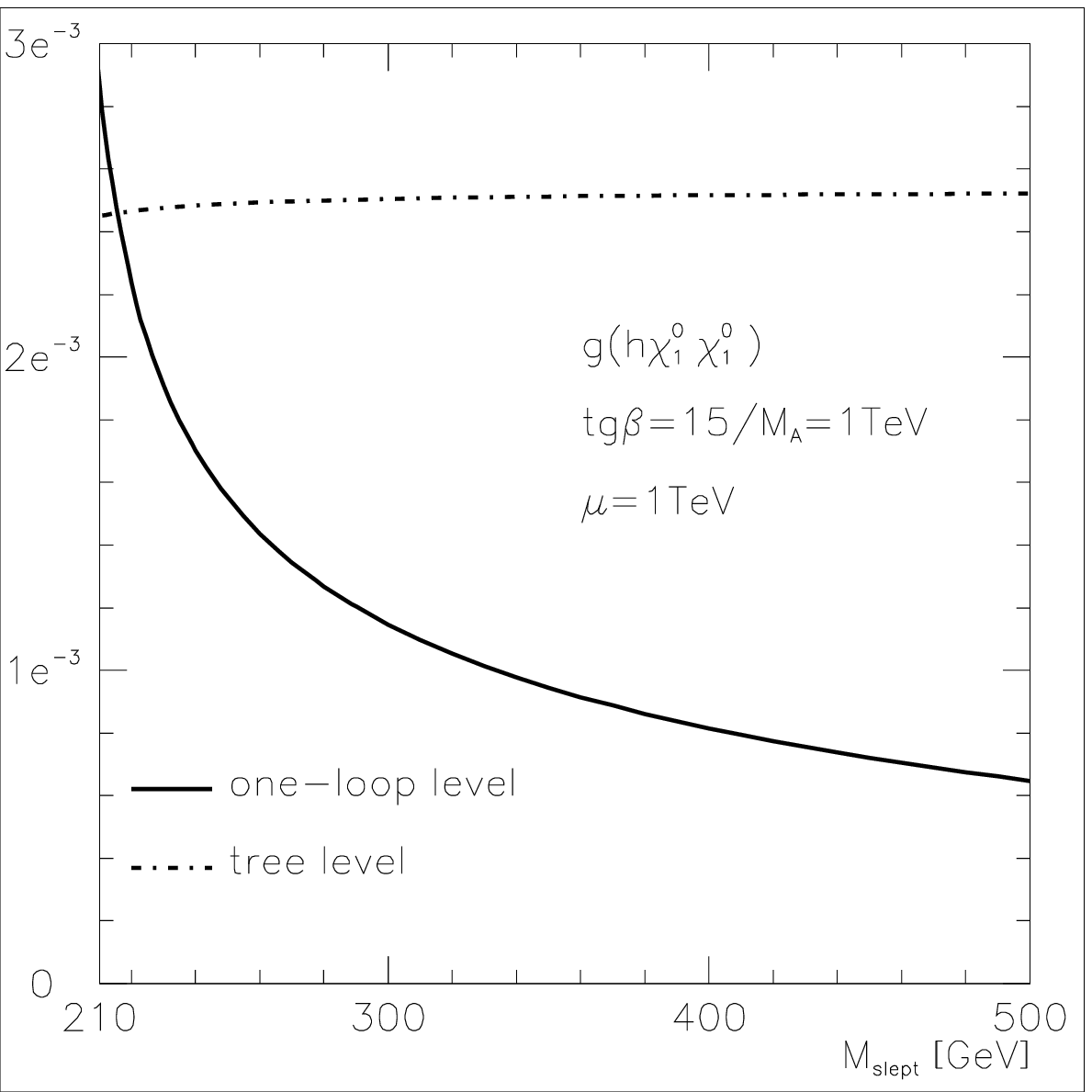}
\includegraphics[width=.5\textwidth]{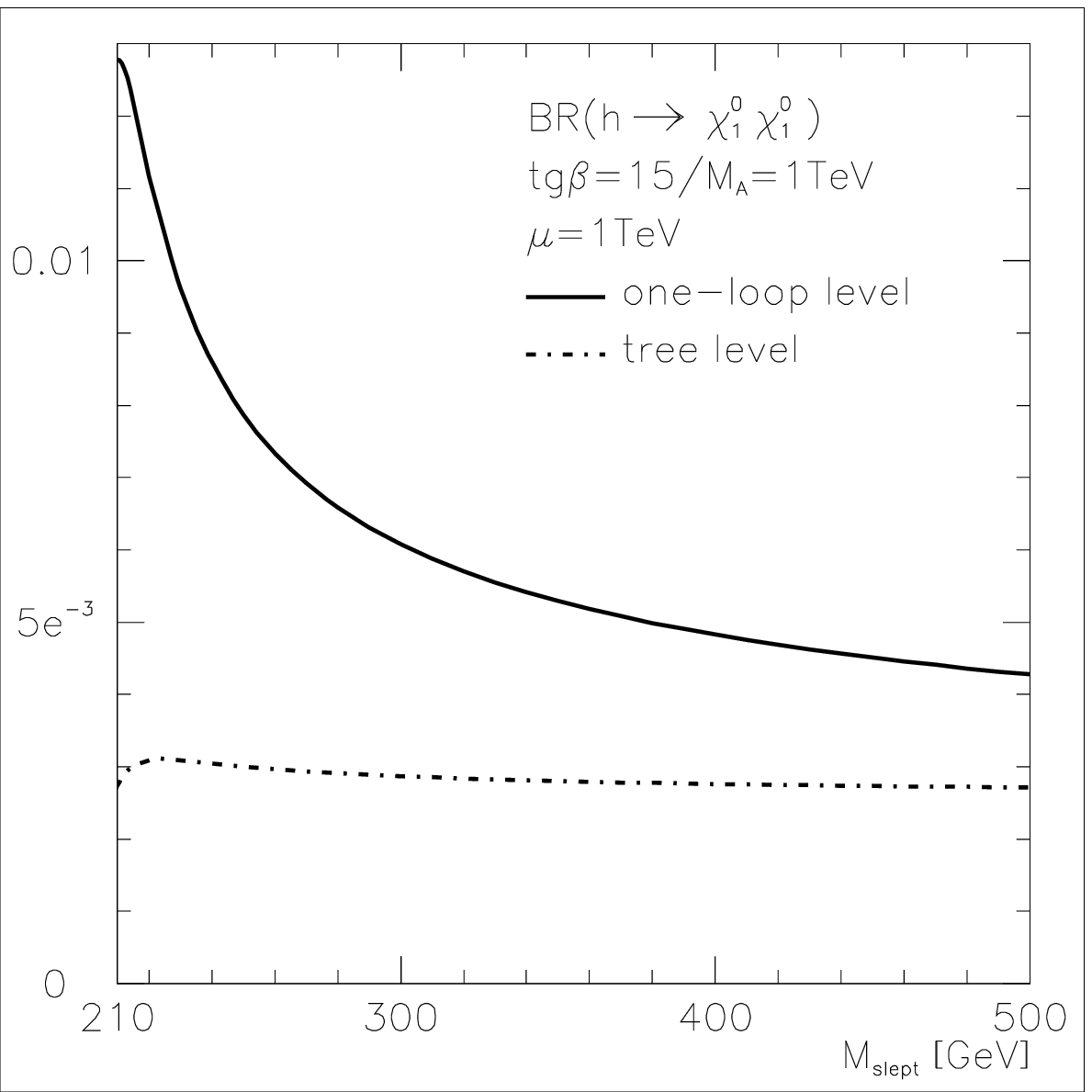}
\caption{\small The lightest $h^0$ boson couplings (top) and branching
ratios (bottom) to pairs of the lightest neutralinos as functions of
the common slepton mass. Most parameters are as in Fig.~4.3, and we took
$M_{A^0} = 1$ TeV}
\end{center}
\end{figure}
%
The variation of the one-loop contribution to the coupling is again
mostly due to the natural decrease with increasing masses of the
sfermions running in the loop, which decouple when they are much
heavier than the $h^0$ boson. For $m_{\tilde q} \simeq 420$ GeV
[i.e. $m_{\tilde l} \simeq 210$ GeV], $m_{\tilde t_1}$ is near its
experimental lower bound of $\sim 100$ GeV, due to strong $\tilde t_L -
\tilde t_R$ mixing. This implies that $m_{\tilde t_1}$ will grow
faster than linearly with increasing $m_{\tilde q}$, which explains
the very rapid decrease of the loop corrections. However, there is
also a variation of the tree--level coupling for $M_{A^0} = 200$ GeV which,
at first sight, is astonishing. It is caused by the variation of the
mixing angle $\alpha$ in the CP--even Higgs sector, and to a lesser
extent by the variation of $M_{h^0}$, due to the strong dependence of
crucial loop corrections in the CP--even Higgs sector on the stop
masses. In fact, for the set of input parameters with $M_{A^0}= 200$ GeV at small
slepton masses, we are in the regime where $\sin \alpha$, which
appears in the $h^0 \tilde{\chi}_1^0 \tilde{\chi}_1^0$ coupling [and
which enters the $h^0 b\bar{b}$ coupling as will be discussed later],
varies very quickly. This ``pathological" region, where the
phenomenology of the MSSM Higgs bosons is drastically affected, has
been discussed in several places in the literature \cite{Carena1,Carena2}.
\\
\\
The branching ratios for the decays of the lightest $h^0$ boson are
shown in the Figs.~4.3 and 4.4, for the same choice of
parameters previously discussed. They have been calculated by
implementing the one--loop Higgs couplings to neutralinos in the
Fortran code {\tt HDECAY} \cite{HDECAY} which calculates all possible
decays of the MSSM Higgs bosons and where all important corrections in
the Higgs sector, in both the spectrum and the various decay widths,
are included. The branching ratio BR$(h^0 \rightarrow
\tilde{\chi}_1^0\tilde{\chi}_1^0)$
can already exceed the one permille level with tree--level couplings.
After including the one-loop corrections, the branching fraction
$BR(h^0 \rightarrow \tilde{\chi}_1^0 \tilde{\chi}_1^0)$ can be
enhanced to reach the level of a few percent.
\\
\\
The branching ratio is especially enhanced if the usually dominant
decay into $b \bar b$ pairs is suppressed, i.e. if $|\sin \alpha|$
is very small; recall that the $h^0 b \bar b$ coupling is $\propto \sin
\alpha / \cos \beta$. In our examples this happens for $M_{A^0}=200$ GeV
and $m_{\tilde{l}} \simeq 250$ GeV. In this case $g^0_{h^0 \tilde \chi_1^0
\tilde \chi_1^0}$ is only about half as large as in the decoupling
limit $M_{A^0} \rightarrow \infty$, but the loop contribution to this
coupling is still sizable for this value of the sfermion masses, and
has the same sign as the tree--level coupling, leading to a quite
large total coupling. The branching ratio falls off quickly for
smaller sfermion masses, since here $\sin \alpha$ becomes sizable (and
positive). Moreover, for $m_{\tilde l} \simeq 210$ GeV the tree--level
and one--loop contributions to the couplings have opposite sign. The
branching ratio also decreases when $m_{\tilde l}$ is raised above 250
GeV, albeit somewhat more slowly; here the rapid decrease of the loop
contribution is compensated by the increase of the tree--level
coupling, which however does not suffice to compensate the
simultaneous increase of $ \sin^2 \alpha$.
\\
\\
Fig.~4.5 shows the dependence of the $h^0 \ \lsp \ \lsp $
coupling and of the corresponding $h^0$ branching ratio on
the mass of the LSP, $m_{\tilde \chi_1^0} \simeq M_1$.
We see that the tree--level contribution to
this coupling depends essentially linearly on the LSP mass.
Eqs.~(\ref{hcoup0}) and (\ref{binostate}) show that, for the given
scenario where $c_\alpha \simeq s_\beta \simeq 1$, this linear
dependence on $M_1$ originates from $N_{14}$, where the contribution
with $M_1$ in the numerator is enhanced by a factor of $\tan\beta$
relative to the contribution with $\mu$ in the numerator. Therefore
the contribution $\propto M_1$ is not negligible even though in Fig.~4.5
we have $M_1 \ll |\mu|$. On the other hand, the one--loop contribution
to this coupling depends only very weakly on $M_1$. The small increase
of this contribution shown in Fig.~4.5 is mostly due to the explicit
$m_{\tilde \chi_1^0}$ dependence of the loop coupling (equation \ref{hloop});
the change of $N_{12}$ with increasing $M_1$, as described by
eq.~(\ref{binostate}), plays a less important role. The increase of the
total coupling with increasing $M_1$ nevertheless remains
significant. However, Fig.~4.5 shows that for
$m_{\tilde \chi_1^0} \geq 15$ GeV this increase of the coupling is
over--compensated by the decrease of the $\beta^3$ threshold factor in
the expression for the $\Gamma (h^0 \to \tilde{\chi_1^0}
\tilde{\chi_1^0})$ partial width.
\\
\\
Once one-loop corrections are included, for certain values of the
MSSM parameters the branching ratio for invisible $h^0$ boson decays can
thus reach the level of several percent even if $\widetilde{\chi}_1^0$ is an almost
purely bino. This would make the detection of these decays possible at
the next generation of $e^+e^-$ linear colliders. At such a collider
it will be possible to isolate $e^+ e^- \rightarrow Z^0 h^0$ production
followed by $Z^0 \rightarrow \ell^+ \ell^-$ decays ($\ell = e$ or $\mu$)
{\em independent} of the $h^0$ decay mode, simply by studying the
distribution of the mass recoiling against the $\ell^+ \ell^-$
pair. This allows accurate measurements of the various $h^0$ decay
branching ratios, including the one for invisible decays, with an
error that is essentially determined by the available statistics
\cite{R11}. Since a collider operating at $\sqrt{s} \sim 300$ to 500
GeV should produce $\sim 10^5$ $Z^0 h^0$ pairs per year if $|\sin(\alpha -
\beta)| \simeq 1$ one should be able to measure an invisible branching
ratio of about 3\% with a relative statistical uncertainty of about
2\%.
%
\begin{figure}
\begin{center}
\includegraphics[width=.5\textwidth]{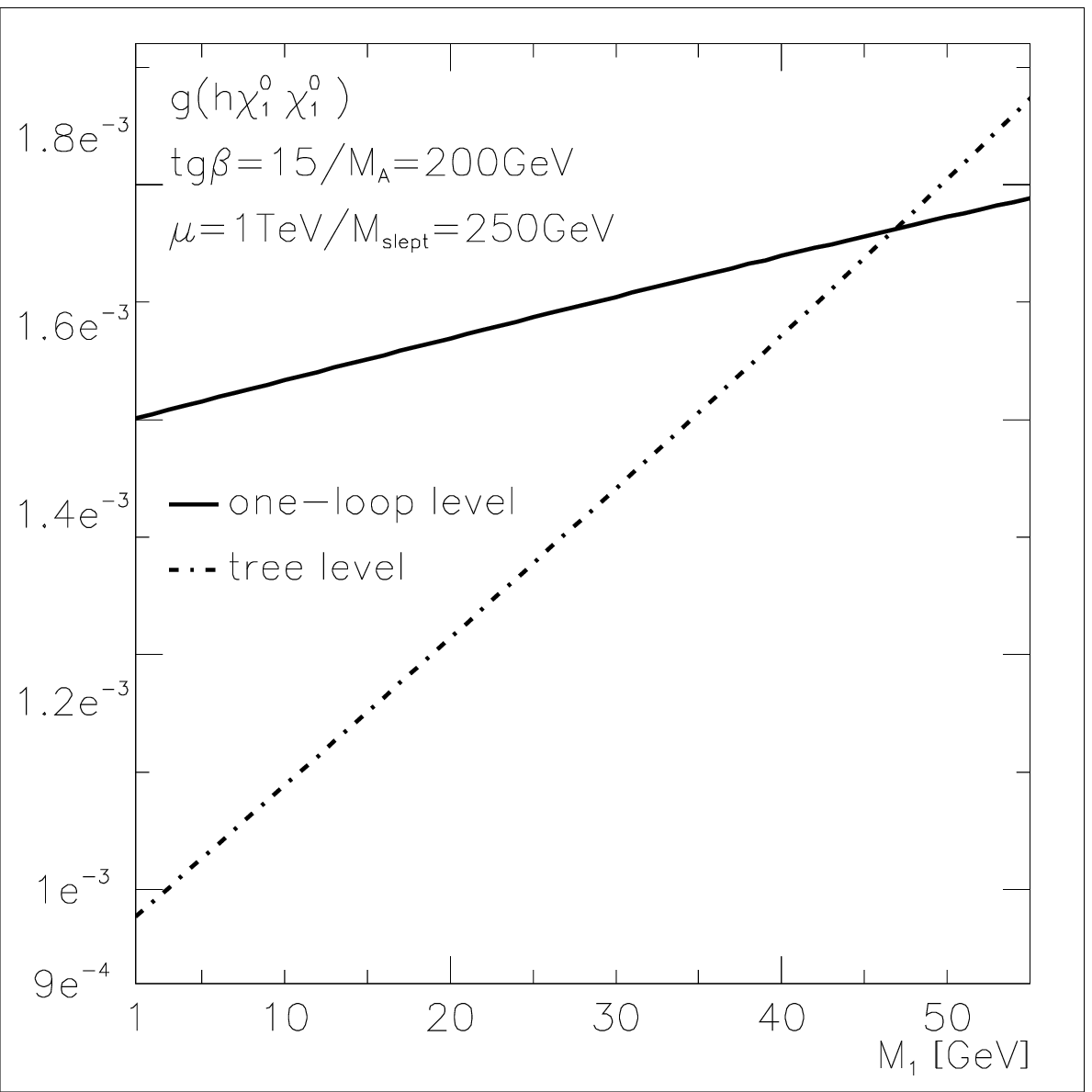}
\includegraphics[width=.5\textwidth]{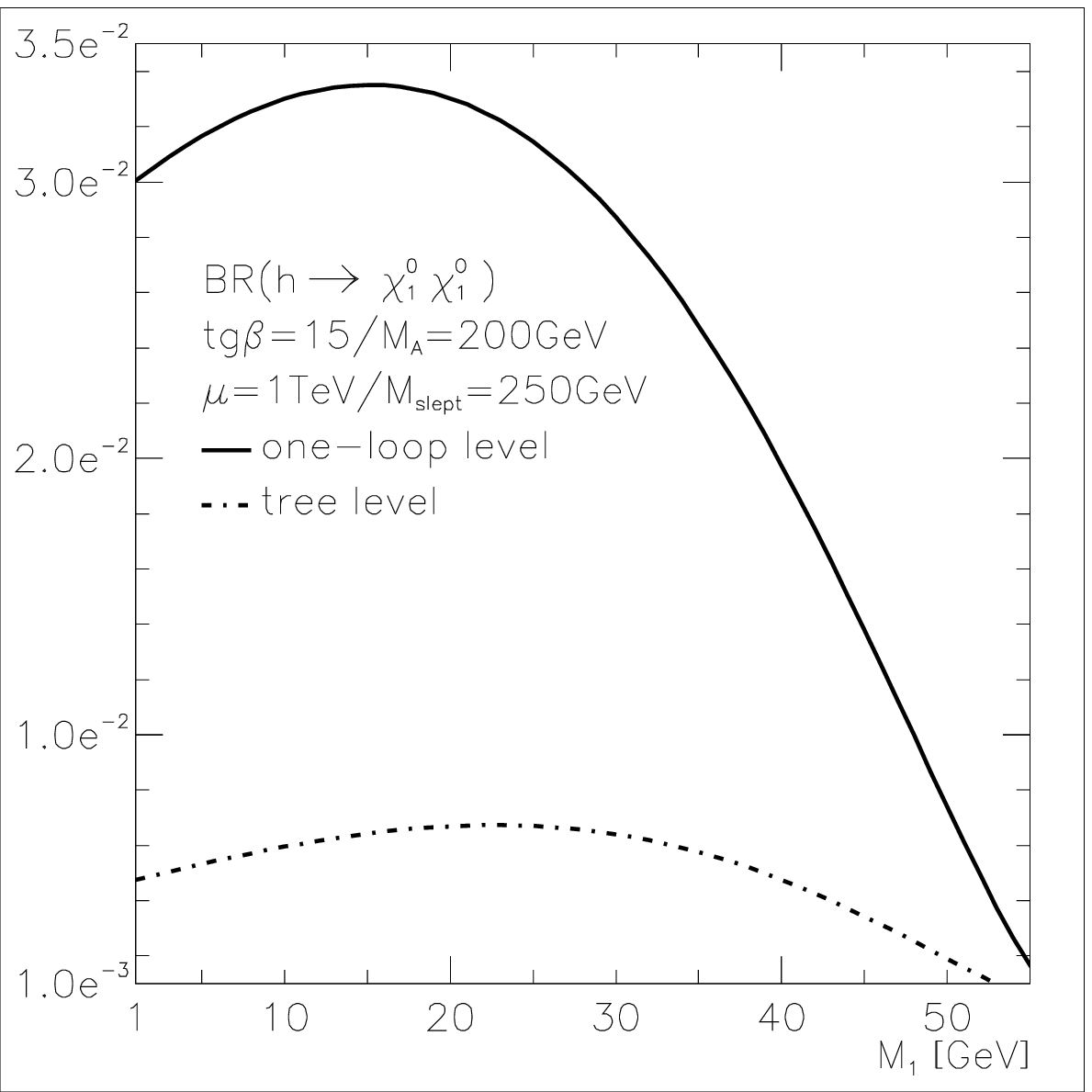}
\caption{\small The lightest $h^0$ boson couplings (top) and branching
ratios (bottom) to pairs of the lightest neutralinos as functions of $M_1$. The parameters are
as in Fig.~4.3 with $m_{\tilde{l}}=250$ GeV and $M_{A^0} = 200$ GeV.}
\end{center}
\end{figure}
%
\begin{figure}
\begin{center}
\includegraphics[width=.5\textwidth]{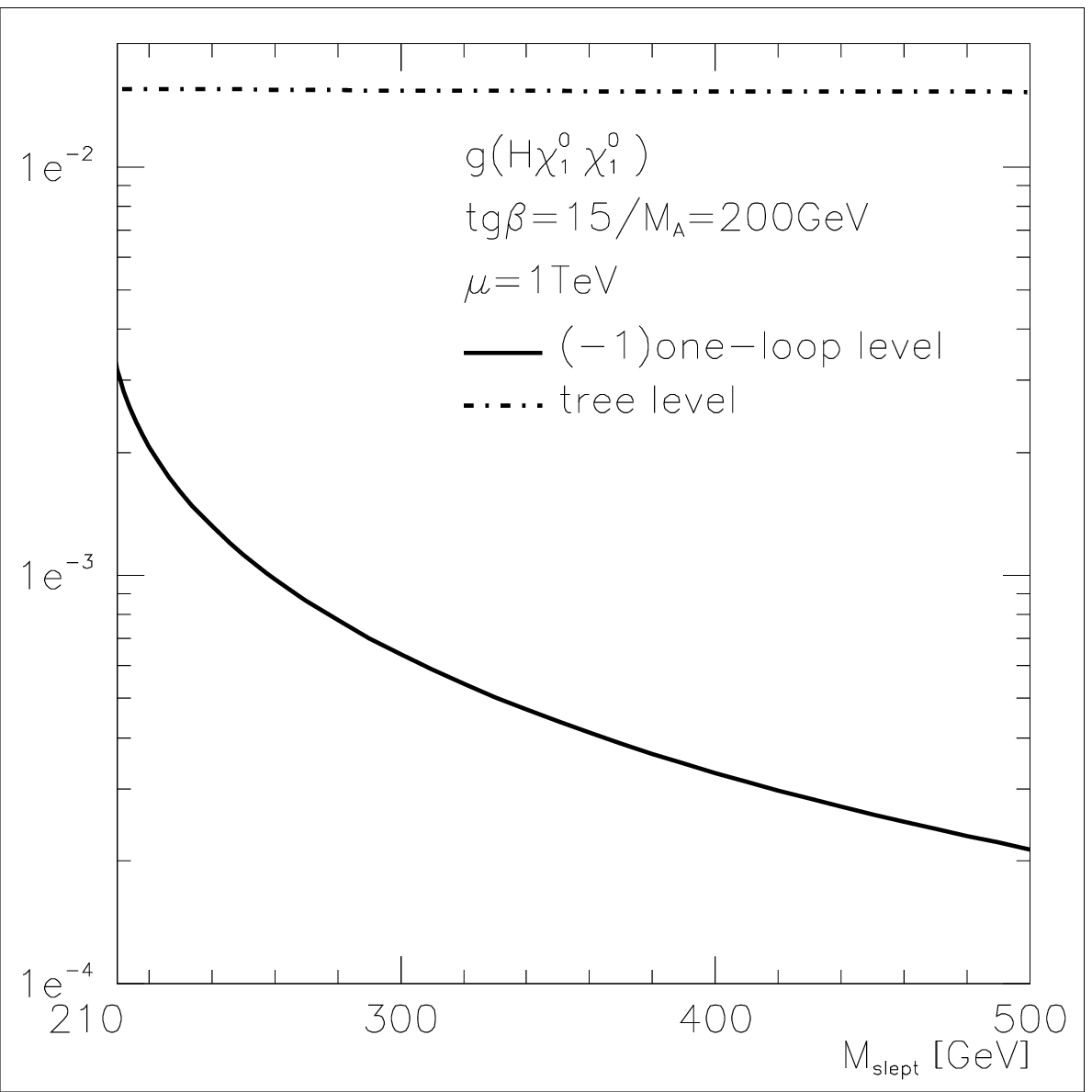}
\includegraphics[width=.5\textwidth]{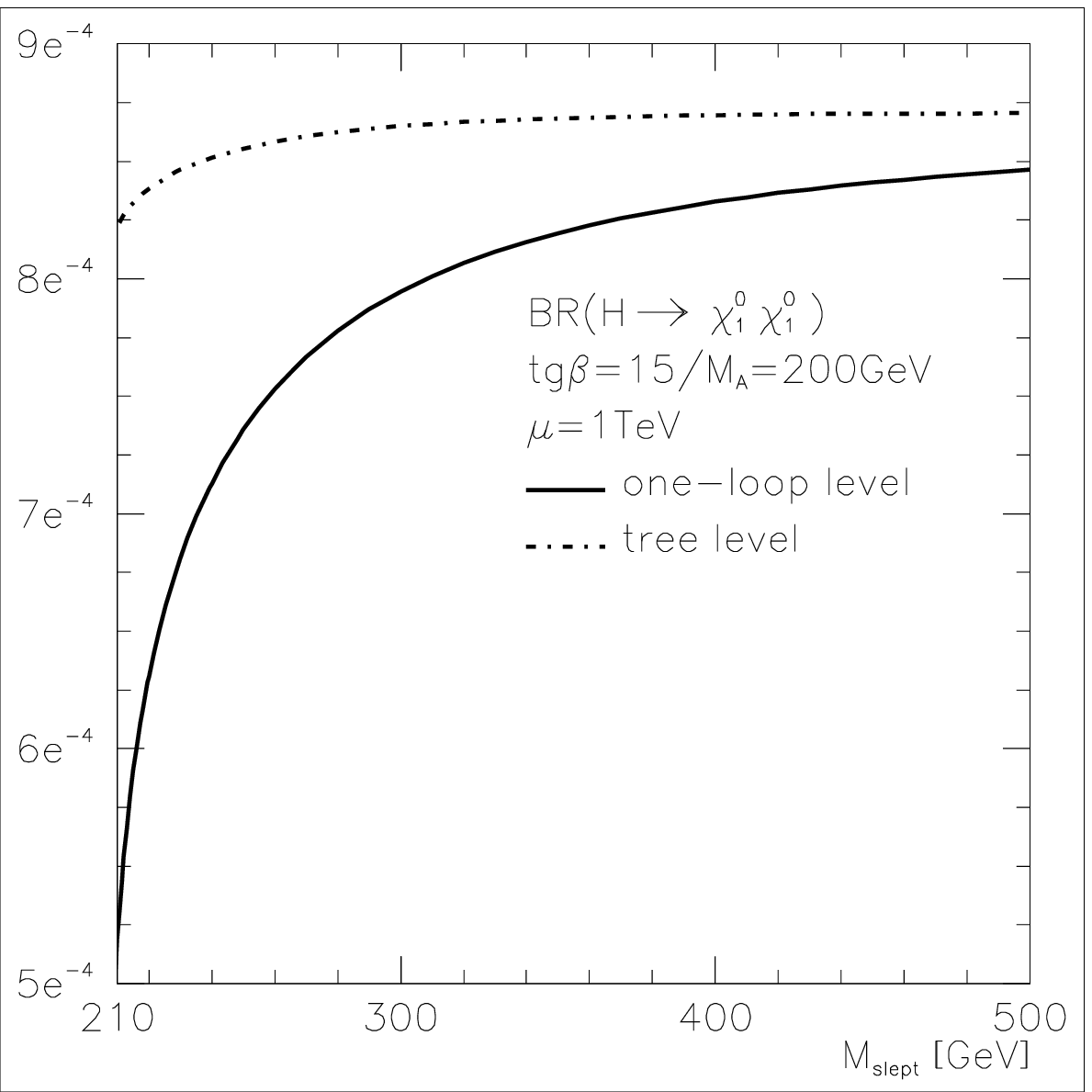}
\caption{\small The heavier CP--even Higgs boson coupling $H^0$ (top) and
branching ratios (bottom) to pairs of the lightest neutralinos
as functions of the common slepton mass.
The parameters are as in Fig.~4.3.}
\end{center}
\end{figure}
%
%
\begin{figure}
\begin{center}
\includegraphics[width=.5\textwidth]{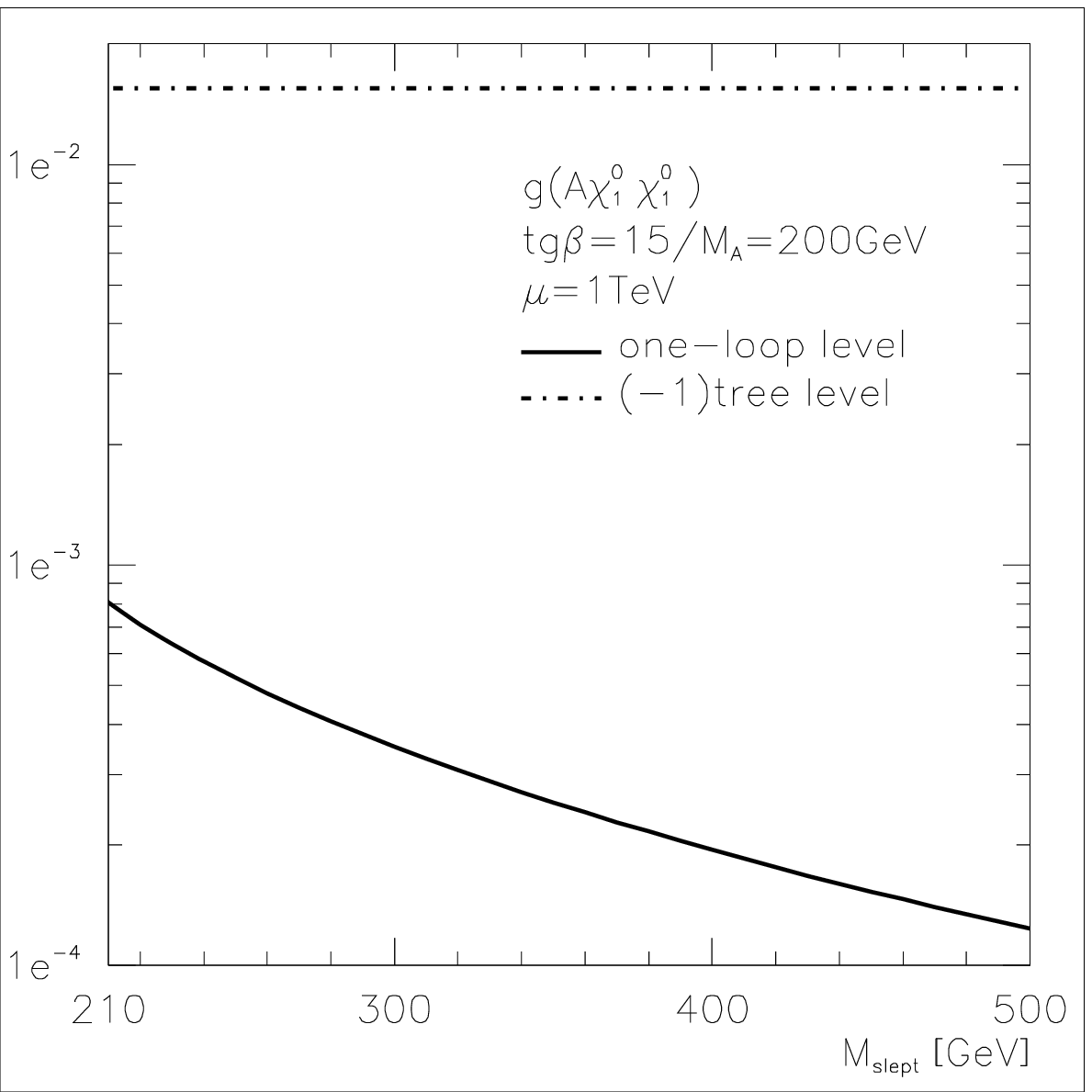}
\includegraphics[width=.5\textwidth]{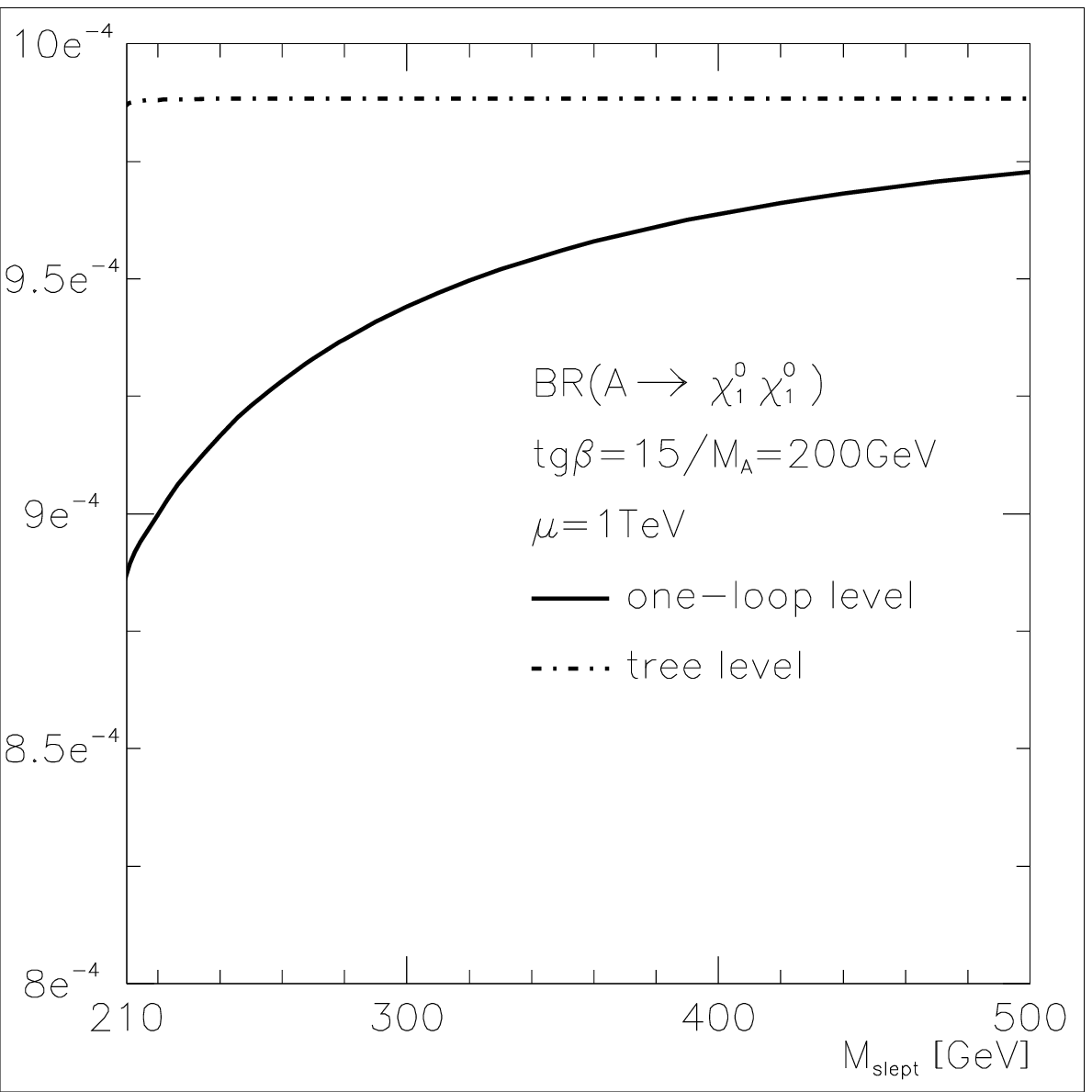}
\caption{\small The pseudoscalar $A^0$ boson couplings (top) and
branching ratios (bottom) to pairs of the lightest neutralinos
as functions of the common slepton mass. The parameters are as in Fig.~4.3.}
\end{center}
\end{figure}
\\
\\
We now turn to the heavier MSSM Higgs bosons $H^0$ and $A^0$.  The
couplings to the lightest neutralinos are shown in Figs.~4.6 and 4.7, for
the same input parameters as in Fig.~4.3. As can be seen, up to a
relative minus sign, the tree--level couplings of these two Higgs bosons are
approximately the same since we are in the decoupling regime where
$d_{A^0} \simeq - d_{H^0}$, and $e_{A^0} \simeq e_{H^0}$ with $|e_{A^0}| \ll 1$, see
eq.~(\ref{deh}). Eqs.~(\ref{hcoup0}--\ref{deh}) and (\ref{binostate})
also show that the tree--level couplings of the heavy Higgs bosons
exceed that of the light Higgs boson $h^0$ by a factor $\tan\beta / 2$
[ignoring contributions to eqs.~(\ref{binostate}) with $M_1$ in the
numerator]. On the other hand, the loop corrections are smaller in
case of the heavy Higgs bosons. The corrections to the
$H^0 \ \lsp \ \lsp $ coupling are reduced by about a factor
of 2 compared to the corrections to the $h^0 \ \lsp \ \lsp$
coupling, mostly due to the relatively smaller coupling 
to $\tilde t$ pairs, see eqs.~(\ref{hsfsfcoup}). 
\\
\\
The corrections to the $A^0 \ \lsp \ \lsp$ coupling
are even smaller, since the CP--odd Higgs boson $A^0$ cannot couple to
two identical squarks. The contribution with two $\tilde t_1$ squarks
and one top quark in the loop, which dominates the corrections to the
couplings of the CP--even Higgs bosons for small $m_{\tilde l}$, does
therefore not exist in case of the $A^0$ boson. As a result, the
corrections to the coupling of $A^0$ are not only smaller, but also
depend less strongly on $m_{\tilde l}$; recall that for our choice of
parameters $m_{\tilde t_1}$ increases very quickly as $m_{\tilde l}$
is increased from its lowest allowed value of $\sim 210$ GeV, which
comes from the requirement $m_{\tilde t_1} \geq 100$ GeV. Note also
that the $H^0 t t$ and $A^0 t t$ couplings are suppressed by a factor
$\cot\beta$ relative to the $h^0 t t$ coupling, see
eqs.~(\ref{hffcoup}); this becomes important for large squark masses,
where the one-loop corrections are relatively less important.
Altogether we thus see that the one--loop corrections are much less
important for the heavy Higgs bosons. Note also that for the given set
of parameters they tend to {\em reduce} the absolute size of these
couplings.
\\
\\
Again, because in the decoupling regime the CP--even $H^0$ boson and the
pseudoscalar $A^0$ boson have almost the same couplings to Standard
Model particles and to the neutralinos [at the tree level], their
branching ratios are approximately the same. The one--loop
contributions decrease the branching ratios by at most $\sim 10$ to
40\%. Note that the total decay widths of the $A^0$ and $H^0$ bosons are
strongly enhanced by $\tan^2 \beta$ factors [$\Gamma (H^0, A^0 \rightarrow
b\bar{b}) \propto m_b^2 g_{A^0,H^0 \ b \ b}^2$]. This over--compensates the
increase of their couplings to neutralinos, so that their branching
ratios into $\lsp$ pairs are far smaller than that of the light Higgs
boson $h^0$, remaining below the 1 permille level over the entire
parameter range shown. Moreover, the cross section for the production
of heavy Higgs bosons at $e^+e^-$ colliders is dominated by associated
$H^0 A^0$ production, which has a much less clean signature than $Z^0 h^0$
production does.
\\
Branching ratios of the size shown in Fig.~4.6 and 4.7 will
therefore not be measurable at $e^+e^-$ colliders. In fact, they will
probably even be difficult to measure at a $\mu^+ \mu^-$ collider
``Higgs factory''; recall that the $Z^0$ factories LEP and SLC ``only''
determined the invisible decay width of the $Z^0$ boson to $\sim 0.1\%$.
\\
After our analysis was completed a different group studied the one-loop
corrections to neutral Higgs bosons decays into neutralinos in the
general case, where we have the decays $H^0_i \to \widetilde{\chi}_m^0
\widetilde{\chi}_n^0$ (i=1,2,3) \cite{Eberl}. They confirmed our results
computing the branching ratios for similar values of the parameters.
\\
\\
As we mentioned before we consider all our parameters to be real,
however in the general case there are new CP violating phases in the MSSM. Due to the
potentially important role of the decays analyzed above, reference 
\cite{Drees2} studied the dependence of the
branching ratios on the CP violating phases. Moreover a large
correlation between the spins of the $\widetilde{\chi}$ states
produced in the decays of heavy neutral Higgs bosons was found.

\chapter{The Minimal Supersymmetric $SU(5)$}
\section{SUSY and Unification of the gauge couplings}
%
In the Standard Model for each group a gauge coupling constant is
defined. It has been known for a long time how the gauge couplings 
change with energy \cite{couplings}. If we study the
evolution of the gauge couplings we see that in the context of the
Standard Model the gauge couplings never meet. 
\\
\\
The meeting of the gauge couplings in the Minimal Supersymmetric
Standard Model is an impressive prediction
\cite{Dimop1,Ibanez1,Einhorn1,Marciano} (see Fig. 5.1), which
tells us that at the high scale $M_{GUT}\sim 10^{16}$ GeV, all the 
interactions are unified. Above the GUT scale the gauge couplings 
remain together only if new particles are present, this is the case 
of SUSY $SU(5)$. 

\begin{figure}
\begin{center}
\leavevmode
\epsfxsize=12cm
\epsffile{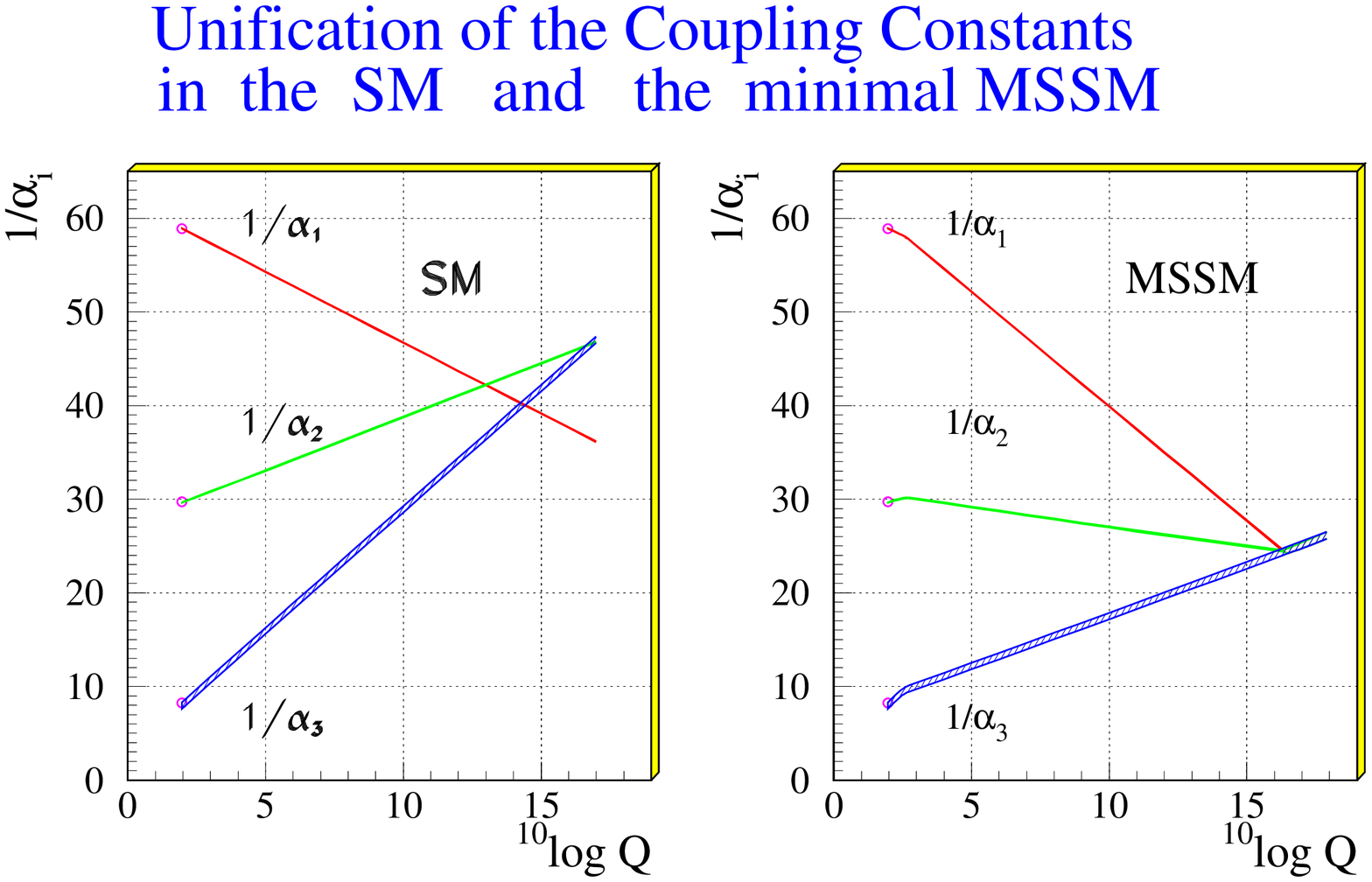}
\end{center}
\vspace{-1.0cm}
\caption{\small Evolution of the inverse of the three coupling 
constants in the Standard Model (left) and in the supersymmetric 
extension of the SM (MSSM) (right). Only in the latter case
unification is obtained. The SUSY particles are assumed to 
contribute only above the effective SUSY scale $\Lambda_{SUSY}$ of 
about  1 TeV, which causes a change in the slope 
in the evolution of couplings. The thickness
of the lines represents the error in the coupling constants as of 1991
\cite{Amaldi}.}
 \end{figure}

\section{Particle assignment}

The minimal Supersymmetric $SU(5)$ model is the simplest framework
where the unification of the Standard Model Interactions is
realized. Using the quantum numbers of the SM particles Georgi and
Glashow \cite{SU(5)} showed how the matter is unified partially in 
two irreducible representations
$\mathbf{\bar{5}}$ and $\mathbf{10}$. Using the $SU(3) \times SU(2)$ decomposition  of
these representations, the fermions of one family are accommodated as:

\begin{displaymath}
\mathbf{10} = (\bar{3},1) \bigoplus (3,2) \bigoplus (1,1) = (u^C_i)_L
  \bigoplus (u_i, d_i)_L \bigoplus (e^C)_L
\end{displaymath}

\vspace{0.5cm}

\begin{displaymath}
\mathbf{10}= \frac{1}{\sqrt{2}} \left( \begin{array} {ccccc}
  0    & u^C_3  & -u^C_2   &  u_1 & d_1 \\
-u^C_3 & 0      & u^C_1    &  u_2 & d_2 \\
u^C_2  & -u^C_1 & 0        &  u_3 & d_3 \\
-u_1   &  -u_2  & -u_3     &  0   & e^C \\
-d_1   & -d_2   & -d_3     & -e^C & 0
\end{array} \right)_L
\end{displaymath}

\begin{displaymath}
\mathbf{\overline{5}} = (\bar{3},1) \bigoplus (1,2) = (d^C_i)_L
\bigoplus (\nu, e)_L
\end{displaymath}
\vspace{0.5cm}
\begin{displaymath}
\mathbf{\overline{5}}=
 \left( \begin{array}{c} d^C_1 \\ d^C_2 \\ d^C_3
\\ e^{-} \\ -\nu
\end{array} \right)_L
\end{displaymath}
$SU(5)$ is a Lie group of rank $4$, with $24$ generators. Therefore we
will have $24$ gauge fields in our model, the usual Standard Model
gauge bosons plus 12 additional gauge bosons:

\beq
A_{\mu}(24)= \frac{1}{2} \lambda^a A^a_{\mu}, \ \ \ \ a=1,2 \ldots 24.
\eeq
\newpage

where the $\lambda^a$ are given by:
\\
\\
\begin{displaymath}
\lambda_1 = \left( \begin{array} {ccccc}
   0  &  1  &  0 &  0 & 0 \\
   1  &  0  &  0 &  0 & 0 \\
   0  &  0  &  0 &  0 & 0 \\
   0  &  0  &  0 &  0 & 0 \\
   0  &  0  &  0 &  0 & 0
\end{array} \right) \qquad  \lambda_2 = \left( \begin{array} {ccccc}
   0  & -i  &  0 &  0 & 0 \\
   i  &  0  &  0 &  0 & 0 \\
   0  &  0  &  0 &  0 & 0 \\
   0  &  0  &  0 &  0 & 0 \\
   0  &  0  &  0 &  0 & 0
\end{array} \right) 
\end{displaymath}

\begin{displaymath}
\lambda_3 = \left( \begin{array} {ccccc}
   1  &  0  &  0 &  0 & 0 \\
   0  & -1  &  0 &  0 & 0 \\
   0  &  0  &  0 &  0 & 0 \\
   0  &  0  &  0 &  0 & 0 \\
   0  &  0  &  0 &  0 & 0
\end{array} \right) \qquad \lambda_4 = \left( \begin{array} {ccccc}
   0  &  0  &  1 &  0 & 0 \\
   0  &  0  &  0 &  0 & 0 \\
   1  &  0  &  0 &  0 & 0 \\
   0  &  0  &  0 &  0 & 0 \\
   0  &  0  &  0 &  0 & 0
\end{array} \right) 
\end{displaymath}

\begin{displaymath}
\lambda_5 = \left( \begin{array} {ccccc}
   0  &  0  & -i &  0 & 0 \\
   0  &  0  &  0 &  0 & 0 \\
   i  &  0  &  0 &  0 & 0 \\
   0  &  0  &  0 &  0 & 0 \\
   0  &  0  &  0 &  0 & 0
\end{array} \right) \qquad  \lambda_6 = \left( \begin{array} {ccccc}
   0  &  0  &  0 &  0 & 0 \\
   0  &  0  &  1 &  0 & 0 \\
   0  &  1  &  0 &  0 & 0 \\
   0  &  0  &  0 &  0 & 0 \\
   0  &  0  &  0 &  0 & 0
\end{array} \right) \qquad
\end{displaymath}

\begin{displaymath}
\lambda_7 = \left( \begin{array} {ccccc}
   0  &  0  &  0 &  0 & 0 \\
   0  &  0  & -i &  0 & 0 \\
   0  &  i  &  0 &  0 & 0 \\
   0  &  0  &  0 &  0 & 0 \\
   0  &  0  &  0 &  0 & 0
\end{array} \right) \qquad  \lambda_8 = \frac{1}{\sqrt{3}}\left( \begin{array} {ccccc}
   1  &  0  &  0 &  0 & 0 \\
   0  &  1  &  0 &  0 & 0 \\
   0  &  0  & -2 &  0 & 0 \\
   0  &  0  &  0 &  0 & 0 \\
   0  &  0  &  0 &  0 & 0
\end{array} \right) 
\end{displaymath}

\begin{displaymath}
\lambda_9 = \left( \begin{array} {ccccc}
   0  &  0  &  0  &  1 & 0 \\
   0  &  0  &  0  &  0 & 0 \\
   0  &  0  &  0  &  0 & 0 \\
   1  &  0  &  0  &  0 & 0 \\
   0  &  0  &  0  &  0 & 0
\end{array} \right)\qquad \lambda_{10} = \left( \begin{array} {ccccc}
   0  &  0  &  0  &  -i & 0 \\
   0  &  0  &  0  &   0 & 0 \\
   0  &  0  &  0  &   0 & 0 \\
   i  &  0  &  0  &   0 & 0 \\
   0  &  0  &  0  &   0 & 0
\end{array} \right) 
\end{displaymath}

\begin{displaymath}
\lambda_{11} = \left( \begin{array} {ccccc}
   0  &  0  &  0  &   0 & 1 \\
   0  &  0  &  0  &   0 & 0 \\
   0  &  0  &  0  &   0 & 0 \\
   0  &  0  &  0  &   0 & 0 \\
   1  &  0  &  0  &   0 & 0
\end{array} \right) \qquad \lambda_{12} = \left( \begin{array} {ccccc}
   0  &  0  &  0  &  0 & -i \\
   0  &  0  &  0  &  0 &  0 \\
   0  &  0  &  0  &  0 &  0 \\
   0  &  0  &  0  &  0 &  0 \\
   i  &  0  &  0  &  0 &  0
\end{array} \right)
\end{displaymath}

\begin{displaymath}
\lambda_{13} = \left( \begin{array} {ccccc}
   0  &  0  &  0  &   0 & 0 \\
   0  &  0  &  0  &   1 & 0 \\
   0  &  0  &  0  &   0 & 0 \\
   0  &  1  &  0  &   0 & 0 \\
   0  &  0  &  0  &   0 & 0
\end{array} \right) \qquad \lambda_{14} = \left( \begin{array} {ccccc}
   0  &  0  &  0  &   0 & 0 \\
   0  &  0  &  0  &  -i & 0 \\
   0  &  0  &  0  &   0 & 0 \\
   0  &  i  &  0  &   0 & 0 \\
   0  &  0  &  0  &   0 & 0
\end{array} \right) 
\end{displaymath}

\begin{displaymath}
\lambda_{15} = \left( \begin{array} {ccccc}
   0  &  0  &  0  &  0 &  0 \\
   0  &  0  &  0  &  0 &  1 \\
   0  &  0  &  0  &  0 &  0 \\
   0  &  0  &  0  &  0 &  0 \\
   0  &  1  &  0  &  0 &  0
\end{array} \right)
\qquad \lambda_{16} = \left( \begin{array} {ccccc}
   0  &  0  &  0  &   0 &  0 \\
   0  &  0  &  0  &   0 & -i \\
   0  &  0  &  0  &   0 &  0 \\
   0  &  0  &  0  &   0 &  0 \\
   0  &  i  &  0  &   0 &  0
\end{array} \right) 
\end{displaymath}

\begin{displaymath}
\lambda_{17} = \left( \begin{array} {ccccc}
   0  &  0  &  0  &   0 & 0 \\
   0  &  0  &  0  &   0 & 0 \\
   0  &  0  &  0  &   1 & 0 \\
   0  &  0  &  1  &   0 & 0 \\
   0  &  0  &  0  &   0 & 0
\end{array} \right) \qquad \lambda_{18} = \left( \begin{array} {ccccc}
   0  &  0  &  0  &   0 &  0 \\
   0  &  0  &  0  &   0 &  0 \\
   0  &  0  &  0  &  -i &  0 \\
   0  &  0  &  i  &   0 &  0 \\
   0  &  0  &  0  &   0 &  0
\end{array} \right)
\end{displaymath}

\begin{displaymath}
\lambda_{19} = \left( \begin{array} {ccccc}
   0  &  0  &  0  &   0 &  0 \\
   0  &  0  &  0  &   0 &  0 \\
   0  &  0  &  0  &   0 &  1 \\
   0  &  0  &  0  &   0 &  0 \\
   0  &  0  &  1  &   0 &  0
\end{array} \right) \qquad \lambda_{20} = \left( \begin{array} {ccccc}
   0  &  0  &  0  &   0 &  0 \\
   0  &  0  &  0  &   0 &  0 \\
   0  &  0  &  0  &   0 & -i \\
   0  &  0  &  0  &   0 &  0 \\
   0  &  0  &  i  &   0 &  0
\end{array} \right) 
\end{displaymath}

\begin{displaymath}
\lambda_{21} = \left( \begin{array} {ccccc}
   0  &  0  &  0  &   0 &  0 \\
   0  &  0  &  0  &   0 &  0 \\
   0  &  0  &  0  &   0 &  0 \\
   0  &  0  &  0  &   0 &  1 \\
   0  &  0  &  0  &   1 &  0
\end{array} \right) \qquad \lambda_{22} = \left( \begin{array} {ccccc}
   0  &  0  &  0  &   0 &  0 \\
   0  &  0  &  0  &   0 &  0 \\
   0  &  0  &  0  &   0 &  0 \\
   0  &  0  &  0  &   0 & -i \\
   0  &  0  &  0  &   i &  0
\end{array} \right) 
\end{displaymath}

\begin{displaymath}
\lambda_{23} = \left( \begin{array} {ccccc}
   0  &  0  &  0  &   0 &  0 \\
   0  &  0  &  0  &   0 &  0 \\
   0  &  0  &  0  &   0 &  0 \\
   0  &  0  &  0  &   1 &  0 \\
   0  &  0  &  0  &   0 & -1
\end{array} \right) \qquad \lambda_{24} = 
\frac{1}{\sqrt{15}}\left( \begin{array} {ccccc}
  -2  &  0  &  0  &   0 &  0 \\
   0  & -2  &  0  &   0 &  0 \\
   0  &  0  & -2  &   0 &  0 \\
   0  &  0  &  0  &   3 &  0 \\
   0  &  0  &  0  &   0 &  3
\end{array} \right)
\end{displaymath}
The $SU(3) \times SU(2) $ decomposition of the 24-plet is
given by:

\begin{displaymath}
24=(8,1) \bigoplus (3,2) \bigoplus (\bar{3},2) \bigoplus (1,3) \bigoplus (1,1)
\end{displaymath}

\begin{displaymath}
A(24)= G_{ij} \bigoplus (X_i, Y_i) \bigoplus (\bar{Y}_i,\bar{X}_i) \bigoplus (W^{+}, W^3, W^-) \bigoplus B^0
\end{displaymath}
\\
The $SU(3)$ octet $G_{ij}$ are identified with the gluons. The $SU(2)$
doublet $(X_i,Y_i)$ represents the two superheavy
$SU(3)$ triplets $X$ and $Y$ gauge bosons with electric charges $4/3$ and $1/3$
respectively. The $SU(2)$ triplet $(W^{+}, W^3, W^-)$ are identified
with the SM $SU(2)$ vector bosons and finally $B^0$ is the $U(1)$
vector boson.
\\
\\
Now in order to know if our model reproduces the well known low-energy
physics, we must break the symmetry spontaneously to the Standard Model gauge group
$SU(3)_C \times SU(2)_L \times U(1)_Y$. In order to achieve this we
define the \textit{minimal} Higgs sector, which is composed of three
representations, $\bf{5}_H$, $\bf{\bar{5}}_H$ and the adjoint
representation $\bf{\Sigma(24)}$:

\begin{displaymath}
\bf{5}_H = \left( \begin{array}{c} T_1 \\ T_2 \\ T_3 \\ H_2^+ \\ H_2^0 \end{array} \right)
\qquad \bf{\bar{5}}_H = \left( \begin{array}{c}
\overline{T}_1 \\ \overline{T}_2 \\ \overline{T}_3 \\
H_1^- \\ - H_1^0\end{array} \right)
\end{displaymath}

\begin{displaymath}
\bf{\Sigma(24)} = \left( \begin{array} {cc}
   \Sigma_8             &  \Sigma_{(3,2)} \\
   \Sigma_{(\bar{3}, 2)}  &  \Sigma_3     \\
  \end{array} \right) + \frac{1}{2 \sqrt{15}}
\left( \begin{array} {cc}
   2   &  0 \\
   0   & -3 \\
\end{array} \right) \Sigma_{24}
\end{displaymath}
Knowing all the particles of our model we are ready to write down 
the interactions and analyze the possible predictions 
coming from new interactions.

\section{The $SU(5)$ Lagrangian}

Using the tools given in Chapter 1 and introducing a superfield for
each representation of $SU(5)$, we can write the lagrangian of our 
model \cite{SUSYSU(5),MohapatraBook}:

\begin{eqnarray}
{\cal L}_{SUSY}^{SU(5)} & = & \frac{1}{4}\int d^2 \theta
~Tr(W^{\alpha}_5 W_{5, {\alpha}})   \ + \ \frac{1}{4}\int d^2 \bar{\theta}
~Tr(\bar{W}^{\alpha}_5 \bar{W}_{5, {\alpha}})  \label{nonab} \nonumber
\\ &+& \sum_{\Psi = \bar{5}, 5_H, \bar{5}_H} \int d^2 \theta \ d^2 \bar \theta
\ \Psi^{\dagger}_i \ e^{g_5 A(24)} \ \Psi_i \nonumber \\
&+& \sum_{\Phi = 10, \Sigma} Tr \int d^2 \theta \ d^2 \bar \theta
\ \Phi^{\dagger}_i \ e^{g_5 A(24)} \ \Phi_i \nonumber \\
&+& \int d^2 \theta \ {\cal W}_5   + \int d^2 \bar{\theta}
\ \bar{{\cal W}}_5
 \end{eqnarray}
The most general (in the renormalizable limit) superpotential ${\cal W}_5$ of
$SU(5)$ which is R-parity invariant, ${\cal W}_5$ has two important
pieces, the corresponding to the Higgs self-interactions and other
describing Yukawa couplings:
\beq
{\cal W}_5 = {\cal W}_H + {\cal W}_Y
\eeq
The superpotential of the Higgs sector reads as:

\beq
\label{wh}
{\cal W}_H = \frac{m_{\Sigma}}{2} \ Tr \Sigma^2 \ + \ \frac{\lambda}{3}
\ Tr\Sigma^3 \ + \ \eta \ \bar{5}_H \Sigma 5_H \ + \ m_{H} \bar{5}_H 5_H
\eeq
while the Yukawa superpotential is:

\beq
\label{wy}
{\cal W}_Y = 10 \Gamma_U 10 5_H + 10 \Gamma_D \bar{5} \bar{5}_H
\eeq
where $\Gamma's$ are $3\times 3$ Yukawa matrices.
\\
\\
In the supersymmetric standard model language the Yukawa sector can be
rewritten as
\begin{eqnarray}
\label{wmssm}
{\cal W}_Y = &&  Q Y_U U^C H \ + \ \bar{H} Q Y_D D^C \ + \ \bar{H} L Y_E E^C \nonumber\\
&+& Q \underline A Q T \ + \ U^C \underline B E^C T \ + \ Q \underline
C L \bar{T} + U^C \underline D D^C \bar{T}
\end{eqnarray}
\noindent
where except for the heavy triplets $T$ and $\bar T$ the rest are
the MSSM superfields in the usual notation. The generation matrices
$Y_{U,D,E}$ and $\underline A$, $\underline B$, $\underline C$ and
$\underline D$ can in general be arbitrary. In the minimal SU(5)
defined above one finds $\underline A=\underline B=Y_U=Y_U^T=\Gamma_U$, and
$\underline C=\underline D=Y_D=Y^T_E=\Gamma_D$ at the GUT scale.
\\
\\
From these interactions we can find the different effective operators 
contributing to the decay of the proton. These are LLLL and RRRR operators:

\beq
\frac{1}{M_T} \int d^2 \theta \ (Q \ \underline{A} \ Q) \ (Q \
\underline{C} \ L )
\eeq

\beq
\frac{1}{M_T} \int  d^2  \theta
\ ({U^C} \ \underline{B} \ E^C)  \ ({U^C} \ \underline{D} \ D^C)
\eeq
In the next Chapter we will study all the properties of these operators, 
and we will analyze the predictions in the minimal model.
 
\section{Symmetry Breaking}

We need the following symmetry breaking:

\begin{displaymath}
SU(5) \times SUSY \Longrightarrow SU(3)_C \times SU(2)_L \times U(1)_Y
\times SUSY
\end{displaymath}
To study this we have to use ${\cal W}_{H}$ and calculate the relevant
F-terms and set them to zero to maintain supersymmetry down to the
electroweak scale. Computing the F-terms and using the condition 
$Tr \Sigma =0 $ we find the possible solutions which give 
us the symmetry breaking preserving SUSY:
\\
\\
\underline{Case 1}.

\beq
\langle \Sigma \rangle = 0
\eeq
In this case the $SU(5)$ symmetry remains unbroken.
\\
\\
\underline{Case 2}.

\beq
\langle \Sigma \rangle = \frac{m_{\Sigma}}{3 \lambda}
\left( \begin{array} {ccccc}
   1  &  0  &  0 &  0 & 0 \\
   0  &  1  &  0 &  0 & 0 \\
   0  &  0  &  1 &  0 & 0 \\
   0  &  0  &  0 &  1 & 0 \\
   0  &  0  &  0 &  0 & -4
\end{array} \right) 
\eeq
In this case $SU(5)$ breaks down to $SU(4) \times U(1)$, and the last
possible solution is:
\\
\\
\underline{Case 3}.

\beq
\langle \Sigma \rangle = \frac{ m_{\Sigma}}{\lambda}
\left( \begin{array} {ccccc}
   2  &  0  &  0 &  0 & 0 \\
   0  &  2  &  0 &  0 & 0 \\
   0  &  0  &  2 &  0 & 0 \\
   0  &  0  &  0 &  -3 & 0 \\
   0  &  0  &  0 &  0 & -3
\end{array} \right) 
\eeq
This is the desired vacuum since $SU(5)$ is broken to $SU(3)_C \times
SU(2)_L \times U(1)_Y$. In the supersymmetric limit all vacua are degenerate.
To complete the symmetry breaking, the $G_{SM}$ must be broken to
$SU(3)_C \times U(1)_{em}$. This is caused by the following
expectation values:

\beq
\langle \bf{5}_H \rangle = 
\left( \begin{array}{c} 0 \\ 0 \\ 0 \\ 0 \\ \frac{v_2}{\sqrt 2} \end{array} \right)
\qquad \langle \bf{\bar{5}}_H \rangle = \left( \begin{array}{c}
0 \\ 0 \\ 0 \\ 0 \\ - \frac{v_1}{\sqrt 2} \end{array} \right)
\eeq
The fact that from ${\cal W}_Y$ we get $\underline A=\underline B=Y_U$,
$\underline C=Y_E$, $\underline D=Y_D$, is simply
a statement of SU(5) symmetry. On the other hand
$Y_U=Y_U^T$ and $Y_D=Y^T_E$ result from the SU(4) Pati-Salam
like symmetry left unbroken by $\langle 5_H\rangle$ and
$\langle\bar{5}_H\rangle$. Under this symmetry $d^c\leftrightarrow e$,
$u\leftrightarrow u^c$, $d\leftrightarrow e^c$. Of course, this symmetry is broken
by $\langle\Sigma_\alpha^\alpha\rangle\ne\langle\Sigma_4^4\rangle$,
where $\alpha=1,2,3$; this becomes relevant when we include
higher dimensional operators suppressed by
$\langle\Sigma\rangle/M_{Pl}$, which will be considered in the next section.
\\
\\
Knowing how $SU(5)$ is broken to the Standard Model gauge group, we can compute the masses of 
different Higgs superfields in our theory. From the expression of ${\cal W}_H$ we can compute 
the Triplet and the $\mu$ mass:

\beq
M_T = 2 \eta \frac{m_{\Sigma}}{\lambda} + m_H
\eeq
and
\beq
\mu = m_H - 3 \eta \frac{m_{\Sigma}}{\lambda}
\eeq
The triplet mass $M_T$ must be close to the GUT scale, while $\mu$ 
must be close to $m_W$. From the above relations we see that only when 
the parameters of the potential are fine-tuned, we can explain this difference in the masses. 
It is the \textit{Fine Tuning} problem of GUTs. As we see here we need a lot 
of fine-tuning to explain how the Triplet is much heavier than the doublet, 
this is the so-called \textit{Doublet-Triplet Splitting} or \textit{Hierarchy Problem}. 
Supersymmetry only helps us to stabilize the splitting againts
radiative corrections, but by itself does not explain its origin.
\\
\\
When the symmetry is broken the $X$ and $Y$ gauge bosons become
massive,
\beq
M_X = M_Y = 5 \sqrt{2} g_5 \frac{m_{\Sigma}}{\lambda}
\eeq
while we find for the members of $\Sigma$
\beq
m_8 = m_3=\frac{5}{2} m_{\Sigma}
\eeq
Note that in this case the weak triplet and color octet masses are equal.

\section{Fermion masses}

As we mentioned before in the Minimal Supersymmetric $SU(5)$ model we find the
relation $Y_D=Y^T_E=\Gamma_D$ at the GUT scale. When $\bar{5}_H$ gets
the expectation value $ \langle \bar{5}_H 
\rangle $= diag $\left(0, 0, 0, 0, -\frac{v_1}{\sqrt{2}}\right)$
the quark and lepton masses are related as:

\begin{eqnarray}
\label{wmssm}
m_{e} = m_d, \ \ \ m_{\mu} = m_s, \ \ \ m_{\tau} = m_{b}
\end{eqnarray}
The first two relations are wrong. Note that the incorrect relation
$\frac{m_e}{m_{\mu}}=\frac{m_d}{m_s}$ is predicted to be valid at any
scale. On the other hand the relation for the third generation
can be considered a \textit{great} success of the theory. 
\\
\\
We can imagine many ways to improve the mass relations for the 
first two generations \cite{MohapatraBook}, but the simplest and 
most suggestive one is to include $1/M_{Pl}$ suppressed operators which
are likely to be present; after all due to the small size of the
Yukawa couplings, these operators should be more
important for the first two generations where the theory fails, and
they require no change in the structure of the theory. Note that the
value of the ratio $\frac{M_{GUT}}{M_{Pl}} \sim 10^{-3}- 10^{-2}$
is even bigger than the Yukawa couplings of the first generation.
\\
\\
The explicit form of the renormalizable, and all the relevant
non-renormalizable terms are \cite{Bajc2}:

\begin{eqnarray}
{\cal W}_Y&=&
\epsilon_{ijklm} \left(10_a^{ij} f_{ab} 10_b^{kl} 5_H^m +
10_a^{ij} f_{1ab} 10_b^{kl}{\Sigma^m_n\over M_{Pl}} 5_H^n+
10_a^{ij}f_{2ab}10_b^{kn} 5_H^l {\Sigma^m_n\over M_{Pl}}\right)\nonumber\\
&+&\bar{5}_{Hi} 10_a^{ij}g_{ab}\bar 5_{bj}
+\bar{5}_{Hi} {\Sigma^i_j \over M_{Pl}}10_a^{jk}g_{1ab}\bar 5_{bk}
+\bar{5}_{Hi} 10_a^{ij}g_{2ab}{\Sigma_j^k\over M_{Pl}}\bar 5_{bk}\
\end{eqnarray}

\noindent
where $i,j,k,l,m,n$ are SU(5)
indices, and $a,b=1,2,3$ are generation indices.
\\
\\
After taking the SU(5) vev $\langle\Sigma\rangle=
\sigma$ diag$(2,2,2,-3,-3)$ we get at $M_{GUT}$ scale.

\begin{eqnarray}
\label{alleq}
Y_U&=&4\left(f+f^T\right)
-12{\sigma\over M_{Pl}}\left(f_1+f_1^T\right)
-2{\sigma\over M_{Pl}}\left(4f_2-f_2^T\right)\ \nonumber\\
\underline A&=&4\left(f+f^T\right)
+8{\sigma\over M_{Pl}}\left(f_1+f_1^T\right)
+2{\sigma\over M_{Pl}}\left(f_2+f_2^T\right)\ \nonumber\\
\underline B&=&4\left(f+f^T\right)
+8{\sigma\over M_{Pl}}\left(f_1+f_1^T\right)
+4{\sigma\over M_{Pl}}\left(3f_2-2f_2^T\right)\ \nonumber\\
Y_D&=&-g
+3{\sigma\over M_{Pl}}g_1
-2{\sigma\over M_{Pl}}g_2\ \nonumber\\
Y_E&=&-g
+3{\sigma\over M_{Pl}}g_1
+3{\sigma\over M_{Pl}}g_2\ \nonumber\\
\underline C&=&-g
-2{\sigma\over M_{Pl}}g_1
+3{\sigma\over M_{Pl}}g_2\ \nonumber\\
\underline D&=&-g
-2{\sigma\over M_{Pl}}g_1
-2{\sigma\over M_{Pl}}g_2\
\end{eqnarray}
using $\lambda \ \sigma= m_{\Sigma}$.
\\
\\
Note the relation $Y_E - Y_D = \underline{C} - \underline{D}$
\\
\\
In the limit $M_{Pl}\to\infty$ we recover the old relations,
but for finite $\sigma/M_{Pl}\approx 10^{-3}-10^{-2}$ one
can correct the relations between Yukawas and at the same time
have some freedom for the couplings to the heavy triplets.
\\
\\
Clearly, due to SU(5) breaking through $\langle\Sigma\rangle$,
the $T$, $\bar T$ couplings are different from the $H$, $\bar H$
couplings. However, under the SU(4) symmetry discussed before
$\underline A\leftrightarrow\underline B$,
$\underline C\leftrightarrow\underline D$ and
$Y_U\leftrightarrow Y_U^T$. Only the terms that probe
$\langle\Sigma_\alpha^\alpha\rangle -\langle\Sigma_4^4\rangle$
can spoil that; this is why $f_1$ and $g_1$ still keep $Y_U=Y_U^T$,
$\underline A=\underline B$ and $\underline C=\underline D$.
\\
\\
Now we are ready to improve the mass relations for the first
two families. In order to get the correct relation
$\frac{m_e}{m_{\mu}} \simeq \frac{1}{9}\frac{m_d}{m_s}$, we must impose
specific values to the coupling $g_2$ since:

\beq
Y_D - Y_E = - \frac{5}{\lambda} \frac{m_{\Sigma}}{M_{Pl}} g_2
\eeq
If we assume that $Y_D$ and $Y_E$ are diagonals, and 
using the relations $m_e = \frac{m_d}{3}$ and $m_{\mu} = 3 m_s$, 
$g_2$ must satisfy the following relations:

\beq
(g_2)_{11} = - \frac{2 \lambda}{15} \frac{m_{d} M_{Pl}}{\langle \bar H
\rangle \ m_{\Sigma}}
\eeq
and
\beq
(g_2)_{22} = \frac{2 \lambda}{5} \frac{m_{s} M_{Pl}}{\langle \bar H
\rangle \ m_{\Sigma}}
\eeq
Note how the parameters of the scalar potential enter
in the expressions for fermion masses.

\section{$\sin^2 \theta_W$}

In the Standard Model the Weinberg angle $\theta_W$ is a free
parameter, which plays an important role in weak interactions. From 
experiment we know that $\sin^2 {\theta_W}_{\overline{MS}}
(M_z)=0.23117 \pm 0.0016$ \cite{PDG}.
\\
\\
Let us see what happens in $SU(5)$. In the Standard model we define
$\tan \theta_{W} = \frac{g_{U(1)}}{g_{SU(2)}}$, the electromagnetic
charge operator $Q_{em}$ = $T_{3W} + \frac{Y}{2}$ must be part of the
$SU(5)$ operators, therefore $Tr Q_{em}=0$.
\\
\\
Now using for example the fundamental representation 
$\bf{5}_H$ we can predict $Q_{em}(T)=
-\frac{1}{3}$, at the same time using ${\bf \bar 5}$ representation we 
can predict $Q_{em}(d^C)= \frac{1}{3} Q_{em}(e)$. This is one of the most beautiful 
predictions of GUTs, the quantization of the electric charge. 
\\
\\
For any fundamental representation we get:

\begin{displaymath}
Q_{em}({\bf 5})= - \frac{1}{3}\left( \begin{array} {ccccc}
             1  &             0  &             0  &   0 &  0 \\
             0  &             1  &             0  &   0 &  0 \\
             0  &             0  &             1  &   0 &  0 \\
             0  &             0  &             0  &  -3 &  0 \\
             0  &             0  &             0  &   0 &  0
\end{array} \right) 
\end{displaymath}

\begin{displaymath}
 T_{3W}({\bf 5}) = \frac{\lambda_{23}}{2}=\frac{1}{2}\left( \begin{array} {ccccc}
   0  &  0  &  0  &   0 &  0 \\
   0  &  0  &  0  &   0 &  0 \\
   0  &  0  &  0  &   0 &  0 \\
   0  &  0  &  0  &   1 &  0 \\
   0  &  0  &  0  &   0 & -1 
\end{array} \right)
\end{displaymath}
knowing these two operators, we see that the hypercharge operator must be:
\begin{displaymath}
\frac{Y}{2}({\bf 5})= \frac{1}{6}\left( \begin{array} {ccccc}
            -2  &             0  &             0  &             0 &  0 \\
             0  &            -2  &             0  &             0 &  0 \\
             0  &             0  &            -2  &             0 &  0 \\
             0  &             0  &             0  &             3 &  0 \\
             0  &             0  &             0  &             0 &  3
\end{array} \right) = \sqrt{\frac{5}{3}} \frac{\lambda_{24}}{2}
\end{displaymath}
as in $SU(5)$ all the couplings are equal, we can get the relation
between $g_{U(1)}$ and $g_{SU(2)}$. From the relations listed above we
can write the following expressions:

\beq
g_{U(1)} \frac{Y}{2} = g_5 \frac{\lambda_{24}}{2}
\eeq

\beq
g_{SU(2)} \frac{T_{3W}}{2} = g_5 \frac{\lambda_{23}}{2}
\eeq
\\
\\
concluding that $g_{U(1)} = \sqrt{\frac{3}{5}} g_5$ and
$g_{SU(2)} = g_5$, so $\tan^2 \theta_W=\frac{3}{5}$ or $\sin^2
\theta_W = \frac{3}{8}$. It is one of the most important predictions
of supersymmetric Grand Unified Theories and in particular of SUSY $SU(5)$. Note that this
value is at the GUT scale, when we use the renormalization group equations
and compute the value of this quantity at the electroweak scale, we see
that it agrees with the experimental measurements \cite{Langacker,Buras}.

\chapter{Proton Decay in the Superworld}

\section{ B violating operators}

As we know the Baryon (B) and Lepton (L) numbers are conserved in the
Standard Model. It is a consequence of the particle assignment and the
gauge principle. However these symmetries could be broken at a high
scale $M \gg m_{W}$. If at the high scale this happens, then we will
have an effective operator ${\cal L}_{eff}$ which describes the new
possible interactions:

\beq
{\cal L}_{eff} = c_{d} \frac{{\cal O}^d}{M^{d-D}}\label{ope}
\eeq
where ${\cal O}^d$ represents an operator of mass dimension $d$, $c_{d}$ is
a coefficient, and $D$ is the space-time dimension. Note that in our ``\textit{real}''
world we have $D=4$.
\\
If the Baryon number is broken, we have a new prediction,
\textit{the decay of the proton}. This is the case of grand unified
models such as $SU(5)$ and $SO(10)$, where from the matter unification we
get new effective operators of the type \ref{ope} \cite{PatiSalam1,PatiSalam2}. In the case of
non-conservation of Leptonic number, we have the possibility to 
explain the smallness of neutrino masses, due to the presence of a
large Majorana mass term \cite{Gell,Yanagida,Moha1,Moha2}.
\\
\\
Using the superfields of the Minimal Supersymmetric Standard Model
$[ Q=(U, D)$, \ $L=(N, E)$, \ $U^C$, \ $D^C$, \ $E^C ]$, we
can write down all the possible effective operators contributing to
the decay of the proton, which are $SU(3)_C \times SU(2)_L \times U(1)_Y$
invariant \cite{Weinberg1,Weinberg2,Weinberg3,Zee,Sakai,Rabyp}.
\\
\\
\underline{$d=4$ operators}\footnote{Note that these are the operators 
present in ${\cal W}_{NR}$ (see equation \ref{WNR})}:

\beq
\int d^2 \theta \ \epsilon_{\alpha \beta \gamma} \
U_{\alpha}^C \ D_{\beta}^C \ D_{\gamma}^C
\eeq

\beq
\int d^2 \theta \ \epsilon_{m n} \ Q_m \ D^C \ L_n
\eeq

\beq
\int d^2 \theta \ \epsilon_{m n} \ E^C \ L_m \ L_n
\eeq
\underline{$d=5$ operators}:

\beq
\frac{C^{ijkl}_{LLLL}}{M} \int d^2 \theta \ 
\epsilon_{m n} \ \epsilon_{p q} \ \epsilon_{\alpha
\beta \gamma} \ Q_{i m \alpha} \ Q_{j n \beta} \ Q_{k p \gamma} \ L_{l
q}
\eeq

\beq
\frac{C^{ijkl}_{RRRR}}{M} \int  d^2  \theta  \ \epsilon_{\alpha \beta \gamma}
U_{i \alpha}^C  \ D_{j \beta}^C  \ U_{k \gamma}^C  \
E^C_{l }
\eeq
\underline{$d=6$ operators}:

\beq
\frac{1}{M^2} \int d^2 \theta \ d^2  \bar{\theta}
\ \epsilon_{\alpha \beta \gamma} \ \epsilon_{m n} \
Q_{\alpha m}^{\dagger}  \ U_{\beta}^C  \ Q_{\gamma n}^{\dagger} \ E^C
\eeq

\beq
\frac{1}{M^2}  \int  d^2 \theta \ d^2 \bar{\theta}
\ \epsilon_{\alpha \beta \gamma} \ \epsilon_{m n} \
Q_{\alpha m}^{\dagger}  \ U_{\beta}^C  \ L_{n}^{\dagger} \ D_{\gamma}^C
\eeq
where $\alpha$, $\beta$ and $\gamma$ are color indices; m, n, p
and q isospin indices, while i, j, k and l represent generation indices.
\\
\\
Note that the product of two $d=4$ operators and $d=6$ operators lead to
proton decay at tree level. In the case of the dimension $d=4$
operators we have contributions with two fermions and one scalar 
field, the exchange of the scalar field can mediate proton decay. 
The $d=6$ operators directly yield terms with four fermions
contributing to the decay of the proton. However $d=5$  operators 
are quite special, in each term we have two fermions and two scalars 
fields, therefore they contribute only at one-loop once we dress these operators. 
\\
\\
In the case that extra spacetime dimensions are considered we see that there will be many new 
contributions to proton decay, without the suppression factor 
\cite{Goran1}. This is in our opinion the most important
phenomenological problem of many models with extra dimensions.
\\
\\
Knowing all the possible operators contributing to the nucleon decay,
the general expression for the proton lifetime could be written as:

\beq
\tau_p \propto \vert {\cal A}_{d=4} + {\cal A}_{d=5} + {\cal A}_{d=6} \vert^{-2}
\eeq
where ${\cal A}_i$ are the different amplitudes. Note that in four
dimensions ${\cal A}_4$ does not have any suppression factor, therefore
the coefficients related with these contributions must be very small
or maybe it is more natural find a symmetry to forbid these
operators.
\\
\\
The second possibility is realized, if we introduce a symmetry
called Matter Parity (see section 3.6):

\beq
M=(-1)^{3(B-L)}
\eeq
where $M = -1$ for $Q$, $L$, $U^C$, $D^C$, $E^C$ \ and \ $M = 1$
for $H$, $\overline{H}$ and $G_{\mu}$. If we assume that this symmetry is conserved, we remove the $d=4$
contributions, retaining the $d=5$ contributions as the most important
ones.
\\
\\
Note that relation between $R$ and $M$ parities, $R=(-1)^{2S}M$. The
case where the $R$ parity is not an exact symmetry has been analyzed
in reference \cite{Vissani}. However the most interesting case is when 
$R$-parity is conserved. As we mentioned before in this case we have
an ideal candidate to describe the Non-Baryonic Dark Matter present in the
Universe. Also the conservation of this symmetry is 
predicts in a large class of Grand Unified Theories as Minimal SUSY
$SO(10)$ \cite{SO101}.

\section{$d=6$ operators}

Let us analyze in detail the $d=6$ contributions. We can use as an
example the hermitian of the operator 6.7, working in 4 dimensions,
choosing $m=1$, $n=2$, and using the properties of the Grassmannian variables we find the following
four fermions effective operator:

\beq
\frac{1}{M^{2}}  \ \epsilon_{\alpha \beta \gamma} \ \
(e^C)^{\dagger} \ d_{\alpha} \ (u_{\beta}^C)^{\dagger}  \ u_{\gamma}
\eeq
therefore we will have a contribution to proton decay at tree level,
in this case the proton decay into $\pi^0$ and $e^+$, the usual most
important channel coming from the d=6 contributions in grand unified models.

\vspace*{-1.0cm}
\begin{center}
\begin{picture}(210,160)(70,70)
\ArrowLine(100,140)(200,140)
\ArrowLine(100,120)(150,100)
\ArrowLine(100,80)(150,100)
\ArrowLine(150,100)(200,120)
\ArrowLine(150,100)(200,80)
\Text(150,85)[]{$\frac{1}{M^2}$}
\Text(100,145)[]{$u$}
\Text(200,145)[]{$u$}
\Text(100,110)[]{$u$}
\Text(100,90)[]{$d$}
\Text(200,110)[]{$u^C$}
\Text(200,90)[]{$e^C$}
\end{picture}
\end{center}
\begin{center}
{\small $d=6$ contribution to the decay of the proton.}
\end{center}
%
Usually the $d=6$ processes are mediated by new superheavy gauge and Higgs
bosons present in grand unified models. There are many aspects to
be considered when we compute the proton
decay amplitudes. In the first place these operators are given at the GUT
scale, therefore to compute the values of the lifetime, we must
compute the matrix elements for each channel, and study
the evolution of these operators to the proton mass scale $1$ GeV (see
\cite{Nathp1,Nathp2,Hisano1} for more details).
\\
\\
Also there is a very important point related with the fermion masses
and the prediction of proton decay. Assume that the fermion mass 
matrices are diagonalized as:

\begin{equation}
\label{defx}
U^T Y_U U_C = Y_U^d\;\;\;,\;\;\;
D^T Y_D D_C = Y_D^d\;\;\;,\;\;\;
E_C^T Y^T_E E = Y_E^d\;\;\;,
\end{equation}
when $u \to U \ u$,  $u^C \to U_C \ u^C$ and so on.
\\
\\
Now if we write the $d=6$ operators (eq. 6.11) in the physical basis, we get:

\beq
\frac{1}{M^{2}}  \
(e^C)^{\dagger} \ (E_C^{\dagger} \ D) \ d \ (u^C)^{\dagger}
\ (U_C^{\dagger} \ U) \ u
\eeq
As we see in general these operators depend of
the textures for $Y_U$, $Y_D$  and $Y_E$, this means that the proton
decay predictions will be different in each model for fermion masses \cite{Mohapatra1,Nandi}.
\\
\\
In the Minimal Supersymmetric $SU(5)$, where
from the relation of the mass matrices $Y_U=Y^T_U$ and $Y_D=Y_E^T$ 
we have $U_C=U$ and $D=E_C$,
the $d=6$ operators are independent of the explicit form of
textures, however as we mentioned in the last chapter this
is not a realistic case, due to the problem of the mass relation for
the first two families.
\\
\\
We could say that $d=6$ proton decay provides a way to test models of
fermion masses, however as we know $M_{GUT} \sim 10^{16}$ GeV which gives us
$\tau_p (d=6) \sim 10^{35}$ years, a value which is much bigger than the
present experimental bounds \cite{SK}. Therefore it is difficult to
test the $d=6$ predictions at present experiments. See
\cite{Wittenproton1,Wittenproton2} for the predictions in string-derived models. 

\section{$d=5$ operators}

The $d=5$ contributions are the dominant to the proton decay. They
are mediated by the superpartner of the colored Higgses (Triplets), which are
present in SUSY GUT models. In this case we have only $\frac{1}{M_T}$
as suppression, where $M_T$ is the Triplet mass.
\\
Let us understand how from these operators we get the proton decay 
amplitudes. We can use
the following operator as a example:

\beq
\frac{C^{ijkl}_{LLLL}}{M_{T}} \int d^2 \theta \ \epsilon_{m n} \ \epsilon_{p q} \ \epsilon_{\alpha
\beta \gamma} \ Q_{i m \alpha} \ Q_{j n \beta} \ Q_{k p \gamma} \ L_{l
q}
\eeq
in this case using the properties of the Grassmannian variables we see
that for each contribution, there are two fermionic and two
bosonic (or superpartners) fields. For example if $i=j=l=1$ and $k=2$,
 we find the following contribution to the decay into $K^+$ and $\bar{\nu}$:

\beq
\frac{1}{M_T} C^{1121}_{LLLL} \ \widetilde{u} \ \widetilde{d} \  s \ \nu
\eeq
Now we must dress this operator to find the four fermions operator
contributing to proton decay. This is possible using gauginos and
higgsinos, since these are Majorana particles:
\vspace*{-1.5cm}
\begin{center}
\begin{picture}(210,160)(70,70)
\ArrowLine(80,140)(200,140)
\ArrowLine(200,120)(150,100)
\ArrowLine(200,80)(150,100)
\DashArrowLine(100,120)(150,100){4}{}
\DashArrowLine(100,80)(150,100){4}{}
\ArrowLine(100,100)(100,120)
\ArrowLine(100,100)(100,80)
\ArrowLine(80,120)(100,120)
\ArrowLine(80,80)(100,80)
\Text(150,85)[]{$\frac{1}{M_T}$}
\Text(75,145)[]{$u$}
\Text(200,145)[]{$u$}
\Text(110,110)[]{$\widetilde{u}$}
\Text(110,90)[]{$\widetilde{d}$}
\Text(200,110)[]{$s$}
\Text(200,90)[]{$\nu$}
\Text(75,120)[]{$u$}
\Text(75,80)[]{$d$}
\Text(100,100)[]{$\times$}
\Text(90,100)[]{$\widetilde{g}$}
\end{picture}
\end{center}
\begin{center}
{\small $d=5$ contribution to the decay of the proton.}
\end{center}
From this graph we can appreciate that these contributions are present
at one-loop level, where we have superpartners inside the loops, the
loop factor and the suppression $1/M_{T}$  must be considered to compute
the proton decay amplitude. As we mentioned in the last section, these
operators are valid at the GUT scale, therefore we must compute the
matrix elements and study the running down to 1 GeV. However in this
case there are many new factors to be considered. $M_T$  is usually the
Triplet mass, which could be smaller than the GUT scale, therefore we
must compute this in order to estimate the amplitudes.
\\
\\
In the case of $d=6$ operators we showed how the amplitudes could 
depend on the textures of fermion masses. For the $d=5$ contributions 
we see that there is something new, the sfermion masses also appear
in this case. Therefore in general the $d=5$ contributions will depend on
the textures for sfermions and fermions, since in a general SUSY model
these textures are different.
\\
\\
As we see in order to compute the proton decay amplitudes we must
consider many unknown factors: the loop factor, which depends on the SUSY
spectrum and mixings between fermion and sfermions, the Triplet mass
and the matrix elements. Therefore we can conclude that it is very
difficult to test the SUSY GUT models using proton decay, since the
dominant $d=5$ contributions are quite model dependent.

\section{Proton decay in Minimal SUSY $SU(5)$}

In the last chapter we studied the structure of the Minimal
Supersymmetric $SU(5)$ model. We noted that new interactions which
violate the baryon and lepton numbers are present, when the matter
unification is realized. From these new interactions we find the
$d=5$ and $d=6$ operators contributing to the decay of the proton.
\\
\\
As we mentioned above the most important contributions are those with 
$d=5$. From the superpotential of $SU(5)$ we find the LLLL and RRRR
$d=5$ operators, which read as:

\beq
\frac{1}{M_T} {\underline A^{ij}} {\underline C^{kl}} 
\int d^2 \theta \ (Q_i \  Q_j) \ (Q_k \ L_l )
\eeq

\beq
\frac{1}{M_T} {\underline B^{ij}} {\underline D^{kl}} \int  d^2  \theta
({U_i^C} \ E_j^C)  \ ({U_k^C} \ D_l^C)
\eeq
Knowing these operators we can write all the $d=5$ contributions for
each channel. The results are listed in Appendix B. Note that we did not assume any specific SUSY
model or any texture for fermion masses. Our analysis is quite
general.
\\
\\
If the Baryon and Lepton numbers are not conserved, there are many
channels for the decay of the proton:

\begin{displaymath}
p \to (K^+, \pi^+, \rho^+, K^{*+}) \bar{\nu_i}
\end{displaymath}

\begin{displaymath}
n\to (\pi^0, \rho^0, \eta, \omega, K^0, K^{*0}){\bar\nu_i}
\end{displaymath}
where $i=1,2,3$, and
\\
\begin{displaymath}
p\to (K^0, \pi^0, \eta, K^{*0}, \rho^0, \omega) e_j^+
\end{displaymath}

\begin{displaymath}
n\to (K^-, \pi^-, K^{*-}, \rho^-) e_j^+
\end{displaymath}
where
$j=1,2$, while for $K^*$ only $j=1$. Note that there are also new channels for neutron decay.
\\
\\
These operators have been studied on and off for the last 20 years with
culminating conclusion that the minimal SUSY $SU(5)$ is ruled
out \cite{Murayama}. 
\\
\\
In this paper by Murayama and Pierce, the different
constraints on the Triplet mass are studied. Using the unification
of the gauge couplings and the proton decay experimental lower bounds
they found inconsistent limits on this mass.
\\
\\
To give an idea of the procedure used in reference \cite{Murayama},
we can use the renormalization group equations for the gauge couplings at
one-loop (neglecting the Yukawa couplings). 
\\
\\
Now assuming exact unification we get at one loop level:

\begin{eqnarray}
\label{eqn:HiggsRGEMass}
3\alpha_{2}^{-1}(m_{Z})-2\alpha_{3}^{-1}(m_{Z})-\alpha_{1}^{-1}({m_{Z}})= \nonumber \\
\frac{1}{2 \pi}
\Bigl(\frac{12}{5} \log \frac{M_T}{m_{Z}} - 2 \log
\frac{m_{SUSY}}{m_{Z}}\Bigr)
\end{eqnarray}
Therefore it is possible invert the above equation and determine the
colored Higgs mass. For numerical calculation, they used the two-loop
RGEs for the gauge and Yukawa couplings between the SUSY and GUT
scale. 
\\
\\
Knowing the values of the gauge couplings at the scale $m_Z$
\cite{PDG}, they found that the $SU(5)$ prediction of exact
unification agrees with data only for colored Higgs masses of:

\begin{equation}
M_T \leq 3.6 \times 10^{15} \mbox{GeV}
\label{eqn:RGELimit}
\end{equation}
The second limit on the Triplet mass is computed using the
experimental lower bound for the channel $p \to K^+ \bar{\nu_i}$  of
 $6.7 \times 10^{32}$ years \cite{SK}. 
\\
\\
Computing the proton lifetime due to the $d=5$ contributions 
using the methods of reference \cite{Goto}, they found:

\begin{equation}
M_T \geq 7.6 \times 10^{16} \mbox{ GeV}
\end{equation}
in order to satisfy the experimental bounds. It was assumed nearly
degenerate scalars at the weak scale, or order 1 TeV in mass.
\\
\\
Comparing this equation with equation (\ref{eqn:RGELimit}), they claim that the
minimal SUSY SU(5) theory is excluded by a lot.
\\
\\
Now in order to consider a different scenario and see the
possibilities to suppress the $d=5$ operators, they considered a
second case, the so called decoupling
scenario \cite{Pomarol1,Pomarol2,Cohen}. 
\\
In this case the first and second generations of superpartners 
could be heavy without severe fine-tuning because they do not 
affect the Higgs boson self-energy
at one-loop level. 
\\
\\
Since the loop factor goes like
$ m_{\widetilde{G}} \ m^{-2}_{\widetilde{q}}$ (when
$m_{\widetilde{G}} \ll m_{\widetilde{q}}$), where $m_{\widetilde{G}}$
is the gaugino or higgsino mass, while $m_{\widetilde{q}}$ is the
slepton or squark mass. Therefore we can get a large suppression by
making the sfermions of the first two generations very heavy. For
example if we assume $m_{\widetilde{q}_{1,2}} \sim 10$ TeV, we get 
an extra suppression factor of $10^{-2}$ to the amplitude.
\\
\\
Now computing the lower bound for the Triplet mass in the
decoupling scenario, they found that:

\begin{equation} \label{eqn:3genmass}
M_{T} > 5.7 \times 10^{16} \mbox{ GeV}.
\end{equation}
Therefore from this analysis they concluded that \textit{the Minimal 
Supersymmetry $SU(5)$ is ruled out}.       
\\
\\
However in the last analysis, they did not consider the most
general scenario, as we mentioned in the last section the $d=5$
contribution are quite model dependent. 
\\
\\
Murayama and Pierce assume the following in their important analysis:

\begin{itemize}

\item They found the different limits on the Triplet mass in the
minimal SUSY $SU(5)$ model without higher dimensional operators in
${\cal W}_5$, which makes wrong predictions for the fermion masses of the 
first two generations. This is not a realistic model, or we
could say that it is already ruled out.

\item They computed the proton decay amplitudes in a very specific
SUSY model, where the mixings between fermions and sfermions 
are known, however as we mentioned before in a general 
SUSY model the situation may be quite different.

\item Assuming exact unification of the gauge couplings, they computed
the limits on the Triplet mass, however if the non-renormalizable 
contributions to ${\cal W}_H$ are considered, the bounds change.

\end{itemize}
For these reasons we think that the model is not ruled out, in our opinion
before ruling out the minimal realization of the idea of unification, 
it is better to see how constraint the model using the experimental 
bounds. 
\\
\\
In the next two sections we will point out different aspects to be considered
in order to satisfy the experimental bounds on proton decay.

\section{(S)Fermion masses versus Proton decay}

In this section we will show how to satisfy all the experimental
bounds on proton decay. As we mentioned before the
prediction of the $d=5$ contributions will depend on the explicit
form of fermion and sfermions mass matrices
\cite{Berezhiani,Nath1,Nath2}.
\\
\\ 
Now to start our analysis we must write all the $d=5$ operators in the physical
basis, therefore we must go from the flavour to the physical
basis as:

\beq
F_p  \ \to \ U_F \ F_f
\eeq
\beq
\widetilde{F}_p  \ \to \ \widetilde{U}_F \ \widetilde{F}_f
\eeq
where $F$ and $\widetilde{F}$ represent the fermion and sfermion
fields respectively, while $U_F$ and $\widetilde{U}_F$ are unitary
matrices. 
\\
\\
In general there is not relation between these matrices,
therefore we expect that in each $d=5$ contribution we will have a
combination of different mixings between fermions and sfermions, which
are unknown. Note that we only know the mixings between left-handed
fermions, $V_{CKM}= U^{\dagger}D$ and $V_l = N^{\dagger} E$, where $N$
and $E$ are the matrices which rotate $\nu_L$ and $e_L$ respectively.
\\
\\
Now the main point in our analysis is, how to suppress the $d=5$
operators in order to satisfy the experimental bounds. 
\\
\\
Assuming the decoupling scenario, where the most important contributions come from
the third family of superpartners, and studying all the
contributions shown in Appendix B, we see that the longevity
of the proton can be achieved by, say, the following conditions at 1
GeV \cite{Bajc1}:

\begin{eqnarray}
\label{constr}
&&(\tilde U^\dagger \ D)_{3a}\approx 0\;\;\;\;\;\;
(\tilde D^\dagger \ D)_{3a}\approx 0\;\;\;\;\;\;
(\tilde E_C^\dagger \ E_C )_{3a}\approx 0\;\;\;\nonumber\\
&&(\tilde N^\dagger \ E)_{3a}\approx 0\;\;\;\;\;\;
(\tilde D_C^\dagger \ D_C)_{3a}\approx 0\;\;\;\;\;\;
(\tilde E^\dagger \ E)_{3a}\approx 0\;\;\;\nonumber\\
&&(\tilde{U}_C^T \ Y_U^T \ D)_{3a}\approx 0\;\;\;
\end{eqnarray}
and

\begin{equation}
\label{groza}
A_0=\epsilon_{ab}( D^T \ \underline C \ \tilde N)_{a3}
(\tilde{U}^T \ \underline A \ D)_{3b}\approx 0\;\;\;
\end{equation}
where $a, b=1, 2$. Note that $\widetilde{U}$ and $\widetilde{U}_C$ are
the matrices which rotate $\widetilde{u}$ and $\widetilde{u}^C$
respectively. We use the same notation for the rest of the fermions
and sfermions. 
\\
\\
In the above equations we simply mean
that all the terms must be small. How small? It is
hard to quantify this precisely and, honestly speaking, it seems to
us a premature task. 
\\
\\
Our aim was to demonstrate that the theory is
still consistent with data and from the above formulae it is obvious.
If (when) proton decay is discovered and the decay modes
measured, it may be sensible to see how small should the above
terms be. 
\\
\\
Suffice it to say, that a percent suppression of the
results in the minimal SUGRA (or super KM) case should be
enough.
This means that on the average each vertex should be suppressed by
a factor of $1/3$ or so with respect to the minimal supergravity
predictions. It is very difficult to say more: in fact one could be
tempted to estimate that for example the combinations on the
lefthand sides of the above equations need to be at least $10^{-2}$ the same
combinations in super KM. However this is generally neither
necessary nor sufficient. The fact is, that we have to deal with a nonlinear
system, since the total decay in a specified mode is proportional
to the square of a sum of single diagrams, each of them is proportional
to a combination of unknown mixings. Some of these mixings contribute to
different diagrams, and some depend on others, so the task of constraining
them numerically seems exaggerated in view of our complete ignorance of all
these parameters. What we can say for sure is that if each of the diagrams
in Appendix B is suppressed by a factor of $1/100$ with respect to
the minimal supergravity predictions, proton decay is not too fast and
\textit{Minimal Supersymmetric SU(5) is not ruled out}.
\\
\\
Notice further that the so called super KM basis, in which the
mixing angles of fermions and sfermions are equal, for example
$\widetilde{U}_C = U_C$, does not work
for the proton decay, since eqs. (\ref{constr}) and (\ref{groza})
are not satisfied. If you believe in super KM,
you would conclude that the theory is ruled out. It is obvious
though, from our work, that this is not true in general.
\\
\\
Notice even further, that all the relations (\ref{constr})-(\ref{groza})
do not require the extreme minimality conditions:
$\underline{A}=\underline{B}=Y_U=Y^T_U$ and $\underline{C}=\underline{D}=Y_D=Y^T_E$.
 More precisely, one can opt for the improvement of the fermion mass
relations and still save the proton.
\\
\\
One could worry that the above constraints for the sfermion
and fermion mixing matrices could be in contradiction with
the experimental bounds on the flavour violating low energy
processes. Fortunately, this is not true. Namely, the same
conditions (\ref{constr})-(\ref{groza}) suffice to render neutral
current flavour violation in-offensive (of course, the decoupling
is necessary for this to be true). We studied the processes $\mu\to e\gamma$, $b\to s\gamma$, 
$B-\bar B$, $K-\bar K$, etc. It is easy to see, that the combinations 
which appear in (\ref{constr}) are exactly the ones that appear in these
flavour changing processes. So they automatically take care also
of these low-energy experimental data. The only flavour changing
processes that could get sizeable contributions are the ones
which involve up type sfermions like for example $D-\bar D$ or
$c\to u\gamma$. These are not constrained by (\ref{constr}), but
at the same time are not very much constrained by the low-energy
experiments, so they do not represent a real issue at this stage.
\\
\\
Constraints (\ref{constr}) are not unique. One can find other
relations between sfermion and fermion mixing matrices that make
the proton decay amplitude zero or small. However, typically, these
solutions can be dangerous for FCNC processes, since they do
not au\-to\-ma\-ti\-cally cancel their contributions. So one has to
analyze the FCNC processes case by case. At the present day status
(or ignorance) of proton decay and FCNC experiments we believe
that this is premature.
\\
\\
The analysis in Appendix B has been done with the assumption
of no left-right sfermion mixing, and gaugino higgsino mixing in the 
neutralino and chargino sector. This mixing can be included in a 
perturbative way, one can show that, up to two mass insertions, 
the same constraints (\ref{constr})-(\ref{groza})
kill all the contributions to nucleon decay. This is enough to
increase the nucleon lifetime above the experimental limit,
since each mixing multiplies the diagram by at least $1/10$.
\\
Actually we could ignore the left-right mixing for sfermion proportional to the small ratio
$M_Z/m_{\widetilde{q}}$; in fact, as long as $\tan{\beta}>10$,
the LR mixing can be safely put even to zero without contradicting
the experimental constraints on the Higgs mass \cite{CarenaHaber}.
\\
\\
One can also worry about naturalness \cite{Ellis,Barbieri,Dimopoulos1,Feng}.
Through the large top
Yukawa couplings, the formula (\ref{higgs}) becomes here
($i=1,2,3$) (for large $\tan{\beta}$ there are similar
contributions of (s)bottom and (s)tau)

\begin{equation}
\label{higgsu}
m_h^2\approx m_0^2+{3 y_t^2\over 16\pi^2}\left[
(\tilde U^\dagger U)_{i3}(U^\dagger\tilde U)_{3i} m^2_{\tilde q_i}+
(\tilde U_C^\dagger U_C)_{i3}(U_C^\dagger\tilde U_C)_{3i}
\tilde m^2_{\tilde q^C_i}\right]
\end{equation}
\noindent
where $m_{\tilde q_i}$ and $m_{\tilde q^C_i}$ are left-handed and
right-handed squark masses. 
\\
\\
Now, for $m_{\tilde q_a}\approx\tilde m_{\tilde q^C_a}\approx 10$ TeV ($a=1,2$)
in the decoupling regime, large $(\tilde U^\dagger U)_{a3}$ or
$(\tilde U_c^\dagger U_c)_{a3}$ would imply a small amount of
fine-tuning ($\approx 1\%$) in (\ref{higgsu}). Hereafter, we accept
that. No fine-tuning whatsoever, although appealing, to us seems
exaggerated; after all it would eliminate large extra dimensions as
a solution to the hierarchy problem.
\\
\\
Strictly speaking, one could then ask why not simply push $m_{\tilde q_3}$
and $m_{\tilde q^C_3}$ all the way up to $10$ TeV and be safe? A sensible
point, we wish to have as many as possible
superpartners below TeV and thus hopefully detectable at LHC.
In other words, all the gauginos and Higgsinos and the third generation
of sfermions are assumed to have masses lower or equal TeV, we only
take $m_{\tilde q_{1,2}}\approx 10$ TeV.
\\
\\
In this case, we need to worry only about the third generation of
sfermions. We also assume light gauginos and Higgsinos,
$m_{\widetilde{G}}\approx 100$ GeV.
\\
\\
The constraints (\ref{constr}) can clearly be satisfied exactly
by the sfermion mixing matrices at $1$ GeV. It is reassuring that
the sfermionic sector does not break strongly SU(2). This is
consistent with the SU(2) invariance of the soft masses, which
dominate the total sfermion masses. 
\\
\\
The last constraint, eq. (\ref{groza}), can even be satisfied in the approximation
$\underline C=Y_D=Y_E$, which is true in the minimal
renormalizable model, but at $M_{GUT}$, not at 1 GeV. 
\\
\\
The relation $\underline C=Y_D=Y_E$ is however not stable under
running. To get an idea of how big this contribution is
at the electroweak scale, one can take the approximation that
the Yukawas do not run. In the leading order in small Yukawas
(except for $y_t$) one gets

\begin{equation}
A_0\approx y_cy_\tau V_{33}^*V_{23}V_{32}V_{21}
\left[1-\left(M_Z/M_{GUT}\right)^{y_t^2/16\pi^2}\right]
x_1^{-1/33}x_2^{-3}x_3^{4/3}
\end{equation}

\noindent
where $V$ is the CKM matrix and $x_i=\alpha_i(M_Z)/\alpha_U$.
There is only one non-vanishing diagram (the rest vanishes due
to (\ref{constr})) and it
is proportional to $V_{13}A_0$: fortunately, this seems to be
small enough. On top of this, in the amplitude the combination
(\ref{groza}) gets multiplied with a combination of neutralino
soft masses $m_{\tilde W_3}$ and $m_{\tilde b}$, which can be
fine-tuned to an arbitrary small (or even zero) value.
And, of course, we must keep in mind that $M_T$ can be large 
and $\underline A$ and/or $\underline C$ completely different 
than $Y_U$ ($Y_E$).
\\
\\
In other words, at this point
the proton decay limits provide information on the properties of
sfermions and {\it not} on the structure of the unified theory.

\section{The Higgs Triplet Mass}

The determination of the GUT scale and the masses ($M_T$) of
the heavy triplets $T$ and $\bar T$ responsible for $d=5$ proton
decay is one of the most important tasks in order to estimate the
proton decay amplitude. 
\\
\\
Specifically, if we allow an arbitrary trilinear couplings of the
heavy fields in $\Sigma$ and use higher dimensional terms as a
possible source of their masses \cite{Bachas:1995yt,Chkareuli:1998wi},
it will turn out that $M_T$ may go up
naturally by a factor of thirty, which would increase the proton lifetime
by a factor of $10^3$.
\\
\\
Now to start our calculation, consider the non-renormalizable terms
in the superpotential for the heavy sector (up to terms $1/M_{Pl}$):

\begin{equation}
{\cal W}_H = \frac{m_{\Sigma}}{2} \ Tr \Sigma^2+ \frac{\lambda}{3} Tr\Sigma^3+a{(Tr\Sigma^2)^2\over M_{Pl}}+
b{Tr\Sigma^4\over M_{Pl}}
\end{equation}
Of course, if $\lambda\approx{\cal O}(1)$, we ignore higher-dimensional
terms. However, in su\-per\-sym\-me\-try $\lambda$ is a Yukawa-type coupling,
i.e. self-renormalizable. For small $\lambda$ ($\lambda\ll M_{GUT}/M_{Pl}$),
the opposite becomes true and $a$ and $b$ terms dominate. In this case,
it is a simple exercise to show that

\begin{equation}
\label{m34m8}
m_3=4m_8
\end{equation}
where $m_3$ and $m_8$ are the masses of the weak triplet and color
octet in $\Sigma$. In the renormalizable case $m_3=m_8$.
\\
\\
At the one loop level, the RGE's for the gauge couplings are

\begin{eqnarray}
\label{alfa1}
\alpha_1^{-1}(M_Z)&=&\alpha_U^{-1}+{1\over 2\pi}\left(
-{5\over 2}\ln{\Lambda_{SUSY}\over M_Z}
+{33\over 5}\ln{M_{GUT}\over M_Z}
+{2\over 5}\ln{M_{GUT}\over M_T}\right) \non \\
\alpha_2^{-1}(M_Z)&=&\alpha_U^{-1}+{1\over 2\pi}\left(
-{25\over 6}\ln{\Lambda_{SUSY}\over M_Z}
+\ln{M_{GUT}\over M_Z}
+2\ln{M_{GUT}\over m_3}\right) \non \\
\label{alfa3}
\alpha_3^{-1}(M_Z)&=&\alpha_U^{-1}+{1\over 2\pi}\left(
-4\ln{\Lambda_{SUSY}\over M_Z}
-3\ln{m_8\over M_Z}
+\ln{M_{GUT}\over M_T}\right)
\end{eqnarray}
Here we take for simplicity $M_{GUT}=M_{X,Y}=$ superheavy gauge
bosons masses, while at the one-loop level we could as well take
$\Lambda_{SUSY}=M_Z$. From (\ref{alfa3}) we obtain

\begin{eqnarray}
2\pi\left(3\alpha_2^{-1}-2\alpha_3^{-1}-\alpha_1^{-1}\right)&=&
-2\ln{\Lambda_{SUSY}\over M_Z}
+{12\over 5}\ln{M_T\over M_Z}
+6\ln{m_8\over m_3} \non \\
2\pi\left(5\alpha_1^{-1}-3\alpha_2^{-1}-2\alpha_3^{-1}\right)&=&
8\ln{\Lambda_{SUSY}\over M_Z}
+36\ln{(\sqrt{m_3m_8}M_{GUT}^2)^{1/3}\over M_Z} \non \\
\end{eqnarray}
This gives

\begin{eqnarray}
M_T&=&M_T^0\left({m_3\over m_8}\right)^{5/2}\\
M_{GUT}&=&M_{GUT}^0\left({M_{GUT}^0\over 2m_8}\right)^{1/2}
\end{eqnarray}
Since, in the case (\ref{m34m8}) is valid, $m_8\approx M_{GUT}^2/M_{Pl}$,
we can also write

\begin{equation}
M_{GUT}\approx \left[\left(M_{GUT}^0\right)^3M_{Pl}\right]^{1/4}
\end{equation}
In the above equations the superscript $^0$ denotes the values in
the case $m_3=m_8$. From (\ref{m34m8}) we get

\begin{equation}
\label{ilia}
M_T=32 M_T^0\;\;\;\;\;\;M_{GUT}\approx 10 M_{GUT}^0\;\;\;
\end{equation}
Now, $M_{GUT}^0\approx 10^{16}$ GeV and it was shown last year
\cite{Murayama} that $M_T>7\times 10^{16}$ GeV is sufficiently
large to be in accord with the newest data on proton decay.
On the other hand, since we had

\begin{equation}
M_T^0 < 3.6 \times 10^{15}{\rm GeV}\;
\end{equation}
from (\ref{ilia}) we see that $m_3=4m_8$ is enough to save the
theory, even in the minimal SUGRA model. Obviously, 
an improvement of the measurement of $\tau_p$ is badly needed. 
It is noteworthy that in this case the usual $d=6$ proton 
decay becomes out of reach: $\tau_p(d=6)>10^{38}$ yrs.
\\
\\
As we see there are many reasons to believe that the Minimal 
Supersymmetric $SU(5)$ model is not ruled out. The proton decay
lifetime depends of many unknown parameters, therefore at the same
time using the proton decay constraint we could learn about a sector
which is orthogonal to grand unification. In other words, the improved 
measurements of proton decay will provide information about the nature
of supersymmetry breaking (i.e., the soft masses) and the fermionic
mass textures.
\\
\\
After our analysis different groups have been studied the same crucial
issue in the minimal supersymmetric $SU(5)$ model. Our predictions
about how suppress the $d=5$ contributions has been confirmed 
in reference \cite{Costa}. In this case they computed the proton lifetime
in realistic scenarios, where the needed non-renormalizable operators in the
Yukawa and Higgs sector of the theory are considered. They showed several numerical
examples where the experimental bounds are satisfied.
\\
\\
Therefore from our analysis it is clear that \textit{the minimal 
realization of the ideas of supersymmetric unification is not ruled out by the Proton decay experiments}.

\chapter{Conclusions}

In this thesis we studied two important phenomenological issues in the
context of supersymmetric gauge theories. Invisible Higgs boson decays into
neutralinos and proton decay.
\\
\\
We started with the study of the invisible Higgs decays into two neutralinos in
the context of the minimal supersymmetric version of the Standard
Model, considering new one-loop corrections to the neutral Higgs boson
couplings to neutralinos in the gaugino limit. We computed these
important quantum corrections and included our results in the Fortran
code HDECAY \cite{HDECAY} to compute the branching ratios for the
neutral Higgs bosons. We focused on the phenomenologically most 
interesting case of a bino--like lightest neutralino \lsp\ as lightest
supersymmetric particle, but our analytical results are valid for a
more general gaugino--like neutralino, irrespective of whether it is
the lightest supersymmetric particle. We found that these
corrections can completely dominate the tree--level contribution to
the coupling of the lightest CP--even Higgs boson. The corrections to
the couplings of heavy CP--even Higgs boson are somewhat
less significant, but can still amount to about a factor of 2. Since the
CP--odd Higgs boson cannot couple to two identical sfermions the
corrections are suppressed in this case. In all cases the corrections
can be significant only if some sfermion masses are considerably
smaller than the supersymmetric higgsino mass $|\mu|$. The Higgs 
couplings receive their potentially largest corrections from loops 
involving third generation quarks and their superpartners. In the 
latter case the corrections are also quite sensitive to the size 
of the trilinear soft breaking parameter $A_t$ (and $A_b$, if $\tan\beta \gg 1$). 
\\
\\
The possible impact of the corrections on the invisible width of the 
lightest CP-even Higgs boson is dramatic, it could be enhanced to a
level that should be easily measurable at future high-energy $e^+ e^-$ 
colliders, even if the neutralino is an almost perfect
bino. \textit{This would open a new window for testing the Minimal 
Supersymmetric Standard Model at the quantum level} 
\\
\\
Turning to applications of these calculations, our results motivated
us to investigate the effect of these quantum corrections in the elastic
neutralino-nucleon cross section. In this case we found \cite{loop} that these
corrections might change the predicted detection rate of Dark Matter
LSPs by up to a factor of two even for scenarios where the rate is
close to the sensitivity of the next round of direct Dark Matter
detection experiments. 
\\
\\
A new analysis has been performed after our results were published, where also the effect 
of the subleading loop corrections has been analyzed. In
reference \cite{Eberl} a different group studied the one-loop
corrections to neutral Higgs bosons decays into neutralinos in the
general case, where we have the decays $H^0_i \to \tilde{\chi}_m^0 
\tilde{\chi}_n^0$ (i=1,2,3). They confirmed our results
computing the branching ratios for similar values of the parameters.
\\
\\
In the second part of our work we focused on proton decay in the
Minimal Supersymmetric $SU(5)$ model. We studied the
$d=5$ operators contributing to the decay of
the proton, writing all the possible contributions for each
decay channel (see Appendix B) in a general SUSY scenario. 
We pointed out the major sources of uncertainties in estimating 
the proton decay lifetime, as the ignorance of the masses of the 
color octet and weak triplet supermultiplets in the adjoint Higgs, 
and the unknown mixings between fermions and sfermions. 
Non-renormalizable operators are considered in order to correct the
relation between the fermion masses.
\\
\\
We found that the Higgs triplet mass may goes up naturally by a factor of
thirty, when we allow for arbitrary trilinear coupling in the presence
of non-renormalizable operators, therefore in this case it is 
possible to satisfy the proton decay bounds, when the 
mixings between fermions and sfermions are known. This is the case of
the minimal SUGRA model. Alternatively, if the standard value of $M_T$
is adopted, the conditions for the sfermion and fermion mass matrices 
in order to achieve the proton longevity were found. Knowing all the aspects mentioned before we
can conclude that the \textit{minimal version of the ideas of
Supersymmetric Grand Unification is not ruled out} as claimed before.
We could say that if proton decay is found, it could provide
indirect information about the nature of supersymmetry breaking 
(i.e., the soft masses) and the fermionic mass textures. 
Thus opening a new way to test models for fermion and sfermion
masses. We could say that our analysis
is also valid for any Supersymmetric version of Grand Unified
Theories as minimal SUSY $SO(10)$ \cite{SO102}. 
\\
\\
Our results motivate the analysis in reference\cite{Costa}, where our predictions
about how to suppress the $d=5$ contributions has been confirmed in
several numerical examples. 
\\
\\
We hope that these results motivate the realization of new studies in
the context of supersymmetric gauge theories. The loop corrections
computed by us could be important for future studies in collider
physics. While our ideas of how to suppress or constrain the $d=5$
operators contributing for proton decay might be useful for
understanding how it is possible test the ideas of supersymmetric unification.

\newpage
\vspace*{\fill}
\newpage
\pagestyle{empty}
\textit{\textbf{{\LARGE Acknowledgments.}}}
\\
\\
In the first place I would like to thank my family and in special my
mother for the inconditional support of any kind during these years
that I was far from home.
\\
\\
I thank my supervisor the Professor Manuel Drees for introduced me in
the field of SUSY phenomenology. He was always ready to discuss my
questions. I would like to thank him for the support during the 
development of this thesis. 
\\   
I would like to thank my supervisor Professor Goran Senjanovi\'c, for
his support during my Ph.D and for provide me the opportunity to work 
in the field of supersymmetric grand unified theories. I thank 
him for his friendship, all the discussions and support in the 
realization of this thesis. At any time he was ready to discuss my
questions and ideas. I would like to thank him for the invitations to
work at the ICTP.
\\
I would like to thank the Max Planck Institut f\"ur Physics
(Werner-Heisenberg-Institut) for support. I will remember forever 
all the moments at the institute. 
\\
I would like to thank the Abdus Salam International Centre for
Theoretical Physics (ICTP) in Trieste for the support. It is a special
place for me, which I will remember forever.  
\\
I thank the Prof. A. Djouadi and M.~Muehlleitner for 
collaboration in the field of Higgs Physics.
\\
I woud like to thank Prof. Borut Bajc for collaboration, support and 
for careful reading of this thesis.
\\
I thank Pushan Majumdar for careful reading of the thesis.
\\
I would like to thank my friends Alex, Shirley, Alina, Ariel (el
Che), Alexei, Silvia, Gabriela and Andrea for their special and strong
support during these years. I will remember their helps and all the 
special moments that we spent together.
\\
Last but not least I would like to express my warmest thanks to Natalia.

\newpage

\newpage
  
\begin{appendix}

\chapter{Passarino-Veltman Loop Formulae}
The explicit form of the Passarino-Veltman loop integrals are:
\begin{displaymath}
C_0 (p_{1}, p_{2}, M_{1}, M_{2}, M_{3}) = (2 \pi \mu)^{4-n} \int \frac{d^{n} l}{i \pi^2} \frac{1}{D_{1}D_{2}D_{3}}
\end{displaymath}
with

\begin{displaymath}
D_1 = l^{2} - M^2_1 + i \epsilon
\end{displaymath}

\begin{displaymath}
D_2 = (l + p_{1})^{2} - M^2_2 + i \epsilon
\end{displaymath}

\begin{displaymath}
D_3 = (l + p_{1}+p_{2})^{2} - M^2_3 + i \epsilon
\end{displaymath}

\begin{displaymath}
\overline{C}_{\mu} (p_{1},p_{2}, M_{1}, M_{2}, M_{3}) =
(2 \pi \mu)^{4-n} \int \frac{d^{n}l}{i \pi^2} \frac{l_{\mu}}{D_{4}D_{5}D_{6}}
\end{displaymath}
where:

\begin{displaymath}
D_4 = (l - p_{1})^{2} - M^2_1 + i \epsilon
\end{displaymath}

\begin{displaymath}
D_5 = (l - p_{2})^{2} - M^2_3  + i \epsilon
\end{displaymath}

\begin{displaymath}
D_6 = l^{2} - M^2_1 + i \epsilon
\end{displaymath}
where $p_{i}$ and $M_{i}$ are the external momenta and the masses 
respectively. For more details see reference \cite{Drees1}.  

\newpage

\chapter{Proton Decay Diagrams}

In this Appendix we present the complete set of diagrams
responsible for d=5 nucleon decay in the minimal
supersymmetric SU(5) theory. In our notation $T$ and $\bar T$
stand for heavy Higgs triplets; $\tilde T$ and $\tilde{\bar T}$
denote their fermionic partners; $\tilde w^{\pm}$ stands for winos,
$\tilde h_{+,0}$ and $\tilde{\bar h}_{-,0}$ are light Higgsinos
and $\tilde V_0$ stand for neutral gauginos.
\vskip 0.5cm
\noindent
\underline{Decay modes}: 
\\
\\
$p \to (K^+, \pi^+, \rho^+, K^{*+}) \bar\nu_i$ \ and 
\ $ n \to (\pi^0, \rho^0, \eta, \omega, K^0, K^{*0})\bar\nu_i$
\\
\\
where $i=1,2,3$.

\begin{center}
\begin{displaymath}
\label{nw1}
\def\trgor{$\tilde T$}
\def\trdol{$\tilde{\bar T}$}
\def\sfgor{$\tilde t$}
\def\sfdol{$\tilde\tau$}
\def\spgor{$\tilde w^+$}
\def\spdol{$\tilde w^-$}
\def\iena{$d_{1,2}$}
\def\idva{$u$}
\def\itri{$\nu_i$}
\def\isti{$d_{2,1}$}
\begin{picture}(80,40)(40,20)
\SetWidth{0.8}
\ArrowLine(0,40)(20,40)
\ArrowLine(0,0)(20,0)
\ArrowLine(80,40)(60,40)
\ArrowLine(80,0)(60,0)
\DashLine(20,40)(60,40)3
\DashLine(20,0)(60,0)3
\ArrowLine(20,20)(20,40)
\ArrowLine(20,20)(20,0)
\Line(17,17)(23,23)
\Line(17,23)(23,17)
\ArrowLine(60,20)(60,40)
\ArrowLine(60,20)(60,0)
\Line(57,17)(63,23)
\Line(57,23)(63,17)
\Text(10,30)[]{\trgor}
\Text(10,10)[]{\trdol}
\Text(40,30)[]{\sfgor}
\Text(40,10)[]{\sfdol}
\Text(75,30)[]{\spgor}
\Text(75,10)[]{\spdol}
\Text(-10,40)[]{\iena}
\Text(-10,0)[]{\idva}
\Text(90,0)[]{\itri}
\Text(90,40)[]{\isti}
\end{picture}

\propto 
(D^T\underline A\tilde U)_{13,23}(\tilde U^\dagger D)_{32,31}
(N^T\tilde E^*)_{i3}(\tilde E^T\underline C^TU)_{31}
\end{displaymath}
\end{center}

\begin{center}
\begin{displaymath}
\def\tr{$T$}
\def\sfgor{$\tilde t$}
\def\sfdol{$\tilde\tau$}
\def\spgor{$\tilde w^+$}
\def\spdol{$\tilde w^-$}
\def\iena{$d_{1,2}$}
\def\idva{$u$}
\def\itri{$\nu_i$}
\def\isti{$d_{2,1}$}
\begin{picture}(80,40)(40,20)
\SetWidth{0.8}
\ArrowLine(0,40)(10,20)
\ArrowLine(0,0)(10,20)
\ArrowLine(80,40)(60,40)
\ArrowLine(80,0)(60,0)
\DashLine(10,20)(40,20)3
\DashLine(40,20)(60,40)3
\DashLine(40,20)(60,0)3
\ArrowLine(60,20)(60,40)
\ArrowLine(60,20)(60,0)
\Line(57,17)(63,23)
\Line(57,23)(63,17)
\Text(25,30)[]{\tr}
\Text(45,35)[]{\sfgor}
\Text(45,5)[]{\sfdol}
\Text(75,30)[]{\spgor}
\Text(75,10)[]{\spdol}
\Text(-10,40)[]{\iena}
\Text(-10,0)[]{\idva}
\Text(90,0)[]{\itri}
\Text(90,40)[]{\isti}
\end{picture}

\propto
(D^T\underline AU)_{11,21}(N^T\tilde E^*)_{i3}
(\tilde E^T\underline C^T\tilde U)_{33}(\tilde U^\dagger D)_{32,31}
\end{displaymath}
\end{center}

\begin{center}
\begin{displaymath}
\def\trgor{$\tilde T$}
\def\trdol{$\tilde{\bar T}$}
\def\sfgor{$\tilde t$}
\def\sfdol{$\tilde b$}
\def\spgor{$\tilde w^+$}
\def\spdol{$\tilde w^-$}
\def\iena{$d_{1,2}$}
\def\idva{$\nu_i$}
\def\itri{$u$}
\def\isti{$d_{2,1}$}
\begin{picture}(80,40)(40,20)
\SetWidth{0.8}
\ArrowLine(0,40)(20,40)
\ArrowLine(0,0)(20,0)
\ArrowLine(80,40)(60,40)
\ArrowLine(80,0)(60,0)
\DashLine(20,40)(60,40)3
\DashLine(20,0)(60,0)3
\ArrowLine(20,20)(20,40)
\ArrowLine(20,20)(20,0)
\Line(17,17)(23,23)
\Line(17,23)(23,17)
\ArrowLine(60,20)(60,40)
\ArrowLine(60,20)(60,0)
\Line(57,17)(63,23)
\Line(57,23)(63,17)
\Text(10,30)[]{\trgor}
\Text(10,10)[]{\trdol}
\Text(40,30)[]{\sfgor}
\Text(40,10)[]{\sfdol}
\Text(75,30)[]{\spgor}
\Text(75,10)[]{\spdol}
\Text(-10,40)[]{\iena}
\Text(-10,0)[]{\idva}
\Text(90,0)[]{\itri}
\Text(90,40)[]{\isti}
\end{picture}

\propto
(D^T\underline A\tilde U)_{13,23}(\tilde U^\dagger D)_{32,31}
(U^T\tilde D^*)_{13}(\tilde D^T\underline CN)_{3i}
\end{displaymath}
\end{center}

\begin{center}
\begin{displaymath}
\def\tr{$\bar T$}
\def\sfgor{$\tilde t$}
\def\sfdol{$\tilde b$}
\def\spgor{$\tilde w^+$}
\def\spdol{$\tilde w^-$}
\def\iena{$d_{1,2}$}
\def\idva{$\nu_i$}
\def\itri{$u$}
\def\isti{$d_{2,1}$}
\begin{picture}(80,40)(40,20)
\SetWidth{0.8}
\ArrowLine(0,40)(10,20)
\ArrowLine(0,0)(10,20)
\ArrowLine(80,40)(60,40)
\ArrowLine(80,0)(60,0)
\DashLine(10,20)(40,20)3
\DashLine(40,20)(60,40)3
\DashLine(40,20)(60,0)3
\ArrowLine(60,20)(60,40)
\ArrowLine(60,20)(60,0)
\Line(57,17)(63,23)
\Line(57,23)(63,17)
\Text(25,30)[]{\tr}
\Text(45,35)[]{\sfgor}
\Text(45,5)[]{\sfdol}
\Text(75,30)[]{\spgor}
\Text(75,10)[]{\spdol}
\Text(-10,40)[]{\iena}
\Text(-10,0)[]{\idva}
\Text(90,0)[]{\itri}
\Text(90,40)[]{\isti}
\end{picture}

\propto
(D^T\underline CN)_{1i,2i}(U^T\tilde D^*)_{13}
(\tilde D^T\underline A\tilde U)_{33}(\tilde U^\dagger D)_{32,31}
\end{displaymath}
\end{center}
\begin{center}
\begin{displaymath}
\def\trgor{$\tilde T$}
\def\trdol{$\tilde{\bar T}$}
\def\sfgor{$\tilde t$}
\def\sfdol{$\tilde b$}
\def\spgor{$\tilde{\bar h}_-^\dagger$}
\def\spdol{$\tilde h_+^\dagger$}
\def\iena{$d_{1,2}$}
\def\idva{$\nu_i$}
\def\itri{${\bar u}^c$}
\def\isti{${\bar d}_{2,1}^c$}
\begin{picture}(80,40)(40,20)
\SetWidth{0.8}
\ArrowLine(0,40)(20,40)
\ArrowLine(0,0)(20,0)
\ArrowLine(80,40)(60,40)
\ArrowLine(80,0)(60,0)
\DashLine(20,40)(60,40)3
\DashLine(20,0)(60,0)3
\ArrowLine(20,20)(20,40)
\ArrowLine(20,20)(20,0)
\Line(17,17)(23,23)
\Line(17,23)(23,17)
\ArrowLine(60,20)(60,40)
\ArrowLine(60,20)(60,0)
\Line(57,17)(63,23)
\Line(57,23)(63,17)
\Text(10,30)[]{\trgor}
\Text(10,10)[]{\trdol}
\Text(40,30)[]{\sfgor}
\Text(40,10)[]{\sfdol}
\Text(75,30)[]{\spgor}
\Text(75,10)[]{\spdol}
\Text(-10,40)[]{\iena}
\Text(-10,0)[]{\idva}
\Text(90,0)[]{\itri}
\Text(90,40)[]{\isti}
\end{picture}

\propto
(D^T\underline A\tilde U)_{13,23}(\tilde U^\dagger Y_D^*D_c^*)_{32,31}
(U_c^\dagger Y_U^\dagger\tilde D^*)_{13}(\tilde D^T\underline CN)_{3i}
\end{displaymath}
\end{center}
\begin{center}
\begin{displaymath}
\label{nh2}
\def\tr{$\bar T$}
\def\sfgor{$\tilde t$}
\def\sfdol{$\tilde b$}
\def\spgor{$\tilde{\bar h}_-^\dagger$}
\def\spdol{$\tilde h_+^\dagger$}
\def\iena{$d_{1,2}$}
\def\idva{$\nu_i$}
\def\itri{${\bar u}^c$}
\def\isti{${\bar d}_{2,1}^c$}
\begin{picture}(80,40)(40,20)
\SetWidth{0.8}
\ArrowLine(0,40)(10,20)
\ArrowLine(0,0)(10,20)
\ArrowLine(80,40)(60,40)
\ArrowLine(80,0)(60,0)
\DashLine(10,20)(40,20)3
\DashLine(40,20)(60,40)3
\DashLine(40,20)(60,0)3
\ArrowLine(60,20)(60,40)
\ArrowLine(60,20)(60,0)
\Line(57,17)(63,23)
\Line(57,23)(63,17)
\Text(25,30)[]{\tr}
\Text(45,35)[]{\sfgor}
\Text(45,5)[]{\sfdol}
\Text(75,30)[]{\spgor}
\Text(75,10)[]{\spdol}
\Text(-10,40)[]{\iena}
\Text(-10,0)[]{\idva}
\Text(90,0)[]{\itri}
\Text(90,40)[]{\isti}
\end{picture}

\propto
(D^T\underline CN)_{1i,2i}(U_c^\dagger Y_U^\dagger\tilde D^*)_{13}
(\tilde D^T\underline A\tilde U)_{33}(\tilde U^\dagger Y_D^*D_c^*)_{32,31}
\end{displaymath}
\end{center}
\begin{center}
\begin{displaymath}
\label{nh3}
\def\trgor{$\tilde T^\dagger$}
\def\trdol{$\tilde{\bar T}^\dagger$}
\def\sfgor{$\tilde\tau^c$}
\def\sfdol{$\tilde t^c$}
\def\spgor{$\tilde{\bar h}_-$}
\def\spdol{$\tilde h_+$}
\def\iena{${\bar u}^c$}
\def\idva{${\bar d}_{2,1}^c$}
\def\itri{$d_{1,2}$}
\def\isti{$\nu_i$}
\begin{picture}(80,40)(40,20)
\SetWidth{0.8}
\ArrowLine(0,40)(20,40)
\ArrowLine(0,0)(20,0)
\ArrowLine(80,40)(60,40)
\ArrowLine(80,0)(60,0)
\DashLine(20,40)(60,40)3
\DashLine(20,0)(60,0)3
\ArrowLine(20,20)(20,40)
\ArrowLine(20,20)(20,0)
\Line(17,17)(23,23)
\Line(17,23)(23,17)
\ArrowLine(60,20)(60,40)
\ArrowLine(60,20)(60,0)
\Line(57,17)(63,23)
\Line(57,23)(63,17)
\Text(10,30)[]{\trgor}
\Text(10,10)[]{\trdol}
\Text(40,30)[]{\sfgor}
\Text(40,10)[]{\sfdol}
\Text(75,30)[]{\spgor}
\Text(75,10)[]{\spdol}
\Text(-10,40)[]{\iena}
\Text(-10,0)[]{\idva}
\Text(90,0)[]{\itri}
\Text(90,40)[]{\isti}
\end{picture}

\propto
(U_c^\dagger\underline B^*\tilde E_c^*)_{13}(\tilde E_c^TY_EN)_{3i}
(D^TY_U\tilde U_c)_{13,23}(\tilde U_c^\dagger\underline D^*D_c^*)_{32,31}
\end{displaymath}
\end{center}
\begin{center}
\begin{displaymath}
\label{nh4}
\def\tr{$\bar T$}
\def\sfgor{$\tilde\tau^c$}
\def\sfdol{$\tilde t^c$}
\def\spgor{$\tilde{\bar h}_-$}
\def\spdol{$\tilde h_+$}
\def\iena{${\bar u}^c$}
\def\idva{${\bar d}_{1,2}^c$}
\def\itri{$d_{2,1}$}
\def\isti{$\nu_i$}
\begin{picture}(80,40)(40,20)
\SetWidth{0.8}
\ArrowLine(0,40)(10,20)
\ArrowLine(0,0)(10,20)
\ArrowLine(80,40)(60,40)
\ArrowLine(80,0)(60,0)
\DashLine(10,20)(40,20)3
\DashLine(40,20)(60,40)3
\DashLine(40,20)(60,0)3
\ArrowLine(60,20)(60,40)
\ArrowLine(60,20)(60,0)
\Line(57,17)(63,23)
\Line(57,23)(63,17)
\Text(25,30)[]{\tr}
\Text(45,35)[]{\sfgor}
\Text(45,5)[]{\sfdol}
\Text(75,30)[]{\spgor}
\Text(75,10)[]{\spdol}
\Text(-10,40)[]{\iena}
\Text(-10,0)[]{\idva}
\Text(90,0)[]{\itri}
\Text(90,40)[]{\isti}
\end{picture}

\propto
(U_c^\dagger\underline D^* D_c^*)_{11,12}(D^TY_U\tilde U_c)_{23,13}
(\tilde U_c^\dagger\underline B^*\tilde E_c^*)_{33}(\tilde E_c^TY_EN)_{3i}
\end{displaymath}
\end{center}
\begin{center}
\begin{displaymath}
\label{nh01}
\def\trgor{$\tilde T$}
\def\trdol{$\tilde{\bar T}$}
\def\sfgor{$\tilde t$}
\def\sfdol{$\tilde b$}
\def\spgor{$\tilde h_0^\dagger$}
\def\spdol{$\tilde{\bar h}_0^\dagger$}
\def\iena{$d_{1,2}$}
\def\idva{$\nu_i$}
\def\itri{${\bar d}_{2,1}^c$}
\def\isti{${\bar u}^c$}
\begin{picture}(80,40)(40,20)
\SetWidth{0.8}
\ArrowLine(0,40)(20,40)
\ArrowLine(0,0)(20,0)
\ArrowLine(80,40)(60,40)
\ArrowLine(80,0)(60,0)
\DashLine(20,40)(60,40)3
\DashLine(20,0)(60,0)3
\ArrowLine(20,20)(20,40)
\ArrowLine(20,20)(20,0)
\Line(17,17)(23,23)
\Line(17,23)(23,17)
\ArrowLine(60,20)(60,40)
\ArrowLine(60,20)(60,0)
\Line(57,17)(63,23)
\Line(57,23)(63,17)
\Text(10,30)[]{\trgor}
\Text(10,10)[]{\trdol}
\Text(40,30)[]{\sfgor}
\Text(40,10)[]{\sfdol}
\Text(75,30)[]{\spgor}
\Text(75,10)[]{\spdol}
\Text(-10,40)[]{\iena}
\Text(-10,0)[]{\idva}
\Text(90,0)[]{\itri}
\Text(90,40)[]{\isti}
\end{picture}

\propto
(D^T\underline A\tilde U)_{13,23}(\tilde U^\dagger Y_U^*U_c^*)_{31}
(D_c^\dagger Y_D^\dagger\tilde D^*)_{23,13}(\tilde D^T\underline CN)_{3i}
\end{displaymath}
\end{center}
\begin{center}
\begin{displaymath}
\def\tr{$\bar T$}
\def\sfgor{$\tilde b$}
\def\sfdol{$\tilde t$}
\def\spgor{$\tilde{\bar h}_0^\dagger$}
\def\spdol{$\tilde h_0^\dagger$}
\def\iena{$d_{1,2}$}
\def\idva{$\nu_i$}
\def\itri{${\bar u}^c$}
\def\isti{${\bar d}_{2,1}^c$}
\begin{picture}(80,40)(40,20)
\SetWidth{0.8}
\ArrowLine(0,40)(10,20)
\ArrowLine(0,0)(10,20)
\ArrowLine(80,40)(60,40)
\ArrowLine(80,0)(60,0)
\DashLine(10,20)(40,20)3
\DashLine(40,20)(60,40)3
\DashLine(40,20)(60,0)3
\ArrowLine(60,20)(60,40)
\ArrowLine(60,20)(60,0)
\Line(57,17)(63,23)
\Line(57,23)(63,17)
\Text(25,30)[]{\tr}
\Text(45,35)[]{\sfgor}
\Text(45,5)[]{\sfdol}
\Text(75,30)[]{\spgor}
\Text(75,10)[]{\spdol}
\Text(-10,40)[]{\iena}
\Text(-10,0)[]{\idva}
\Text(90,0)[]{\itri}
\Text(90,40)[]{\isti}
\end{picture}

\propto
(D^T\underline CN)_{1i,2i}(U_c^\dagger Y_U^\dagger\tilde U^*)_{13}
(\tilde U^T\underline A\tilde D)_{33}(\tilde D^\dagger Y_D^*D_c^*)_{32,31}
\end{displaymath}
\end{center}
\begin{center}
\begin{displaymath}
\def\trgor{$\tilde T$}
\def\trdol{$\tilde{\bar T}$}
\def\sfgor{$\tilde t$}
\def\sfdol{$\tilde b$}
\def\spgor{$\tilde V_0$}
\def\spdol{$\tilde V_0$}
\def\iena{$d_{1,2}$}
\def\idva{$\nu_i$}
\def\itri{$d_{2,1}$}
\def\isti{$u$}
\begin{picture}(80,40)(40,20)
\SetWidth{0.8}
\ArrowLine(0,40)(20,40)
\ArrowLine(0,0)(20,0)
\ArrowLine(80,40)(60,40)
\ArrowLine(80,0)(60,0)
\DashLine(20,40)(60,40)3
\DashLine(20,0)(60,0)3
\ArrowLine(20,20)(20,40)
\ArrowLine(20,20)(20,0)
\Line(17,17)(23,23)
\Line(17,23)(23,17)
\ArrowLine(60,20)(60,40)
\ArrowLine(60,20)(60,0)
\Line(57,17)(63,23)
\Line(57,23)(63,17)
\Text(10,30)[]{\trgor}
\Text(10,10)[]{\trdol}
\Text(40,30)[]{\sfgor}
\Text(40,10)[]{\sfdol}
\Text(75,30)[]{\spgor}
\Text(75,10)[]{\spdol}
\Text(-10,40)[]{\iena}
\Text(-10,0)[]{\idva}
\Text(90,0)[]{\itri}
\Text(90,40)[]{\isti}
\end{picture}

\propto
(D^T\underline A\tilde U)_{13,23}(\tilde U^\dagger U)_{31}
(D^T\tilde D^*)_{23,13}(\tilde D^T\underline CN)_{3i}
\end{displaymath}
\end{center}
\begin{center}
\begin{displaymath}
\def\tr{$\bar T$}
\def\sfgor{$\tilde t$}
\def\sfdol{$\tilde b$}
\def\spgor{$\tilde V_0$}
\def\spdol{$\tilde V_0$}
\def\iena{$d_{1,2}$}
\def\idva{$\nu_i$}
\def\itri{$d_{2,1}$}
\def\isti{$u$}
\begin{picture}(80,40)(40,20)
\SetWidth{0.8}
\ArrowLine(0,40)(10,20)
\ArrowLine(0,0)(10,20)
\ArrowLine(80,40)(60,40)
\ArrowLine(80,0)(60,0)
\DashLine(10,20)(40,20)3
\DashLine(40,20)(60,40)3
\DashLine(40,20)(60,0)3
\ArrowLine(60,20)(60,40)
\ArrowLine(60,20)(60,0)
\Line(57,17)(63,23)
\Line(57,23)(63,17)
\Text(25,30)[]{\tr}
\Text(45,35)[]{\sfgor}
\Text(45,5)[]{\sfdol}
\Text(75,30)[]{\spgor}
\Text(75,10)[]{\spdol}
\Text(-10,40)[]{\iena}
\Text(-10,0)[]{\idva}
\Text(90,0)[]{\itri}
\Text(90,40)[]{\isti}
\end{picture}

\propto
(D^T\underline CN)_{1i,2i}(D^T\tilde D^*)_{23,13}
(\tilde D^T\underline A\tilde U)_{33}(\tilde U^\dagger U)_{31}
\end{displaymath}
\end{center}
\begin{center}
\begin{displaymath}
\def\trgor{$\tilde{\bar T}$}
\def\trdol{$\tilde T$}
\def\sfgor{$\tilde\nu$}
\def\sfdol{$\tilde b$}
\def\spgor{$\tilde V_0$}
\def\spdol{$\tilde V_0$}
\def\iena{$d_{1,2}$}
\def\idva{$u$}
\def\itri{$d_{2,1}$}
\def\isti{$\nu_i$}
\begin{picture}(80,40)(40,20)
\SetWidth{0.8}
\ArrowLine(0,40)(20,40)
\ArrowLine(0,0)(20,0)
\ArrowLine(80,40)(60,40)
\ArrowLine(80,0)(60,0)
\DashLine(20,40)(60,40)3
\DashLine(20,0)(60,0)3
\ArrowLine(20,20)(20,40)
\ArrowLine(20,20)(20,0)
\Line(17,17)(23,23)
\Line(17,23)(23,17)
\ArrowLine(60,20)(60,40)
\ArrowLine(60,20)(60,0)
\Line(57,17)(63,23)
\Line(57,23)(63,17)
\Text(10,30)[]{\trgor}
\Text(10,10)[]{\trdol}
\Text(40,30)[]{\sfgor}
\Text(40,10)[]{\sfdol}
\Text(75,30)[]{\spgor}
\Text(75,10)[]{\spdol}
\Text(-10,40)[]{\iena}
\Text(-10,0)[]{\idva}
\Text(90,0)[]{\itri}
\Text(90,40)[]{\isti}
\end{picture}

\propto
(D^T\underline C\tilde N)_{13,23}(\tilde N^\dagger N)_{3i}
(D^T\tilde D^*)_{23,13}(\tilde D^T\underline AU)_{31}
\end{displaymath}
\end{center}
\begin{center}
\begin{displaymath}
\label{nv4}
\def\tr{$T$}
\def\sfgor{$\tilde\nu$}
\def\sfdol{$\tilde b$}
\def\spgor{$\tilde V_0$}
\def\spdol{$\tilde V_0$}
\def\iena{$u$}
\def\idva{$d_{1,2}$}
\def\itri{$d_{2,1}$}
\def\isti{$\nu_i$}
\begin{picture}(80,40)(40,20)
\SetWidth{0.8}
\ArrowLine(0,40)(10,20)
\ArrowLine(0,0)(10,20)
\ArrowLine(80,40)(60,40)
\ArrowLine(80,0)(60,0)
\DashLine(10,20)(40,20)3
\DashLine(40,20)(60,40)3
\DashLine(40,20)(60,0)3
\ArrowLine(60,20)(60,40)
\ArrowLine(60,20)(60,0)
\Line(57,17)(63,23)
\Line(57,23)(63,17)
\Text(25,30)[]{\tr}
\Text(45,35)[]{\sfgor}
\Text(45,5)[]{\sfdol}
\Text(75,30)[]{\spgor}
\Text(75,10)[]{\spdol}
\Text(-10,40)[]{\iena}
\Text(-10,0)[]{\idva}
\Text(90,0)[]{\itri}
\Text(90,40)[]{\isti}
\end{picture}

\propto
(U^T\underline AD)_{11,12}(D^T\tilde D^*)_{23,13}
(\tilde D^T\underline C\tilde N)_{33}(\tilde N^\dagger N)_{3i}
\end{displaymath}
\end{center}
\begin{center}
\begin{displaymath}
\label{nv5}
\def\trgor{$\tilde{\bar T}$}
\def\trdol{$\tilde T$}
\def\sfgor{$\tilde\nu$}
\def\sfdol{$\tilde t$}
\def\spgor{$\tilde V_0$}
\def\spdol{$\tilde V_0$}
\def\iena{$d_{1,2}$}
\def\idva{$d_{2,1}$}
\def\itri{$u$}
\def\isti{$\nu_i$}
\begin{picture}(80,40)(40,20)
\SetWidth{0.8}
\ArrowLine(0,40)(20,40)
\ArrowLine(0,0)(20,0)
\ArrowLine(80,40)(60,40)
\ArrowLine(80,0)(60,0)
\DashLine(20,40)(60,40)3
\DashLine(20,0)(60,0)3
\ArrowLine(20,20)(20,40)
\ArrowLine(20,20)(20,0)
\Line(17,17)(23,23)
\Line(17,23)(23,17)
\ArrowLine(60,20)(60,40)
\ArrowLine(60,20)(60,0)
\Line(57,17)(63,23)
\Line(57,23)(63,17)
\Text(10,30)[]{\trgor}
\Text(10,10)[]{\trdol}
\Text(40,30)[]{\sfgor}
\Text(40,10)[]{\sfdol}
\Text(75,30)[]{\spgor}
\Text(75,10)[]{\spdol}
\Text(-10,40)[]{\iena}
\Text(-10,0)[]{\idva}
\Text(90,0)[]{\itri}
\Text(90,40)[]{\isti}
\end{picture}

\propto
(D^T\underline C\tilde N)_{13,23}(\tilde N^\dagger N)_{3i}
(U^T\tilde U^*)_{13}(\tilde U^T\underline AD)_{32,31}
\end{displaymath}
\end{center}
\vspace{2.0cm}
\underline{Decay modes}:
\\
\\
$p \to (K^0, \pi^0, \eta, K^{*0}, \rho^0, \omega) e_i^+$
and $n \to (K^-, \pi^-, K^{*-}, \rho^-) e_i^+$
\\
\\
where $i=1, 2$, while for $K^*$ $i=1$.

\begin{center}
\begin{displaymath}
\def\trgor{$\tilde{\bar T}$}
\def\trdol{$\tilde T$}
\def\sfgor{$\tilde\nu$}
\def\sfdol{$\tilde b$}
\def\spgor{$\tilde w^+$}
\def\spdol{$\tilde w^-$}
\def\iena{$d_{1,2}$}
\def\idva{$u$}
\def\itri{$u$}
\def\isti{$e_i$}
\begin{picture}(80,40)(40,20)
\SetWidth{0.8}
\ArrowLine(0,40)(20,40)
\ArrowLine(0,0)(20,0)
\ArrowLine(80,40)(60,40)
\ArrowLine(80,0)(60,0)
\DashLine(20,40)(60,40)3
\DashLine(20,0)(60,0)3
\ArrowLine(20,20)(20,40)
\ArrowLine(20,20)(20,0)
\Line(17,17)(23,23)
\Line(17,23)(23,17)
\ArrowLine(60,20)(60,40)
\ArrowLine(60,20)(60,0)
\Line(57,17)(63,23)
\Line(57,23)(63,17)
\Text(10,30)[]{\trgor}
\Text(10,10)[]{\trdol}
\Text(40,30)[]{\sfgor}
\Text(40,10)[]{\sfdol}
\Text(75,30)[]{\spgor}
\Text(75,10)[]{\spdol}
\Text(-10,40)[]{\iena}
\Text(-10,0)[]{\idva}
\Text(90,0)[]{\itri}
\Text(90,40)[]{\isti}
\end{picture}

\propto
(D^T\underline C\tilde N)_{13,23}(\tilde N^\dagger E)_{3i}
(U^T\tilde D^*)_{13}(\tilde D^T\underline AU)_{31}
\end{displaymath}
\end{center}

\begin{center}
\begin{displaymath}
\def\tr{$T$}
\def\sfgor{$\tilde\nu$}
\def\sfdol{$\tilde b$}
\def\spgor{$\tilde w^+$}
\def\spdol{$\tilde w^-$}
\def\iena{$d_{1,2}$}
\def\idva{$u$}
\def\itri{$u$}
\def\isti{$e_i$}
\begin{picture}(80,40)(40,20)
\SetWidth{0.8}
\ArrowLine(0,40)(10,20)
\ArrowLine(0,0)(10,20)
\ArrowLine(80,40)(60,40)
\ArrowLine(80,0)(60,0)
\DashLine(10,20)(40,20)3
\DashLine(40,20)(60,40)3
\DashLine(40,20)(60,0)3
\ArrowLine(60,20)(60,40)
\ArrowLine(60,20)(60,0)
\Line(57,17)(63,23)
\Line(57,23)(63,17)
\Text(25,30)[]{\tr}
\Text(45,35)[]{\sfgor}
\Text(45,5)[]{\sfdol}
\Text(75,30)[]{\spgor}
\Text(75,10)[]{\spdol}
\Text(-10,40)[]{\iena}
\Text(-10,0)[]{\idva}
\Text(90,0)[]{\itri}
\Text(90,40)[]{\isti}
\end{picture}

\propto
(D^T\underline AU)_{11,21}(U^T\tilde D^*)_{13}
(\tilde D^T\underline C\tilde N)_{33}(\tilde N^\dagger E)_{3i}
\end{displaymath}
\end{center}

\begin{center}
\begin{displaymath}
\def\trgor{$\tilde T$}
\def\trdol{$\tilde{\bar T}$}
\def\sfgor{$\tilde b$}
\def\sfdol{$\tilde t$}
\def\spgor{$\tilde w^-$}
\def\spdol{$\tilde w^+$}
\def\iena{$u$}
\def\idva{$e_i$}
\def\itri{$d_{1,2}$}
\def\isti{$u$}
\begin{picture}(80,40)(40,20)
\SetWidth{0.8}
\ArrowLine(0,40)(20,40)
\ArrowLine(0,0)(20,0)
\ArrowLine(80,40)(60,40)
\ArrowLine(80,0)(60,0)
\DashLine(20,40)(60,40)3
\DashLine(20,0)(60,0)3
\ArrowLine(20,20)(20,40)
\ArrowLine(20,20)(20,0)
\Line(17,17)(23,23)
\Line(17,23)(23,17)
\ArrowLine(60,20)(60,40)
\ArrowLine(60,20)(60,0)
\Line(57,17)(63,23)
\Line(57,23)(63,17)
\Text(10,30)[]{\trgor}
\Text(10,10)[]{\trdol}
\Text(40,30)[]{\sfgor}
\Text(40,10)[]{\sfdol}
\Text(75,30)[]{\spgor}
\Text(75,10)[]{\spdol}
\Text(-10,40)[]{\iena}
\Text(-10,0)[]{\idva}
\Text(90,0)[]{\itri}
\Text(90,40)[]{\isti}
\end{picture}

\propto
(U^T\underline A\tilde D)_{13}(\tilde D^\dagger U)_{31}
(D^T\tilde U^*)_{13,23}(\tilde U^T\underline CE)_{3i}
\end{displaymath}
\end{center}

\begin{center}
\begin{displaymath}
\def\tr{$\bar T$}
\def\sfgor{$\tilde b$}
\def\sfdol{$\tilde t$}
\def\spgor{$\tilde w^-$}
\def\spdol{$\tilde w^+$}
\def\iena{$u$}
\def\idva{$e_i$}
\def\itri{$d_{1,2}$}
\def\isti{$u$}
\begin{picture}(80,40)(40,20)
\SetWidth{0.8}
\ArrowLine(0,40)(10,20)
\ArrowLine(0,0)(10,20)
\ArrowLine(80,40)(60,40)
\ArrowLine(80,0)(60,0)
\DashLine(10,20)(40,20)3
\DashLine(40,20)(60,40)3
\DashLine(40,20)(60,0)3
\ArrowLine(60,20)(60,40)
\ArrowLine(60,20)(60,0)
\Line(57,17)(63,23)
\Line(57,23)(63,17)
\Text(25,30)[]{\tr}
\Text(45,35)[]{\sfgor}
\Text(45,5)[]{\sfdol}
\Text(75,30)[]{\spgor}
\Text(75,10)[]{\spdol}
\Text(-10,40)[]{\iena}
\Text(-10,0)[]{\idva}
\Text(90,0)[]{\itri}
\Text(90,40)[]{\isti}
\end{picture}

\propto
(U^T\underline CE)_{1i}(D^T\tilde U^*)_{13,23}
(\tilde U^T\underline A\tilde D)_{33}(\tilde D^\dagger U)_{31}
\end{displaymath}
\end{center}

\begin{center}
\begin{displaymath}
\def\trgor{$\tilde{\bar T}^\dagger$}
\def\trdol{$\tilde T^\dagger$}
\def\sfgor{$\tilde t^c$}
\def\sfdol{$\tilde b^c$}
\def\spgor{$\tilde h_+$}
\def\spdol{$\tilde{\bar h}_-$}
\def\iena{${\bar e}_i^c$}
\def\idva{${\bar u}^c$}
\def\itri{$u$}
\def\isti{$d_{1,2}$}
\begin{picture}(80,40)(40,20)
\SetWidth{0.8}
\ArrowLine(0,40)(20,40)
\ArrowLine(0,0)(20,0)
\ArrowLine(80,40)(60,40)
\ArrowLine(80,0)(60,0)
\DashLine(20,40)(60,40)3
\DashLine(20,0)(60,0)3
\ArrowLine(20,20)(20,40)
\ArrowLine(20,20)(20,0)
\Line(17,17)(23,23)
\Line(17,23)(23,17)
\ArrowLine(60,20)(60,40)
\ArrowLine(60,20)(60,0)
\Line(57,17)(63,23)
\Line(57,23)(63,17)
\Text(10,30)[]{\trgor}
\Text(10,10)[]{\trdol}
\Text(40,30)[]{\sfgor}
\Text(40,10)[]{\sfdol}
\Text(75,30)[]{\spgor}
\Text(75,10)[]{\spdol}
\Text(-10,40)[]{\iena}
\Text(-10,0)[]{\idva}
\Text(90,0)[]{\itri}
\Text(90,40)[]{\isti}
\end{picture}

\propto
(E_c^\dagger\underline B^\dagger\tilde U_c^*)_{i3}
(\tilde U_c^TY_U^TD)_{31,32}(U^TY_D\tilde D_c)_{13}
(\tilde D_c^\dagger\underline D^\dagger U_c^*)_{31}
\end{displaymath}
\end{center}

\begin{center}
\begin{displaymath}
\def\tr{$T$}
\def\sfgor{$\tilde t^c$}
\def\sfdol{$\tilde b^c$}
\def\spgor{$\tilde h_+$}
\def\spdol{$\tilde{\bar h}_-$}
\def\iena{${\bar e}_i^c$}
\def\idva{${\bar u}^c$}
\def\itri{$u$}
\def\isti{$d_{1,2}$}
\begin{picture}(80,40)(40,20)
\SetWidth{0.8}
\ArrowLine(0,40)(10,20)
\ArrowLine(0,0)(10,20)
\ArrowLine(80,40)(60,40)
\ArrowLine(80,0)(60,0)
\DashLine(10,20)(40,20)3
\DashLine(40,20)(60,40)3
\DashLine(40,20)(60,0)3
\ArrowLine(60,20)(60,40)
\ArrowLine(60,20)(60,0)
\Line(57,17)(63,23)
\Line(57,23)(63,17)
\Text(25,30)[]{\tr}
\Text(45,35)[]{\sfgor}
\Text(45,5)[]{\sfdol}
\Text(75,30)[]{\spgor}
\Text(75,10)[]{\spdol}
\Text(-10,40)[]{\iena}
\Text(-10,0)[]{\idva}
\Text(90,0)[]{\itri}
\Text(90,40)[]{\isti}
\end{picture}

\propto
(E_c^\dagger\underline B^\dagger U_c^*)_{i1}(U^T Y_D\tilde D_c)_{13}
(\tilde D_c^\dagger\underline D^\dagger\tilde U_c^*)_{33}
(\tilde U_c^TY_U^TD)_{31,32}
\end{displaymath}
\end{center}

\begin{center}
\begin{displaymath}
\def\trgor{$\tilde T$}
\def\trdol{$\tilde{\bar T}$}
\def\sfgor{$\tilde b$}
\def\sfdol{$\tilde t$}
\def\spgor{$\tilde h_+^\dagger$}
\def\spdol{$\tilde{\bar h}_-^\dagger$}
\def\iena{$u$}
\def\idva{$e_i$}
\def\itri{${\bar d}_{1,2}^c$}
\def\isti{${\bar u}^c$}
\begin{picture}(80,40)(40,20)
\SetWidth{0.8}
\ArrowLine(0,40)(20,40)
\ArrowLine(0,0)(20,0)
\ArrowLine(80,40)(60,40)
\ArrowLine(80,0)(60,0)
\DashLine(20,40)(60,40)3
\DashLine(20,0)(60,0)3
\ArrowLine(20,20)(20,40)
\ArrowLine(20,20)(20,0)
\Line(17,17)(23,23)
\Line(17,23)(23,17)
\ArrowLine(60,20)(60,40)
\ArrowLine(60,20)(60,0)
\Line(57,17)(63,23)
\Line(57,23)(63,17)
\Text(10,30)[]{\trgor}
\Text(10,10)[]{\trdol}
\Text(40,30)[]{\sfgor}
\Text(40,10)[]{\sfdol}
\Text(75,30)[]{\spgor}
\Text(75,10)[]{\spdol}
\Text(-10,40)[]{\iena}
\Text(-10,0)[]{\idva}
\Text(90,0)[]{\itri}
\Text(90,40)[]{\isti}
\end{picture}

\propto
(U^T\underline A\tilde D)_{13}(\tilde D^\dagger Y_U^*U_c^*)_{31}
(D_c^\dagger Y_D^\dagger\tilde U^*)_{13,23}(\tilde U^T\underline CE)_{3i}
\end{displaymath}
\end{center}

\begin{center}
\begin{displaymath}
\def\tr{$\bar T$}
\def\sfgor{$\tilde b$}
\def\sfdol{$\tilde t$}
\def\spgor{$\tilde h_+^\dagger$}
\def\spdol{$\tilde{\bar h}_-^\dagger$}
\def\iena{$u$}
\def\idva{$e_i$}
\def\itri{${\bar d}_{1,2}^c$}
\def\isti{${\bar u}^c$}
\begin{picture}(80,40)(40,20)
\SetWidth{0.8}
\ArrowLine(0,40)(10,20)
\ArrowLine(0,0)(10,20)
\ArrowLine(80,40)(60,40)
\ArrowLine(80,0)(60,0)
\DashLine(10,20)(40,20)3
\DashLine(40,20)(60,40)3
\DashLine(40,20)(60,0)3
\ArrowLine(60,20)(60,40)
\ArrowLine(60,20)(60,0)
\Line(57,17)(63,23)
\Line(57,23)(63,17)
\Text(25,30)[]{\tr}
\Text(45,35)[]{\sfgor}
\Text(45,5)[]{\sfdol}
\Text(75,30)[]{\spgor}
\Text(75,10)[]{\spdol}
\Text(-10,40)[]{\iena}
\Text(-10,0)[]{\idva}
\Text(90,0)[]{\itri}
\Text(90,40)[]{\isti}
\end{picture}

\propto
(U^T\underline CE)_{1i}(D_c^\dagger Y_D^\dagger\tilde U^*)_{13,23}
(\tilde U^T\underline A\tilde D)_{33}(\tilde D^\dagger Y_U^*U_c^*)_{31}
\end{displaymath}
\end{center}

\begin{center}
\begin{displaymath}
\def\trgor{$\tilde{\bar T}$}
\def\trdol{$\tilde T$}
\def\sfgor{$\tilde\nu$}
\def\sfdol{$\tilde b$}
\def\spgor{$\tilde{\bar h}_-^\dagger$}
\def\spdol{$\tilde h_+^\dagger$}
\def\iena{$d_{1,2}$}
\def\idva{$u$}
\def\itri{${\bar u}^c$}
\def\isti{${\bar e}_i^c$}
\begin{picture}(80,40)(40,20)
\SetWidth{0.8}
\ArrowLine(0,40)(20,40)
\ArrowLine(0,0)(20,0)
\ArrowLine(80,40)(60,40)
\ArrowLine(80,0)(60,0)
\DashLine(20,40)(60,40)3
\DashLine(20,0)(60,0)3
\ArrowLine(20,20)(20,40)
\ArrowLine(20,20)(20,0)
\Line(17,17)(23,23)
\Line(17,23)(23,17)
\ArrowLine(60,20)(60,40)
\ArrowLine(60,20)(60,0)
\Line(57,17)(63,23)
\Line(57,23)(63,17)
\Text(10,30)[]{\trgor}
\Text(10,10)[]{\trdol}
\Text(40,30)[]{\sfgor}
\Text(40,10)[]{\sfdol}
\Text(75,30)[]{\spgor}
\Text(75,10)[]{\spdol}
\Text(-10,40)[]{\iena}
\Text(-10,0)[]{\idva}
\Text(90,0)[]{\itri}
\Text(90,40)[]{\isti}
\end{picture}

\propto
(D^T\underline C\tilde N)_{13,23}(\tilde N^\dagger Y_E^\dagger E_c^*)_{3i}
(U_c^\dagger Y_U^\dagger\tilde D^*)_{13}(\tilde D^T\underline AU)_{31}
\end{displaymath}
\end{center}

\begin{center}
\begin{displaymath}
\def\tr{$T$}
\def\sfgor{$\tilde\nu$}
\def\sfdol{$\tilde b$}
\def\spgor{$\tilde{\bar h}_-^\dagger$}
\def\spdol{$\tilde h_+^\dagger$}
\def\iena{$d_{1,2}$}
\def\idva{$u$}
\def\itri{${\bar u}^c$}
\def\isti{${\bar e}_i^c$}
\begin{picture}(80,40)(40,20)
\SetWidth{0.8}
\ArrowLine(0,40)(10,20)
\ArrowLine(0,0)(10,20)
\ArrowLine(80,40)(60,40)
\ArrowLine(80,0)(60,0)
\DashLine(10,20)(40,20)3
\DashLine(40,20)(60,40)3
\DashLine(40,20)(60,0)3
\ArrowLine(60,20)(60,40)
\ArrowLine(60,20)(60,0)
\Line(57,17)(63,23)
\Line(57,23)(63,17)
\Text(25,30)[]{\tr}
\Text(45,35)[]{\sfgor}
\Text(45,5)[]{\sfdol}
\Text(75,30)[]{\spgor}
\Text(75,10)[]{\spdol}
\Text(-10,40)[]{\iena}
\Text(-10,0)[]{\idva}
\Text(90,0)[]{\itri}
\Text(90,40)[]{\isti}
\end{picture}

\propto
(D^T\underline AU)_{11,21}(U_c^\dagger Y_U^\dagger\tilde
D^*)_{13}(\tilde D^T\underline C\tilde N)_{33}
(\tilde N^\dagger Y_E^\dagger E_c^*)_{3i}
\end{displaymath}
\end{center}

\begin{center}
\begin{displaymath}
\def\trgor{$\tilde T$}
\def\trdol{$\tilde{\bar T}$}
\def\sfgor{$\tilde b$}
\def\sfdol{$\tilde t$}
\def\spgor{$\tilde{\bar h}_0^\dagger$}
\def\spdol{$\tilde h_0^\dagger$}
\def\iena{$u$}
\def\idva{$e_i$}
\def\itri{${\bar u}^c$}
\def\isti{${\bar d}_{1,2}^c$}
\begin{picture}(80,40)(40,20)
\SetWidth{0.8}
\ArrowLine(0,40)(20,40)
\ArrowLine(0,0)(20,0)
\ArrowLine(80,40)(60,40)
\ArrowLine(80,0)(60,0)
\DashLine(20,40)(60,40)3
\DashLine(20,0)(60,0)3
\ArrowLine(20,20)(20,40)
\ArrowLine(20,20)(20,0)
\Line(17,17)(23,23)
\Line(17,23)(23,17)
\ArrowLine(60,20)(60,40)
\ArrowLine(60,20)(60,0)
\Line(57,17)(63,23)
\Line(57,23)(63,17)
\Text(10,30)[]{\trgor}
\Text(10,10)[]{\trdol}
\Text(40,30)[]{\sfgor}
\Text(40,10)[]{\sfdol}
\Text(75,30)[]{\spgor}
\Text(75,10)[]{\spdol}
\Text(-10,40)[]{\iena}
\Text(-10,0)[]{\idva}
\Text(90,0)[]{\itri}
\Text(90,40)[]{\isti}
\end{picture}

\propto
(U^T\underline A\tilde D)_{13}(\tilde D^\dagger Y_D^* D_c^*)_{31,32}
(U_c^\dagger Y_U^\dagger\tilde U^*)_{13}(\tilde U^T\underline CE)_{3i}
\end{displaymath}
\end{center}

\begin{center}
\begin{displaymath}
\def\tr{$\bar T$}
\def\sfgor{$\tilde t$}
\def\sfdol{$\tilde b$}
\def\spgor{$\tilde h_0^\dagger$}
\def\spdol{$\tilde{\bar h}_0^\dagger$}
\def\iena{$u$}
\def\idva{$e_i$}
\def\itri{${\bar d}_{1,2}^c$}
\def\isti{${\bar u}^c$}
\begin{picture}(80,40)(40,20)
\SetWidth{0.8}
\ArrowLine(0,40)(10,20)
\ArrowLine(0,0)(10,20)
\ArrowLine(80,40)(60,40)
\ArrowLine(80,0)(60,0)
\DashLine(10,20)(40,20)3
\DashLine(40,20)(60,40)3
\DashLine(40,20)(60,0)3
\ArrowLine(60,20)(60,40)
\ArrowLine(60,20)(60,0)
\Line(57,17)(63,23)
\Line(57,23)(63,17)
\Text(25,30)[]{\tr}
\Text(45,35)[]{\sfgor}
\Text(45,5)[]{\sfdol}
\Text(75,30)[]{\spgor}
\Text(75,10)[]{\spdol}
\Text(-10,40)[]{\iena}
\Text(-10,0)[]{\idva}
\Text(90,0)[]{\itri}
\Text(90,40)[]{\isti}
\end{picture}

\propto
(U^T\underline CE)_{1i}(D_c^\dagger Y_D^\dagger\tilde D^*)_{13,23}
(\tilde D^T\underline A\tilde U)_{33}(\tilde U^\dagger Y_U^*U_c^*)_{31}
\end{displaymath}
\end{center}

\begin{center}
\begin{displaymath}
\def\trgor{$\tilde{\bar T}$}
\def\trdol{$\tilde T$}
\def\sfgor{$\tilde t^c$}
\def\sfdol{$\tilde\tau^c$}
\def\spgor{$\tilde h_0$}
\def\spdol{$\tilde{\bar h}_0$}
\def\iena{${\bar d}_{1,2}^c$}
\def\idva{${\bar u}^c$}
\def\itri{${\bar e}_i$}
\def\isti{$u$}
\begin{picture}(80,40)(40,20)
\SetWidth{0.8}
\ArrowLine(0,40)(20,40)
\ArrowLine(0,0)(20,0)
\ArrowLine(80,40)(60,40)
\ArrowLine(80,0)(60,0)
\DashLine(20,40)(60,40)3
\DashLine(20,0)(60,0)3
\ArrowLine(20,20)(20,40)
\ArrowLine(20,20)(20,0)
\Line(17,17)(23,23)
\Line(17,23)(23,17)
\ArrowLine(60,20)(60,40)
\ArrowLine(60,20)(60,0)
\Line(57,17)(63,23)
\Line(57,23)(63,17)
\Text(10,30)[]{\trgor}
\Text(10,10)[]{\trdol}
\Text(40,30)[]{\sfgor}
\Text(40,10)[]{\sfdol}
\Text(75,30)[]{\spgor}
\Text(75,10)[]{\spdol}
\Text(-10,40)[]{\iena}
\Text(-10,0)[]{\idva}
\Text(90,0)[]{\itri}
\Text(90,40)[]{\isti}
\end{picture}

\propto
(D_c^\dagger\underline D^\dagger\tilde U_c^*)_{13,23}
(\tilde U_c^TY_U^TU)_{31}(E^TY_E^T\tilde E_c)_{i3}
(\tilde E_c^\dagger\underline B^\dagger U_c^*)_{31}
\end{displaymath}
\end{center}

\begin{center}
\begin{displaymath}
\def\tr{$\bar T$}
\def\sfgor{$\tilde t^c$}
\def\sfdol{$\tilde\tau^c$}
\def\spgor{$\tilde h_0$}
\def\spdol{$\tilde{\bar h}_0$}
\def\iena{${\bar d}_{1,2}^c$}
\def\idva{${\bar u}^c$}
\def\itri{${\bar e}_i$}
\def\isti{$u$}
\begin{picture}(80,40)(40,20)
\SetWidth{0.8}
\ArrowLine(0,40)(10,20)
\ArrowLine(0,0)(10,20)
\ArrowLine(80,40)(60,40)
\ArrowLine(80,0)(60,0)
\DashLine(10,20)(40,20)3
\DashLine(40,20)(60,40)3
\DashLine(40,20)(60,0)3
\ArrowLine(60,20)(60,40)
\ArrowLine(60,20)(60,0)
\Line(57,17)(63,23)
\Line(57,23)(63,17)
\Text(25,30)[]{\tr}
\Text(45,35)[]{\sfgor}
\Text(45,5)[]{\sfdol}
\Text(75,30)[]{\spgor}
\Text(75,10)[]{\spdol}
\Text(-10,40)[]{\iena}
\Text(-10,0)[]{\idva}
\Text(90,0)[]{\itri}
\Text(90,40)[]{\isti}
\end{picture}

\propto
(D_c^\dagger\underline D^\dagger U_c^*)_{11,21}
(E^TY_E^T\tilde E_c)_{i3}(\tilde E_c^\dagger\underline B^\dagger
\tilde U_c^*)_{33}(\tilde U_c^TY_U^TU)_{31}
\end{displaymath}
\end{center}

\begin{center}
\begin{displaymath}
\def\trgor{$\tilde{\bar T}$}
\def\trdol{$\tilde T$}
\def\sfgor{$\tilde b^c$}
\def\sfdol{$\tilde t^c$}
\def\spgor{$\tilde{\bar h}_0$}
\def\spdol{$\tilde h_0$}
\def\iena{${\bar u}^c$}
\def\idva{${\bar e}_i^c$}
\def\itri{$u$}
\def\isti{$d_{1,2}$}
\begin{picture}(80,40)(40,20)
\SetWidth{0.8}
\ArrowLine(0,40)(20,40)
\ArrowLine(0,0)(20,0)
\ArrowLine(80,40)(60,40)
\ArrowLine(80,0)(60,0)
\DashLine(20,40)(60,40)3
\DashLine(20,0)(60,0)3
\ArrowLine(20,20)(20,40)
\ArrowLine(20,20)(20,0)
\Line(17,17)(23,23)
\Line(17,23)(23,17)
\ArrowLine(60,20)(60,40)
\ArrowLine(60,20)(60,0)
\Line(57,17)(63,23)
\Line(57,23)(63,17)
\Text(10,30)[]{\trgor}
\Text(10,10)[]{\trdol}
\Text(40,30)[]{\sfgor}
\Text(40,10)[]{\sfdol}
\Text(75,30)[]{\spgor}
\Text(75,10)[]{\spdol}
\Text(-10,40)[]{\iena}
\Text(-10,0)[]{\idva}
\Text(90,0)[]{\itri}
\Text(90,40)[]{\isti}
\end{picture}

\propto
(U_c^\dagger\underline D^*\tilde D_c^*)_{13}
(\tilde D_c^TY_D^TD)_{31,32}(U^TY_U\tilde U_c)_{13}
(\tilde U_c^\dagger\underline B^* E_c^*)_{3i}
\end{displaymath}
\end{center}

\begin{center}
\begin{displaymath}
\def\tr{$T$}
\def\sfgor{$\tilde t^c$}
\def\sfdol{$\tilde b^c$}
\def\spgor{$\tilde h_0$}
\def\spdol{$\tilde{\bar h}_0$}
\def\iena{${\bar e}_i^c$}
\def\idva{${\bar u}^c$}
\def\itri{$d_{1,2}$}
\def\isti{$u$}
\begin{picture}(80,40)(40,20)
\SetWidth{0.8}
\ArrowLine(0,40)(10,20)
\ArrowLine(0,0)(10,20)
\ArrowLine(80,40)(60,40)
\ArrowLine(80,0)(60,0)
\DashLine(10,20)(40,20)3
\DashLine(40,20)(60,40)3
\DashLine(40,20)(60,0)3
\ArrowLine(60,20)(60,40)
\ArrowLine(60,20)(60,0)
\Line(57,17)(63,23)
\Line(57,23)(63,17)
\Text(25,30)[]{\tr}
\Text(45,35)[]{\sfgor}
\Text(45,5)[]{\sfdol}
\Text(75,30)[]{\spgor}
\Text(75,10)[]{\spdol}
\Text(-10,40)[]{\iena}
\Text(-10,0)[]{\idva}
\Text(90,0)[]{\itri}
\Text(90,40)[]{\isti}
\end{picture}

\propto
(E_c^\dagger\underline B^\dagger U_c^*)_{i1}(D^TY_D\tilde D_c)_{13,23}
(\tilde D_c^\dagger\underline D^\dagger\tilde U_c^*)_{33}
(\tilde U_c^TY_U^TU)_{31}
\end{displaymath}
\end{center}

\begin{center}
\begin{displaymath}
\def\trgor{$\tilde{\bar T}$}
\def\trdol{$\tilde T$}
\def\sfgor{$\tilde t$}
\def\sfdol{$\tilde\tau$}
\def\spgor{$\tilde h_0^\dagger$}
\def\spdol{$\tilde{\bar h}_0^\dagger$}
\def\iena{$d_{1,2}$}
\def\idva{$u$}
\def\itri{${\bar e}_i^c$}
\def\isti{${\bar u}^c$}
\begin{picture}(80,40)(40,20)
\SetWidth{0.8}
\ArrowLine(0,40)(20,40)
\ArrowLine(0,0)(20,0)
\ArrowLine(80,40)(60,40)
\ArrowLine(80,0)(60,0)
\DashLine(20,40)(60,40)3
\DashLine(20,0)(60,0)3
\ArrowLine(20,20)(20,40)
\ArrowLine(20,20)(20,0)
\Line(17,17)(23,23)
\Line(17,23)(23,17)
\ArrowLine(60,20)(60,40)
\ArrowLine(60,20)(60,0)
\Line(57,17)(63,23)
\Line(57,23)(63,17)
\Text(10,30)[]{\trgor}
\Text(10,10)[]{\trdol}
\Text(40,30)[]{\sfgor}
\Text(40,10)[]{\sfdol}
\Text(75,30)[]{\spgor}
\Text(75,10)[]{\spdol}
\Text(-10,40)[]{\iena}
\Text(-10,0)[]{\idva}
\Text(90,0)[]{\itri}
\Text(90,40)[]{\isti}
\end{picture}

\propto
(D^T\underline A\tilde U)_{13,23}
(\tilde U^\dagger Y_U^*U_c^*)_{31}(E_c^\dagger Y_E^*\tilde E^*)_{i3}
(\tilde E^T\underline C^T U)_{31}
\end{displaymath}
\end{center}

\begin{center}
\begin{displaymath}
\def\tr{$T$}
\def\sfgor{$\tilde\tau$}
\def\sfdol{$\tilde t$}
\def\spgor{$\tilde{\bar h}_0^\dagger$}
\def\spdol{$\tilde h_0^\dagger$}
\def\iena{$u$}
\def\idva{$d_{1,2}$}
\def\itri{${\bar u}^c$}
\def\isti{${\bar e}_i^c$}
\begin{picture}(80,40)(40,20)
\SetWidth{0.8}
\ArrowLine(0,40)(10,20)
\ArrowLine(0,0)(10,20)
\ArrowLine(80,40)(60,40)
\ArrowLine(80,0)(60,0)
\DashLine(10,20)(40,20)3
\DashLine(40,20)(60,40)3
\DashLine(40,20)(60,0)3
\ArrowLine(60,20)(60,40)
\ArrowLine(60,20)(60,0)
\Line(57,17)(63,23)
\Line(57,23)(63,17)
\Text(25,30)[]{\tr}
\Text(45,35)[]{\sfgor}
\Text(45,5)[]{\sfdol}
\Text(75,30)[]{\spgor}
\Text(75,10)[]{\spdol}
\Text(-10,40)[]{\iena}
\Text(-10,0)[]{\idva}
\Text(90,0)[]{\itri}
\Text(90,40)[]{\isti}
\end{picture}

\propto
(U^T\underline AD)_{11,12}
(U_c^\dagger Y_U^\dagger\tilde U^*)_{13}(\tilde U^T\underline
C\tilde E)_{33}(\tilde E^\dagger Y_E^\dagger E_c^*)_{3i}
\end{displaymath}
\end{center}

\begin{center}
\begin{displaymath}
\def\trgor{$\tilde T$}
\def\trdol{$\tilde{\bar T}$}
\def\sfgor{$\tilde b$}
\def\sfdol{$\tilde t$}
\def\spgor{$\tilde V_0$}
\def\spdol{$\tilde V_0$}
\def\iena{$u$}
\def\idva{$e_i$}
\def\itri{$u$}
\def\isti{$d_{1,2}$}
\begin{picture}(80,40)(40,20)
\SetWidth{0.8}
\ArrowLine(0,40)(20,40)
\ArrowLine(0,0)(20,0)
\ArrowLine(80,40)(60,40)
\ArrowLine(80,0)(60,0)
\DashLine(20,40)(60,40)3
\DashLine(20,0)(60,0)3
\ArrowLine(20,20)(20,40)
\ArrowLine(20,20)(20,0)
\Line(17,17)(23,23)
\Line(17,23)(23,17)
\ArrowLine(60,20)(60,40)
\ArrowLine(60,20)(60,0)
\Line(57,17)(63,23)
\Line(57,23)(63,17)
\Text(10,30)[]{\trgor}
\Text(10,10)[]{\trdol}
\Text(40,30)[]{\sfgor}
\Text(40,10)[]{\sfdol}
\Text(75,30)[]{\spgor}
\Text(75,10)[]{\spdol}
\Text(-10,40)[]{\iena}
\Text(-10,0)[]{\idva}
\Text(90,0)[]{\itri}
\Text(90,40)[]{\isti}
\end{picture}

\propto
(U^T\underline A\tilde D)_{13}(\tilde D^\dagger D)_{31,32}
(U^T\tilde U^*)_{13}(\tilde U^T\underline CE)_{3i}
\end{displaymath}
\end{center}

\begin{center}
\begin{displaymath}
\def\tr{$\bar T$}
\def\sfgor{$\tilde b$}
\def\sfdol{$\tilde t$}
\def\spgor{$\tilde V_0$}
\def\spdol{$\tilde V_0$}
\def\iena{$u$}
\def\idva{$e_i$}
\def\itri{$u$}
\def\isti{$d_{1,2}$}
\begin{picture}(80,40)(40,20)
\SetWidth{0.8}
\ArrowLine(0,40)(10,20)
\ArrowLine(0,0)(10,20)
\ArrowLine(80,40)(60,40)
\ArrowLine(80,0)(60,0)
\DashLine(10,20)(40,20)3
\DashLine(40,20)(60,40)3
\DashLine(40,20)(60,0)3
\ArrowLine(60,20)(60,40)
\ArrowLine(60,20)(60,0)
\Line(57,17)(63,23)
\Line(57,23)(63,17)
\Text(25,30)[]{\tr}
\Text(45,35)[]{\sfgor}
\Text(45,5)[]{\sfdol}
\Text(75,30)[]{\spgor}
\Text(75,10)[]{\spdol}
\Text(-10,40)[]{\iena}
\Text(-10,0)[]{\idva}
\Text(90,0)[]{\itri}
\Text(90,40)[]{\isti}
\end{picture}

\propto
(U^T\underline CE)_{1i}(U^T\tilde U^*)_{13}
(\tilde U^T\underline A\tilde D)_{33}(\tilde D^\dagger D)_{31,32}
\end{displaymath}
\end{center}

\begin{center}
\begin{displaymath}
\def\trgor{$\tilde{\bar T}$}
\def\trdol{$\tilde T$}
\def\sfgor{$\tilde\tau$}
\def\sfdol{$\tilde t$}
\def\spgor{$\tilde V_0$}
\def\spdol{$\tilde V_0$}
\def\iena{$u$}
\def\idva{$d_{1,2}$}
\def\itri{$u$}
\def\isti{$e_i$}
\begin{picture}(80,40)(40,20)
\SetWidth{0.8}
\ArrowLine(0,40)(20,40)
\ArrowLine(0,0)(20,0)
\ArrowLine(80,40)(60,40)
\ArrowLine(80,0)(60,0)
\DashLine(20,40)(60,40)3
\DashLine(20,0)(60,0)3
\ArrowLine(20,20)(20,40)
\ArrowLine(20,20)(20,0)
\Line(17,17)(23,23)
\Line(17,23)(23,17)
\ArrowLine(60,20)(60,40)
\ArrowLine(60,20)(60,0)
\Line(57,17)(63,23)
\Line(57,23)(63,17)
\Text(10,30)[]{\trgor}
\Text(10,10)[]{\trdol}
\Text(40,30)[]{\sfgor}
\Text(40,10)[]{\sfdol}
\Text(75,30)[]{\spgor}
\Text(75,10)[]{\spdol}
\Text(-10,40)[]{\iena}
\Text(-10,0)[]{\idva}
\Text(90,0)[]{\itri}
\Text(90,40)[]{\isti}
\end{picture}

\propto
(U^T\underline C\tilde E)_{13}(\tilde E^\dagger E)_{3i}
(U^T\tilde U^*)_{13}(\tilde U^T\underline AD)_{31,32}
\end{displaymath}
\end{center}

\begin{center}
\begin{displaymath}
\def\tr{$T$}
\def\sfgor{$\tilde\tau$}
\def\sfdol{$\tilde t$}
\def\spgor{$\tilde V_0$}
\def\spdol{$\tilde V_0$}
\def\iena{$u$}
\def\idva{$d_{1,2}$}
\def\itri{$u$}
\def\isti{$e_i$}
\begin{picture}(80,40)(40,20)
\SetWidth{0.8}
\ArrowLine(0,40)(10,20)
\ArrowLine(0,0)(10,20)
\ArrowLine(80,40)(60,40)
\ArrowLine(80,0)(60,0)
\DashLine(10,20)(40,20)3
\DashLine(40,20)(60,40)3
\DashLine(40,20)(60,0)3
\ArrowLine(60,20)(60,40)
\ArrowLine(60,20)(60,0)
\Line(57,17)(63,23)
\Line(57,23)(63,17)
\Text(25,30)[]{\tr}
\Text(45,35)[]{\sfgor}
\Text(45,5)[]{\sfdol}
\Text(75,30)[]{\spgor}
\Text(75,10)[]{\spdol}
\Text(-10,40)[]{\iena}
\Text(-10,0)[]{\idva}
\Text(90,0)[]{\itri}
\Text(90,40)[]{\isti}
\end{picture}

\propto
(U^T\underline AD)_{11,12}(U^T\tilde U^*)_{13}
(\tilde U^T\underline C\tilde E)_{33}(\tilde E^\dagger E)_{3i}
\end{displaymath}
\end{center}

\begin{center}
\begin{displaymath}
\def\trgor{$\tilde{\bar T}^\dagger$}
\def\trdol{$\tilde T^\dagger$}
\def\sfgor{$\tilde b^c$}
\def\sfdol{$\tilde t^c$}
\def\spgor{$\tilde V_0^\dagger$}
\def\spdol{$\tilde V_0^\dagger$}
\def\iena{${\bar u}^c$}
\def\idva{${\bar e}_i^c$}
\def\itri{${\bar u}^c$}
\def\isti{${\bar d}_{1,2}^c$}
\begin{picture}(80,40)(40,20)
\SetWidth{0.8}
\ArrowLine(0,40)(20,40)
\ArrowLine(0,0)(20,0)
\ArrowLine(80,40)(60,40)
\ArrowLine(80,0)(60,0)
\DashLine(20,40)(60,40)3
\DashLine(20,0)(60,0)3
\ArrowLine(20,20)(20,40)
\ArrowLine(20,20)(20,0)
\Line(17,17)(23,23)
\Line(17,23)(23,17)
\ArrowLine(60,20)(60,40)
\ArrowLine(60,20)(60,0)
\Line(57,17)(63,23)
\Line(57,23)(63,17)
\Text(10,30)[]{\trgor}
\Text(10,10)[]{\trdol}
\Text(40,30)[]{\sfgor}
\Text(40,10)[]{\sfdol}
\Text(75,30)[]{\spgor}
\Text(75,10)[]{\spdol}
\Text(-10,40)[]{\iena}
\Text(-10,0)[]{\idva}
\Text(90,0)[]{\itri}
\Text(90,40)[]{\isti}
\end{picture}

\propto
(U_c^\dagger\underline D^*\tilde D_c^*)_{13}(\tilde D_c^T D_c^*)_{31,32}
(U_c^\dagger\tilde U_c)_{13}(\tilde U_c^\dagger\underline B^*E_c^*)_{3i}
\end{displaymath}
\end{center}

\begin{center}
\begin{displaymath}
\def\tr{$T$}
\def\sfgor{$\tilde b^c$}
\def\sfdol{$\tilde t^c$}
\def\spgor{$\tilde V_0^\dagger$}
\def\spdol{$\tilde V_0^\dagger$}
\def\iena{${\bar u}^c$}
\def\idva{${\bar e}_i^c$}
\def\itri{${\bar u}^c$}
\def\isti{${\bar d}_{1,2}^c$}
\begin{picture}(80,40)(40,20)
\SetWidth{0.8}
\ArrowLine(0,40)(10,20)
\ArrowLine(0,0)(10,20)
\ArrowLine(80,40)(60,40)
\ArrowLine(80,0)(60,0)
\DashLine(10,20)(40,20)3
\DashLine(40,20)(60,40)3
\DashLine(40,20)(60,0)3
\ArrowLine(60,20)(60,40)
\ArrowLine(60,20)(60,0)
\Line(57,17)(63,23)
\Line(57,23)(63,17)
\Text(25,30)[]{\tr}
\Text(45,35)[]{\sfgor}
\Text(45,5)[]{\sfdol}
\Text(75,30)[]{\spgor}
\Text(75,10)[]{\spdol}
\Text(-10,40)[]{\iena}
\Text(-10,0)[]{\idva}
\Text(90,0)[]{\itri}
\Text(90,40)[]{\isti}
\end{picture}

\propto
(U_c^\dagger\underline B^*E_c^*)_{1i}(U_c^\dagger\tilde U_c)_{13}
(\tilde U_c^\dagger\underline D^*\tilde D_c^*)_{33}
(\tilde D_c^TD_c^*)_{31,32}
\end{displaymath}
\end{center}

\begin{center}
\begin{displaymath}
\def\trgor{$\tilde T^\dagger$}
\def\trdol{$\tilde{\bar T}^\dagger$}
\def\sfgor{$\tilde\tau^c$}
\def\sfdol{$\tilde t^c$}
\def\spgor{$\tilde V_0^\dagger$}
\def\spdol{$\tilde V_0^\dagger$}
\def\iena{${\bar u}^c$}
\def\idva{${\bar d}_{1,2}^c$}
\def\itri{${\bar u}^c$}
\def\isti{${\bar e}_i^c$}
\begin{picture}(80,40)(40,20)
\SetWidth{0.8}
\ArrowLine(0,40)(20,40)
\ArrowLine(0,0)(20,0)
\ArrowLine(80,40)(60,40)
\ArrowLine(80,0)(60,0)
\DashLine(20,40)(60,40)3
\DashLine(20,0)(60,0)3
\ArrowLine(20,20)(20,40)
\ArrowLine(20,20)(20,0)
\Line(17,17)(23,23)
\Line(17,23)(23,17)
\ArrowLine(60,20)(60,40)
\ArrowLine(60,20)(60,0)
\Line(57,17)(63,23)
\Line(57,23)(63,17)
\Text(10,30)[]{\trgor}
\Text(10,10)[]{\trdol}
\Text(40,30)[]{\sfgor}
\Text(40,10)[]{\sfdol}
\Text(75,30)[]{\spgor}
\Text(75,10)[]{\spdol}
\Text(-10,40)[]{\iena}
\Text(-10,0)[]{\idva}
\Text(90,0)[]{\itri}
\Text(90,40)[]{\isti}
\end{picture}

\propto
(U_c^\dagger\underline B^*\tilde E_c^*)_{13}(\tilde E_c^TE_c^*)_{3i}
(U_c^\dagger\tilde U_c)_{13}
(\tilde U_c^\dagger\underline D^*D_c^*)_{31,32}
\end{displaymath}
\end{center}
\pagestyle{empty}
\begin{center}
\begin{displaymath}
\def\tr{$\bar T$}
\def\sfgor{$\tilde\tau^c$}
\def\sfdol{$\tilde t^c$}
\def\spgor{$\tilde V_0^\dagger$}
\def\spdol{$\tilde V_0^\dagger$}
\def\iena{${\bar u}^c$}
\def\idva{${\bar d}_{1,2}^c$}
\def\itri{${\bar u}^c$}
\def\isti{${\bar e}_i^c$}
\begin{picture}(80,40)(40,20)
\SetWidth{0.8}
\ArrowLine(0,40)(10,20)
\ArrowLine(0,0)(10,20)
\ArrowLine(80,40)(60,40)
\ArrowLine(80,0)(60,0)
\DashLine(10,20)(40,20)3
\DashLine(40,20)(60,40)3
\DashLine(40,20)(60,0)3
\ArrowLine(60,20)(60,40)
\ArrowLine(60,20)(60,0)
\Line(57,17)(63,23)
\Line(57,23)(63,17)
\Text(25,30)[]{\tr}
\Text(45,35)[]{\sfgor}
\Text(45,5)[]{\sfdol}
\Text(75,30)[]{\spgor}
\Text(75,10)[]{\spdol}
\Text(-10,40)[]{\iena}
\Text(-10,0)[]{\idva}
\Text(90,0)[]{\itri}
\Text(90,40)[]{\isti}
\end{picture}

\propto
(U_c^\dagger\underline D^*\tilde D_c^*)_{11,12}
(U_c^\dagger\tilde U_c)_{13}
(\tilde U_c^\dagger\underline B^*\tilde E_c^*)_{33}
(\tilde E_c^T E_c^*)_{3i}
\end{displaymath}
\end{center}
\end{appendix}
\newpage

\pagestyle{empty}
\newpage

\bibliographystyle{plain}
\pagestyle{headings}

\end{document}